\documentclass[11pt]{article}

\usepackage{amsmath,amssymb}
\usepackage[table]{xcolor}% http://ctan.org/pkg/xcolor

\usepackage{url}
\usepackage{graphicx}
\usepackage{subfigure}
\usepackage{color}
\usepackage{cite}
\usepackage{epsfig}
\usepackage{amssymb}
\usepackage{latexsym}
\newcommand\given[1][]{\:#1\vert\:}
\usepackage{mathtools}

\newcommand {\C} {{\rm I\kern-5.5pt C}}

\usepackage{cite}

\usepackage{tcolorbox}

\allowdisplaybreaks
\begin{document}

%\preprint{APS/123-QED}

\title{The Effects of Evolutionary Adaptations on Spreading Processes in Complex Networks}% Force line breaks with \\

\author{
Rashad~Eletreby${}^1$, Yong~Zhuang${}^1$, Kathleen~M.~Carley${}^2$, Osman~Ya\u{g}an${}^1$, and H.~Vincent~Poor${}^3$
         \\
         \\
${}^1$Department of Electrical and Computer Engineering\\
Carnegie Mellon University, Pittsburgh, PA 15213 USA\\ \\
${}^2$Institute for Software Research, School of Computer Science\\ Carnegie Mellon University, Pittsburgh, PA 15213 USA\\ \\
${}^3$Department of Electrical Engineering, \\Princeton University, Princeton, NJ 08544 USA
}
%
%\author{Rashad Eletreby}
%%\email{reletreby@cmu.edu}
%\affiliation{%
% Department of Electrical and Computer Engineering, Carnegie Mellon University \\Pittsburgh, PA 15213 USA
%}%
%\author{Yong Zhuang}%
%\affiliation{%
% Department of Electrical and Computer Engineering, Carnegie Mellon University \\Pittsburgh, PA 15213 USA
%}%
%\author{Kathleen M. Carley}
%\affiliation{%
% Institute for Software Research, School of Computer Science, Carnegie Mellon University, \\Pittsburgh, PA 15213 USA
%}%
%\author{Osman Ya\u{g}an}%
%\email{oyagan@ece.cmu.edu}
%\affiliation{%
% Department of Electrical and Computer Engineering, Carnegie Mellon University \\Pittsburgh, PA 15213 USA
% }%
% \author{H. Vincent Poor}%
%\affiliation{%
% Department of Electrical Engineering, Princeton University, Princeton, NJ 08544 USA
% }%

\date{\today}% It is always \today, today,
             %  but any date may be explicitly specified
\maketitle

\begin{abstract}
A common theme among the proposed models for {\em network epidemics} is the assumption that the propagating object, i.e., a pathogen (in the context of infectious disease propagation) or a piece of information (in the context of information propagation), is transferred across the nodes without going through any modification or {\em evolution}. However, in real-life spreading processes, pathogens often {\em evolve} in response to changing environments and medical interventions and information is often modified by individuals before being forwarded. In this paper, we investigate the {\em evolution} of spreading processes, such as infectious diseases or information, in complex networks with the aim of i) revealing the role of evolutionary adaptations on the threshold, probability, and final size of epidemics; and ii) exploring the interplay between the structural properties of the network and the process of evolution. We start by considering the case where {\em co-infection} with different pathogen strains (respectively, different variations of information) is not possible, i.e., a susceptible individual may only be infected with a {\em single} pathogen strain (respectively, a single variant of the information). In this case, we develop a mathematical theory that accurately predicts the epidemic threshold and the expected epidemic size as functions of the characteristics of the spreading process, the evolutionary pathways of the pathogen (respectively, information), and the structure of the underlying contact network. In addition to the mathematical theory, we perform extensive simulations on random and real-world contact networks to verify our theory and reveal the significant shortcomings of the classical mathematical models that do not capture evolution. Our results reveal that the classical, single-type bond-percolation models may accurately predict the threshold and final size of epidemics, but their predictions on the probability of emergence are {\em inaccurate} on both random and real-world networks. This inaccuracy sheds the light on a fundamental disconnect between the classical bond-percolation models and real-life spreading processes that entail evolution. Then, we consider the case when {\em co-infection} is possible, i.e., a susceptible individual who receives {\em simultaneous} infections with multiple pathogen strains (respectively, multiple variations of information) becomes co-infected. We show by computer simulations that co-infection gives rise to a rich set of  dynamics: it can amplify or inhibit the spreading dynamics, and more remarkably {\em lead the order of phase transition to change from second-order to first-order}. We investigate the delicate interplay between the characteristics of co-infection, the structure of the underlying contact network, and the evolutionary pathways of the pathogen (respectively, information) and reveal the cases where such interplay induces a {\em first-order} phase transition for the expected epidemic size.
\end{abstract}


\section{Introduction}
\label{sec:introduction}

What causes an outbreak of a disease? How can we predict its emergence and control its progression? Over the past several decades, multidisciplinary research efforts were converging to tackle the above questions, aiming for providing a better understanding of the intricate dynamics of disease propagation and accurate predictions on its course \cite{barabasi2016network,fraser2004factors, newman2002spread, lloyd2005superspreading, anderson1992infectious, pastor2001epidemic, pastor2001epidemic, moreno2002epidemic, granell2014competing,morens2004challenge,wolfe2007origins, daszak1999emerging, wei1995viral}. At the heart of these research efforts is the development of mathematical models that provide insights on predicting, assessing, and controlling potential outbreaks \cite{brauer2012mathematical, siettos2013mathematical, diekmann2000mathematical, keeling2011modeling}. The early mathematical models relied on the {\em homogeneous mixing} assumption, meaning that an infected individual is equally likely to infect any other individual in the population, without regard to her location, age, or the people with whom she interacts. Homogeneity allowed writing a set of differential equations that characterize the speed and scale of propagation (in the limit of large population size), providing insights on how the parameters of a disease, e.g., its {\em basic reproductive number}, indicate whether a disease will die out, or an epidemic will emerge \cite{anderson1992infectious,keeling2011modeling}.

In real-life, however, the spread of a disease is highly dependent on the contact patterns between individuals. In particular, a person may only infect those with whom she interacts, and the number of contacts people have, varies dramatically between individuals. These basic observations render the homogeneous mixing models inaccurate, as they tend to underestimate the epidemic size in the initial stages of the outbreak and overestimate it towards the end \cite{bansal2007individual}. As a result of the these shortcomings, {\em network epidemics} has emerged as a mathematical modeling approach that takes the underlying contact network into consideration \cite{keeling2005networks, newman2002spread, barabasi2016network, pastor2015epidemic, miller2014epidemic}. Since then, a large body of research has looked into the delicate interplay between the structural properties of the contact network and the dynamics of propagation, leading to accurate predictions of the spatio-temporal progression of disease outbreaks. In addition to diseases, opinions and information also propagate through networks in patterns similar to those of epidemics \cite{durrett2010some}. Hence, research efforts on {\em information propagation} draw on the theory of infectious diseases to model the dynamics of propagation \cite{zhuang2016information, yagan2013conjoining, huang2006information, moreno2004dynamics, granell2014competing}. Throughout, we use the term {\em spreading processes} to denote a general class of processes that propagate in contact networks, such as infectious diseases and information. 

%Although infectious disease propagation and information propagation do share similarities \cite{durrett2010some}, they are likely to have their own peculiar features, hence in Section~\ref{sec:realworld}, we 

A common theme among the proposed models for network epidemics is the assumption that the propagating object, i.e., a virus or a piece of information, is transferred across the nodes without going through any modification or {\em evolution} \cite{zhuang2016information, dodds2004universal, anderson1992infectious, sahneh2013generalized, yagan2013conjoining, YaganGligor, newman2002email, balthrop2004technological,qian2012diffusion}. However, in real-life spreading processes, pathogens often {\em evolve} in response to changing environments and medical interventions \cite{leventhal2015evolution,alexander2010risk, morens2004challenge, antia2003role, pfennig2001evolution}, and information is often modified by individuals before being forwarded \cite{Adamic2016, zhang2013rumor}. In fact, 60\% of the (approximately) 400 emerging infectious diseases that have been identified since 1940 are zoonotic \footnote{A zoonosis is any disease or infection that is naturally transmissible from vertebrate animals to humans \cite{WHO}.} \cite{MORSE20121956, jones2008global}. A zoonotic disease is initially poorly adapted, poorly replicated, and inefficiently transmitted \cite{parrish2008cross}, hence its ability to go from animal-to-human transmissions to human-to-human transmissions depends on the pathogen {\em evolving} to a strain that is well-adapted to the human host. For instance, genetic variations in some critical genes were reported to be essential for the transition from animal-to-human transmission to human-to-human transmission in the severe acute respiratory syndrome (SARS) outbreak of 2002-2003 \cite{Song2430}. 

Similar patterns of evolution are observed in the way information propagates among individuals. Needless to say, one observes, on a daily basis, how information mutates unintentionally, or perhaps intentionally by an adversary, on social media platforms \cite{Adamic2016}. At a high-level, an individual may mutate the information by exaggeration, hoping for her variant to go viral. Mutations may also occur unintentionally. In particular, Dawkins \cite{dawkins2016selfish} argued that ideas and information spread and evolve between individuals with patterns similar to genes, in a sense that they self-replicate, mutate, and respond to selective pressure as they interact with their host. Concluding, if we are to ignore evolution, we underestimate the severity of the epidemic and fail to understand the intricate interplay between the dynamics of propagation and evolution. 

In this paper, we aim to bridge the disconnect between how spreading processes propagate {\em and evolve} in real-life, and the current mathematical and simulation models that do not capture evolution. In particular, we investigate the {\em evolution} of spreading processes with the aim of i) revealing the role of evolutionary adaptations on the threshold, probability, and final size of epidemics; and ii) understanding the interplay between the structural properties of the network and the process of evolution. Throughout, we use the term {\em epidemics} to denote disease/information outbreaks that result in a positive fraction of infected individuals in the limit of large network size and {\em self-limited outbreaks} to denote small disease/information outbreaks for which the fraction of infected individuals tends to zero in the limit of large network size. { We also use the term {\em strain} to denote a pathogen strain in the context of infectious disease propagation, or a particular variation of the information in the context of information propagation. At a high level, strains represent homogeneous groups within species \cite{balmer2011prevalence} and they generally possess unique features such as virulence, infectivity, growth rate, etc.}

In modeling the evolution of spreading processes, we adopt the multiple-strain model that was introduced by Alexander and Day in \cite{alexander2010risk}. Their model can be briefly outlined as follows (more details are given in Section~\ref{sec:model}). Consider a multiple-strain spreading process that starts with an individual, i.e., the seed, receiving infection (from an external reservoir) with strain-$1$ of a particular pathogen { (respectively, information)}. The seed infects each of her contacts independently with probability $T_1$, called the {\em transmissibility} of strain-$1$. Once a susceptible individual receives the infection from the seed, the pathogen may evolve within that new host prior to any subsequent infections. In particular, the pathogen may remain as strain-$1$ with probability $\mu_{11}$ or mutate to strain-$2$ (that has transmissibility $T_2$) with probability $\mu_{12}=1-\mu_{11}$. If the pathogen remains as strain-$1$ (respectively, mutates to strain-$2$) { within a newly infected host, then that host} infects each of her susceptible neighbors in the subsequent stages independently with probability $T_1$ (respectively, $T_2$). { As the process continues to grow}, if any susceptible individual receives strain-$1$, the pathogen may remain as strain-$1$ with probability $\mu_{11}$ or mutate to strain-$2$ with probability $\mu_{12}=1-\mu_{11}$ prior to subsequent infections. Similarly, if any susceptible individual receives strain-$2$, the pathogen may remain as strain-$2$ with probability $\mu_{22}$ or mutate to strain-$1$ with probability $\mu_{21}=1-\mu_{22}$ prior to subsequent infections. The process continues to grow until no additional infections are possible. We remark that it is straightforward to extend the model to the general case, where there are $m$ possible strains for some finite integer $m \geq 2$. More details are given in Section~\ref{sec:analysis}.

{
Note that as multiple strains propagate throughout the population, a susceptible individual may simultaneously get into infectious contact with neighbors infected with strain-$1$ as well as neighbors infected with strain-$2$. This gives rise to the possibility of a susceptible individual becoming {\em co-infected} with multiple pathogen strains. Indeed, co-infection with multiple pathogen strains is prevalent in disease-causing protozoa, helminths, bacteria, fungi, and viruses and is known to cause significant implications \cite{susi2015co, balmer2011prevalence, read2001ecology, cohen2012mixed, alizon2013multiple}. However, from a mathematical standpoint, the possibility of co-infections creates phase discontinuities (see Section~\ref{sec:coinfection}) that render the process mathematically intractable.

We start by considering the case when co-infection is ignored, meaning that a susceptible individual may {\em only} be infected with a single strain. In particular, a susceptible individual who simultaneously receives $x$ infections of strain-$1$ and $y$ infections of strain-$2$ becomes infected by strain-$1$ (respectively, strain-$2$) with probability $x/(x+y)$ (respectively, $y/(x+y)$). In this case, we develop a mathematical theory that draws on the tools developed for analyzing the zero-temperature random-field Ising model on Bethe lattices \cite{Sethna1993} as well as on random graphs \cite{Gleeson2007seed, Gleeson2008cascades}}. Our theory fully characterizes the process and accurately predicts the epidemic threshold, expected epidemic size and the expected fraction of individuals infected by each strain (all at steady state). These metrics are computed as functions of the characteristics of the spreading process (i.e., $T_1$ and $T_2$), evolutionary adaptations (i.e., $\mu_{11} $ and $\mu_{22}$), and the structure of the underlying contact network (e.g., its degree distribution). 

In addition to the mathematical theory, we perform extensive simulations on random graphs with arbitrary degree distributions (generated by the configuration model \cite{molloy1995critical, Bollobas, newman2001random}) as well as with real-world networks (obtained from SNAP dataset \cite{snapnets}) to verify our theory and reveal the significant shortcomings of the classical mathematical models that do not capture evolution. In particular, we show that the classical, single-type bond-percolation models \cite{allen2008mathematical, newman2002spread, moore2000exact, meyers2007contact} may accurately predict the threshold and final size of epidemics, but their predictions on the probability of emergence are {\em significantly inaccurate} on both random and real-world networks. This inaccuracy sheds the light on a fundamental disconnect between the classical single-type, bond-percolation models and real-life spreading processes that entail evolution. 

{
We then focus on the case where co-infection is possible. Although recent studies have shown that co-infection with multiple pathogen strains is prevalent in nature \cite{susi2015co, balmer2011prevalence, read2001ecology, cohen2012mixed, alizon2013multiple}, there has been a lack of models that explain its occurrence, reveal its implications, and investigate its delicate interplay with the underlying contact network. Note that a considerable amount of literature has examined the case where co-infection with multiple {\em diseases} is possible \cite{cai2015avalanche, PhysRevEAzimi, Grassberger2016, CuiPRE2017}, yet multiple-disease co-infection is fundamentally different from multiple-strain co-infection (see Section~\ref{sec:related}). In this paper, we use computer simulations to explore the case where multiple-strain co-infection is possible. In particular, a susceptible individual who gets infected with strain-$1$ and strain-$2$ {\em simultaneously} becomes co-infected, and starts to transmit the co-infection, i.e., the mixture of the two strains, with a transmissibility $T_{co}$. 

The transmissibility $T_{co}$ could be larger than the maximum of $T_1$ and $T_2$ (e.g., modeling a synergistic cooperation between the two resident strains) or smaller than their minimum (e.g., modeling a negative competition among the two resident strains), and it may also fall anywhere in between. We show that co-infection gives rise to a rich set of dynamics: it can amplify or inhibit the spreading dynamics, and more remarkably {\em lead the order of phase transition to change from second-order to first-order}. We investigate the interplay between the characteristics of co-infection, the structure of the underlying contact network, and evolutionary adaptations and reveal the cases where such interplay induces a {\em first-order} phase transition for the expected epidemic size.

{\bf Summary:}
We consider the evolution of spreading processes in complex networks. We start with the case where co-infection is ignored. In this case, we develop a mathematical theory that unravels the relationship between the characteristics of the spreading process, the structure of the underlying contact network, and the process of evolution, thereby, providing accurate predictions on the epidemic threshold, expected epidemic size, and the expected fraction of individuals infected by each strain at steady state. In addition to the mathematical theory, we perform extensive simulations on random and real-world networks to verify our theory and reveal the significant shortcomings of the classical mathematical models that do not capture evolution. Then, we use computer simulations to explore the case where co-infection is possible and show that co-infection could {\em lead the order of phase transition to change from second-order to first-order}. We investigate the interplay between the characteristics of co-infection, the structure of the underlying contact network, and evolutionary adaptations and explain how such interplay controls the order of phase transition for the expected epidemic size.}

{
{\bf Structure:}} The rest of the paper is organized as follows. In Section~\ref{sec:related}, we survey the related work on evolution and co-infection. In Section~\ref{sec:model}, we present the multiple-strain model for evolution and demonstrate how we model the underlying contact network. In Section~\ref{sec:analysis}, we present and derive the main results of this work, while in Section~\ref{sec:numerical}, we confirm our theoretical results via computer simulations. We empirically consider the case where co-infection is possible in Section~\ref{sec:coinfection}. In Section~\ref{sec:realworld}, we consider evolution on real-world networks obtained from SNAP dataset \cite{snapnets} and reveal the significant shortcomings of the classical mathematical models that do not capture evolution. Finally, Section~\ref{sec:conclusion} concludes the paper.

%We start by considering the case where {\em co-infection} is not possible, i.e., each infected host either carries strain-$1$ or strain-$2$, but not both. Existing research on similar model \cite{alexander2010risk} only explores the {\em probability of epidemics}, but lacks any insights on the expected epidemic size (denoted by $S$) or, more precisely, the expected fraction of individuals infected by each strain (denoted by $S_1$ and $S_2$, respectively). We present a {\em mathematical theory} that accurately predicts the epidemic threshold, expected epidemic size and the expected fraction of individuals infected by each strain as functions of the characteristics of the spreading process (i.e., $T_1$ and $T_2$), evolution (i.e., $\mu_{11} $ and $\mu_{22}$), and the structure of the underlying contact network (e.g., its degree distribution). We perform extensive simulations on random graphs with arbitrary degree distributions (generated by the configuration model \cite{molloy1995critical, Bollobas, newman2001random}) as well as with real-world networks (obtained from SNAP dataset \cite{snapnets}) to verify our theory and reveal the significant shortcomings of the classical mathematical models that do not account for evolution.

\section{Related Work}
\label{sec:related}

{
\subsection{Evolution of Infectious Diseases}
}

A large body of research has investigated the role of evolutionary adaptations in enabling pathogen establishment in human populations \cite{woolhouse2005emerging, jones2008global, morse2012prediction, pfennig2001evolution, Song2430, morens2004challenge}. A pronounced example of such evolutionary adaptations is the emergence of zoonoses. In particular, zoonotic diseases are poorly adapted and inefficiently transmitted at first \cite{parrish2008cross}, yet they may eventually (through evolutionary adaptations) cross the {\em species barrier} and start to spread from human to human. In fact, a key event that is thought to have caused the emergence of the $1918$ H1N1 pandemic is a {\em recombination} in the hemagglutinin gene that resulted in a novel virus with increased virulence \cite{klempner2004crossing}. Other evolutionary adaptations include genetic changes (e.g., {\em Salmonella enterica}), recombination or reassortment (e.g., {\em H5N1 influenza}), and hybridization (e.g., {\em Phytophthora alni}) \cite{woolhouse2005emerging}.

To date, most of the research studies on the evolution of infectious diseases either assume a homogeneous-mixing host population, or focus entirely on the ecological or environmental factors of pathogen evolution. Indeed, the recent advances in {\em network epidemics} pave the way for exploring new depths and revealing new insights on the delicate interplay between the structural properties of the host contact network and the process of evolution. In what follows, we review the recent progress in creating a modeling framework that captures the spread and evolution of infectious diseases on realistic host contact networks.

In \cite{alexander2010risk}, Alexander and Day proposed a network-based framework that characterizes the spread and evolution of an introduced pathogen on a contact network. Their main objective was to investigate the probability of emergence, and its relation to mutation probabilities, pathogens' transmissibilities, and the structure of the underlying contact network. Using a multi-type branching process \cite{mode1971multitype, haccou2005branching}, they derived recursive relations governing the probability of emergence for a given initial strain of the pathogen. The initial strain was assumed to have a poor transmissibility, hence, evolution to a strain with sufficient transmissibility was necessary for emergence. Alexander and Day explored the potential risk factors that could lead to such evolutionary emergence of the pathogen. In particular, they showed that for a given transmissibility, heterogeneity in network structure can significantly increase the risk of emergence. Moreover, certain mutational schemes (e.g., reverse mutation) have limited impact on the probability of emergence, while others (e.g., simultaneous point mutations or recombination) have a dramatic effect on the probability of emergence. 

{ 
The framework proposed by Alexander and Day in \cite{alexander2010risk} represents a crucial first step towards understanding the role of evolutionary adaptations in driving the emergence of infectious diseases, but it lacks any insights on the expected epidemic size (denoted by $S$) or, more precisely, the expected fraction of individuals infected by each strain (denoted by $S_1$ and $S_2$, respectively). Also, the multi-type branching formalism inherently assumes a tree structure of the underlying graph, hence co-infection (which mainly occurs due to the existence of {\em cycles}) is essentially ignored in their framework.  Finally, the results presented in \cite{alexander2010risk} were neither verified on theoretical, nor real-world contact networks. Our paper addresses those limitations by means of i) developing a mathematical theory that characterizes the epidemic threshold, expected epidemic size and the expected fraction of individuals infected by each strain; ii) validating our results (as well as Alexander and Day's results) on theoretical and real-world contact networks; and iii) investigating the case when co-infection is possible.

}

When the timescale of evolution is much longer than the timescale of propagation, mutations might occur after the original pathogen has invaded the population. In \cite{leventhal2015evolution}, Leventhal et al. considered an SIS process that starts with a pathogen (of single-strain) invading the population. As the disease reaches an endemic equilibrium, a second strain of the disease appears in a random infected individual.  Authors assumed that co-infection is not possible, i.e., an infected host carries either strain-1 or strain-2, but not both. Moreover, hosts infected by either strain have perfect immunity against the other strain. Authors investigated the probability that the second strain invades the population and drives the resident strain to extinction, i.e., the {\em fixation probability}. Results from both theoretical and real-world networks suggested that the heterogeneity in network structure (which facilitated the spread of the resident strain) {\em lowers} the fixation probability, hence enhancing the resiliency of the resident strain to invasion by new variants.

{ In contrast to \cite{leventhal2015evolution}, our paper considers the case when the epidemiological and evolutionary processes occur on a similar time scale. In particular, each new infection event entails an opportunity for mutation, leading to an entirely different model (with different scope) than the one proposed by Leventhal et al. in \cite{leventhal2015evolution}. The model considered in our paper is reasonable for pathogens with long infectious periods, e.g., HIV, or pathogens with short infectious periods but high mutation rates, large population sizes, and short generation times, e.g., RNA viruses \cite{grenfell2004unifying}. Furthermore, Leventhal et al. \cite{leventhal2015evolution} ignore the case where co-infection is possible. However, recent studies revealed the prevalence of multiple-strain co-infection in disease-causing protozoa, bacteria, and viruses \cite{susi2015co, balmer2011prevalence, read2001ecology, cohen2012mixed, alizon2013multiple}.}

\begin{figure}[t]
    \centering
    \subfigure[]{\hspace{-0.4cm} 
    \includegraphics[scale=0.18] {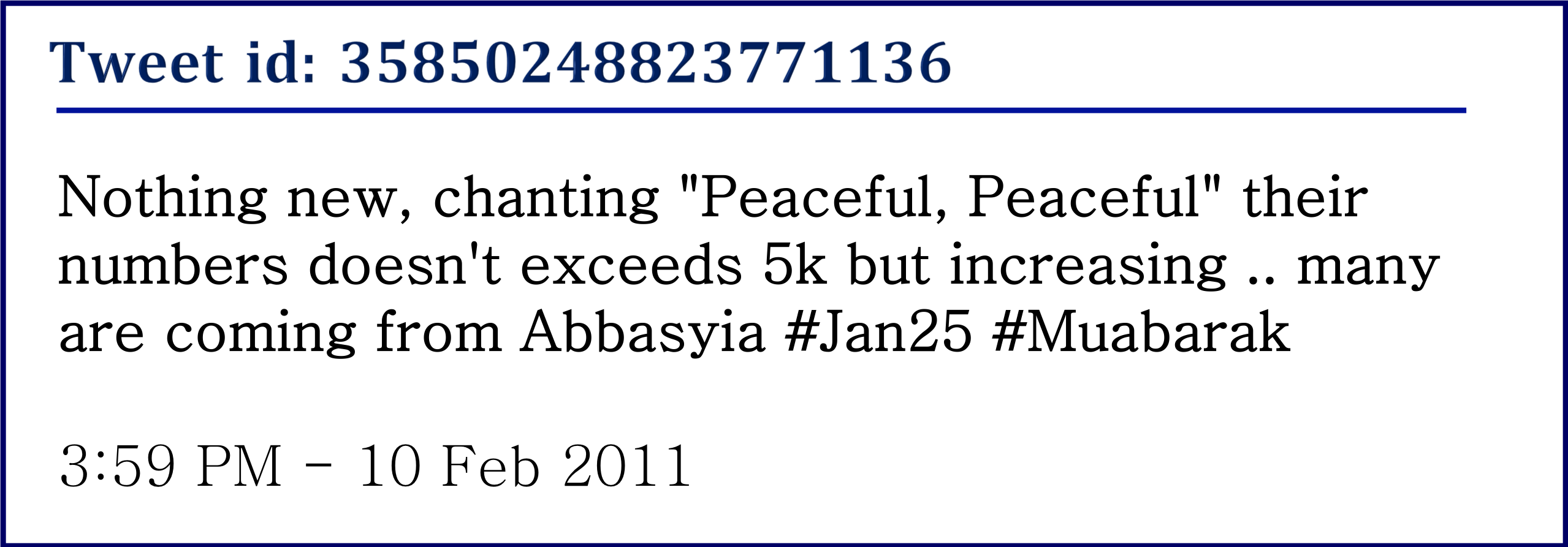}}
    \subfigure[]{\hspace{0.4cm} 
    \includegraphics[scale=0.18] {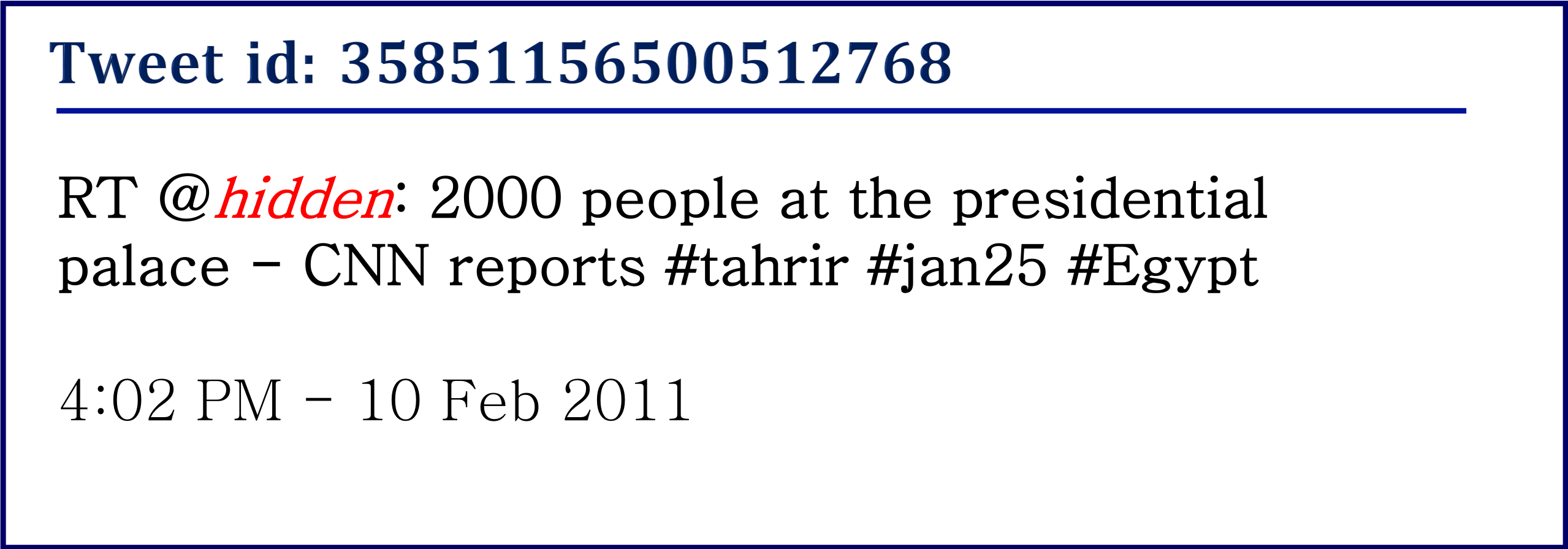}}
    \subfigure[]{\hspace{-0.4cm} 
    \includegraphics[scale=0.18] {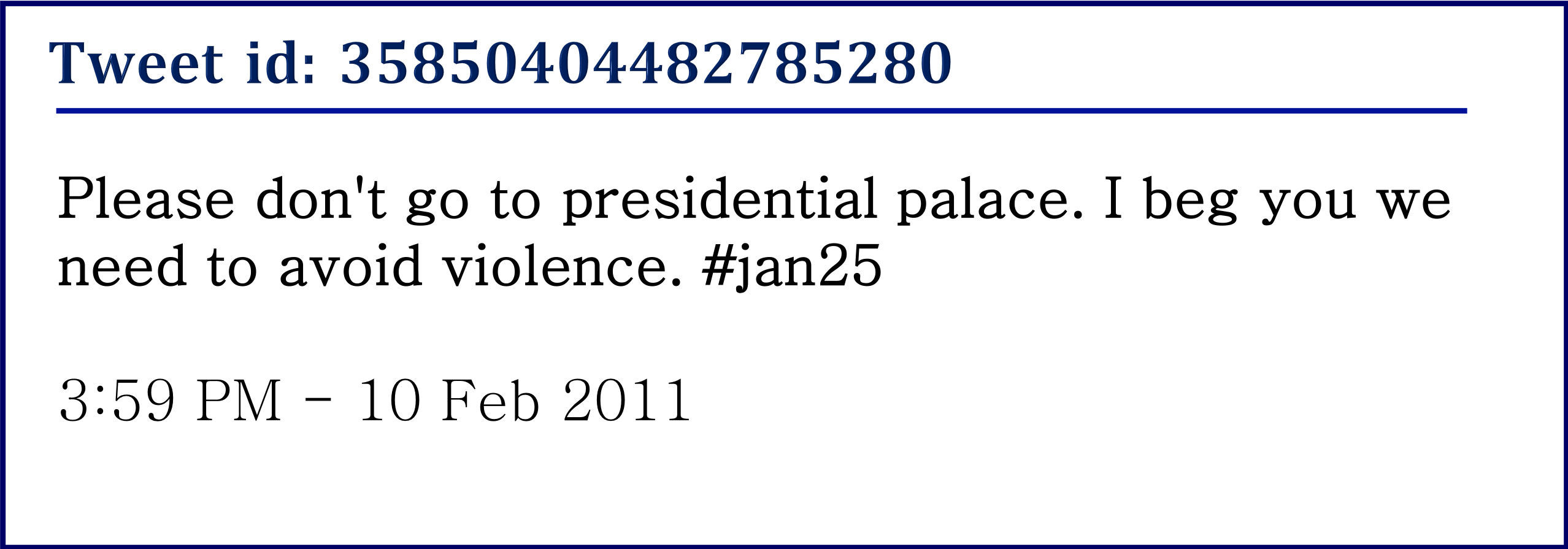}}
    \subfigure[]{\hspace{0.4cm}
    \includegraphics[scale=0.18] {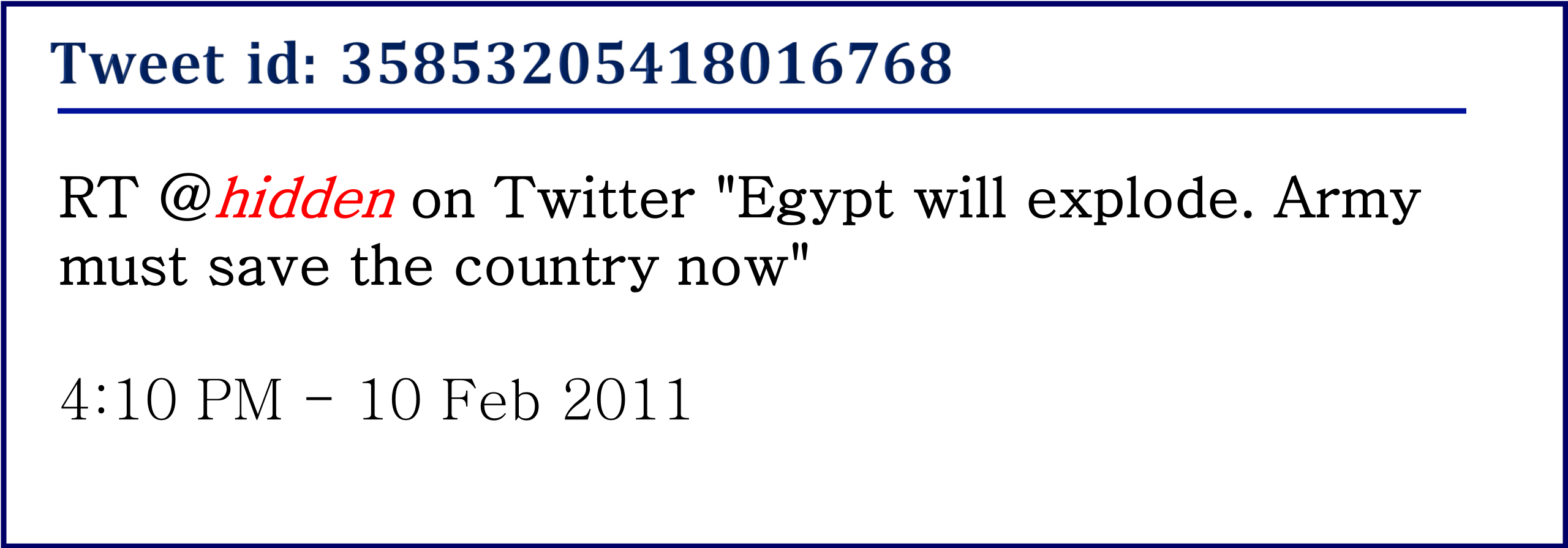}}
    \vspace{4mm}
    \caption{\sl {\bf Information mutation on Twitter}. 
    A collection of four tweets posted within a 10-minute window during the $2011$ Arab Spring in Cairo, Egypt. The tweets were posted in response to the same underlying event, namely, the marching of protesters towards the presidential palace in order to force the then president, Mubarak, to resign. Information mutation gave rise to several variants with potentially different consequences. Observe that (a) reports peaceful, traditional demonstrations while (d) suggests that the country is on a brink of collapse. User names are hidden for anonymity and tweet ids are given instead.
    }
\label{fig:tweets}
\end{figure}

{
Since humans, animals, plants, and other organisms may become {\em co-infected} with multiple diseases, a growing body of research has attempted to explore the emergence of this phenomenon and its consequences on complex networks \cite{cai2015avalanche, PhysRevEAzimi, Grassberger2016, CuiPRE2017}. However, most of the research studies focus on the case where co-infection results from simultaneous exposure to multiple diseases (or pathogen species), rather than multiple-strains of the same pathogen.} In \cite{cai2015avalanche}, Cai et al. considered the case when two { diseases} are spreading on the same contact network. A susceptible host that has not been exposed to either disease has probability $p$ to get infected by an infective neighbor. Note that the infection probabilities are the same for both diseases. Infected hosts recover after exactly one time step, and gain immunity against the disease that they were infected with, but {\em not} the other disease. A host that has been infected by one disease (being still active or has already recovered) has a probability $q$ (with $q>p$) to get infected by the other disease, i.e., an infection with one disease weakens the immune system of the infected individual and makes her more susceptible to the second disease. Cai et al. revealed that co-infection dynamics could give rise to a {\em hybrid} phase transition, where the probability of emergence exhibits a second-order transition, while the fraction of doubly infected nodes exhibits a {\em first-order} transition. 

{
In Section~\ref{sec:coinfection}, we consider the case where co-infection with multiple strains of the {\em same} pathogen is possible, giving rise to a different class of epidemiological processes than those considered in \cite{cai2015avalanche, PhysRevEAzimi, Grassberger2016, CuiPRE2017}. Our model is motivated by the recent research findings that revealed the prevalence of multiple-strain co-infection in disease-causing protozoa, helminths, bacteria, fungi, and viruses \cite{susi2015co, balmer2011prevalence, read2001ecology, cohen2012mixed, alizon2013multiple}. From a modeling standpoint, the key difference between the two processes is that evolution is a {\em perquisite} for co-infection in our model. In particular, the epidemic process in \cite{cai2015avalanche} i) does not entail any mutation events and ii) starts with a {\em doubly} infected seed, i.e., an infected host that initially carries both diseases. However, our epidemic process starts with a host receiving infection with only one strain of the pathogen, e.g., strain-$1$, hence the emergence of other strains (which is dictated by the underlying mutational scheme, transmissibility, and network structure) is a perquisite for co-infection. Moreover, our co-infection process differs fundamentally in the way a host becomes co-infected. Unlike the model given in \cite{cai2015avalanche}, we assume a perfect {\em cross-immunity}, i.e., a host that has recovered from strain-$1$ develops immunity against {\em both} strain-$1$ and strain-$2$. Hence, the only pathway for co-infection is when a susceptible host is exposed {\em simultaneously} to one or more infections of strain-$1$ and one or more infections of strain-$2$.} 

%Finally, our model differs in the stage that follows co-infection. Namely, once a host becomes co-infected, she starts to spread the co-infection, i.e., the mixture of the two pathogen strains. In contrast, a co-infected host in \cite{cai2015avalanche} spreads each strain independently with a probability that corresponds to the state of her neighbor; see  \cite{cai2015avalanche} for more details.

%In particular, if her neighbor is already infected with (or has been recovered from) strain-$1$ (respectively, strain-$2$), then the co-infected host transmits strain-$2$ (respectively, strain-$1$) with probability $q$. If her neighbor is still susceptible, then the co-infected host transmits one of the strains first with probability $p$, then transmits the other strain with probability $p$ (assuming synchronous state update).
%or $q$ (under asynchronous state update and assuming that the transmission of the first strain was successful, otherwise the transmission occurs with probability $p$).

{
\subsection{Evolution of Information}
Evolution and co-infection are two key phenomena of significant relevance to epidemiological processes. However, we are also beginning to observe their emergence and roles in the context of information propagation. We notice on daily basis how news is mutated intentionally, e.g., by adversaries, or unintentionally, e.g., by exaggeration, on social media platforms. A single underlying event could be expressed very differently by different people, creating several variants of information with different implications (see Figure~\ref{fig:tweets}). 

A few research studies have recently explored information evolution on complex networks \cite{zhang2013rumor, Adamic2016}. In \cite{zhang2013rumor}, Zhang et al. investigated the evolution of rumors on homogeneous and scale-free social networks. In their model, each individual could be in one of three different states, namely, ignorant, spreader, or stifler. These states resemble the susceptible, infected, and recovered states that we have in our model. A fraction $F$ of ignorant individuals are deemed as {\em forwarders}, i.e., they forward the received rumor to their neighbors without any modifications. The remaining $1-F$ fraction is deemed as {\em modifiers}, i.e., they modify the received rumor before forwarding it to their friends. Each modification increments the version number by one. Note that as the process continues to grow, different individuals would receive different versions of the rumor before they turn into stiflers. The main objective of \cite{zhang2013rumor} was to determine the average version number of a rumor as a function of time (and degree, for scale-free networks).

Although our paper is essentially motivated by the same observation of information evolution in social contexts, our approach and contributions are significantly different from those of \cite{zhang2013rumor, Adamic2016}. From a modeling perspective, the model presented in \cite{zhang2013rumor} is a special case of the multiple-strain model \cite{alexander2010risk} that we utilize in our paper. In particular, the model proposed by Zhang et al. essentially assumes that i) $T_i=1$ for all $i=1,2,\ldots$; and ii) the evolutionary pathways are only limited to one-step irreversible mutations. As for the contributions, we focus on the final epidemic size and final fraction of individuals infected by each version of information, in contrast to \cite{zhang2013rumor} where authors only focus on the average revision frequency. Another weakness of \cite{zhang2013rumor} is that authors made no attempt to provide closed-form expressions for the final epidemic size, the fraction of individuals infected by each version of the rumor, or the average version number of the rumor (only the corresponding differential equations were given). A closed-form expression of the average version number of the rumor at steady state was given only for networks with homogeneous degree distributions. 

In \cite{Adamic2016}, Adamic et al. explored the propagation and evolution of memes on Facebook. Authors considered a dataset of Facebook posts which were spread using a copy-and-paste mechanism (prior to the introduction of the {\em ``Share"} functionality in Facebook). The mutation rate of a particular meme was defined as the proportion of copies which introduce new edits as opposed to creating exact replicas. Authors revealed that individuals preferentially transmit a specific variant of a meme that matches their beliefs or culture. Moreover, authors showed that the distribution of variant popularity (the number of copies of that variant posted as Facebook status update) behaves as a power-law distribution for low-mutation rates, yet it deviates from the power-law behavior for high mutation rates. Theoretical predictions based on {\em Yule} processes \cite{yule1925ii} (in the limits of very low and very large mutation rates) were shown to have a close resemblance to the empirically observed distributions. 

The scope of \cite{Adamic2016} was limited to one type of propagation, i.e., copy-paste mechanism, and mutations were only characterized by the edit distance \footnote{The {\em edit distance} was defined in \cite{Adamic2016} as the number of character additions and deletions that must be performed in order to obtain one variant of the meme from another.} between a given variant of the meme and its original version. Indeed, the copy-paste mechanism is no longer sensible in modern social networks where individuals have the option to {\em ``Share"} a post rather than copying and pasting it. In addition, using the edit distance as the sole metric for mutation essentially ignores the {\em semantic} differences between two different versions of the meme. The theoretical model presented in \cite{Adamic2016} is technically different than the multiple-strain model \cite{alexander2010risk}, yet it resembles a very special case of the latter when i) $T_i=1$ for all $i=1,2,\ldots$; and ii) the evolutionary pathways are only limited to one-step irreversible mutations. Even then, Yule model was considered in \cite{Adamic2016} only in the limit of very low and very high mutation rates. In contrast to \cite{Adamic2016}, our paper attempts to explore information propagation and evolution from a {\em mathematical modeling} perspective aiming to lay down the foundations for creating a universal model for information propagation and evolution across a wide variety of social media and different possible evolutionary pathways. }

%Finally, we focus on deriving closed-form equations of the final epidemic size and the final fraction of individuals who received each version of information rather than the behaviour o 

\section{Model Definitions}
\label{sec:model}
%\subsection{A multiple-strain SIR model for evolution}
%Consider a SIR spreading process with $m$ strains. The process starts with a single individual receiving infection with strain-$1$ from an external reservoir. In the context of the multiple-strain SIR model, an individual is either {\em susceptible} (S), meaning that she has not yet received the infection, or {\em infectious and type-$i$} meaning that she was already infected and is currently spreading strain-$i$, for $i=1,\ldots,m$, or {\em recovered} (R) meaning that she is no longer spreading the infection. 

\subsection{A multiple-strain model for evolution}
In \cite{alexander2010risk}, Alexander and Day proposed a {\em multiple-strain model} that accounts for evolution. Their model is captured by two matrices, namely, the transmissibility matrix $\pmb{T}$ and the mutation matrix $\pmb{\mu}$, both with dimensions $m \times m$ for a finite integer $m \geq 2$ denoting the number of possible strains. The transmissibility matrix $\pmb{T}$ is a $m \times m$ diagonal matrix, with $[T_i]$ representing the transmissibility of strain-$i$, i.e., 
\begin{equation}\nonumber
\pmb{T} = \left[\begin{matrix}
T_1 & 0 & \ldots &  0 \\
 0 & T_2 & \ldots &  0\\ 
 \vdots & \vdots & \ddots & \vdots \\
0 & 0 & \ldots &  T_m \\
\end{matrix} \right].
\end{equation}

The mutation matrix $\pmb{\mu}$ is a $m \times m$ matrix with $\mu_{ij} $ denoting the probability that strain-$i$ mutates to strain-$j$. Note that $\sum_{j} \mu_{ij}=1$, hence $\pmb{\mu}$ is a row-stochastic matrix. One example for the transmissibility and mutation matrices was given by Antia et al. in \cite{antia2003role}, where the fitness landscape consisted of $m$ strains, with strain-$1$ through $m-1$ having identical transmissibility such that $R_{0,i}<1$ for $i=1,\ldots,m-1$, with $R_{0,i}$ denoting the basic reproductive number of strain-$i$. Strain-$m$ has transmissibility $T_m$ such that $R_{0,m}>1$, hence the emergence of the pathogen requires evolution from strain-$1$ to strain-$m$. Antia et al. considered the the so-called {\em one-step irreversible mutation} \cite{alexander2010risk, antia2003role} where the pathogen must acquire $m-1$ mutations (in order and one at a time) to evolve to strain-$m$ , i.e.,
\begin{equation}\nonumber
\pmb{T} = 
\left[ 
\begin{matrix}
T_1 & 0 & 0& \ldots & 0 \\
0 & T_1 & 0&\ldots & 0 \\
0&0&T_1&\ldots&0 \\
\vdots & \vdots &\vdots &  \ddots & \vdots  \\
0 & 0 & \ldots & 0&T_m \\
\end{matrix}
\right]
\end{equation}  
and
\begin{equation}\nonumber
\pmb{\mu} = 
\left[ 
\begin{matrix}
1-\mu & \mu & 0 & \ldots & 0 & 0 & 0 \\
0 & 1-\mu & \mu & \ldots  & 0 & 0 & 0 \\
\vdots & \vdots & \vdots & \ddots & \vdots & \vdots & \vdots \\
0 & 0 & 0 & \ldots & 0 & 1-\mu & \mu \\
0& 0&0& \ldots&0 & 0 & 1\\
\end{matrix}
\right]
\end{equation}  

\begin{figure*}[t]
    \centering
    \subfigure[]{\hspace{-0.4cm} 
    \includegraphics[scale=0.07] {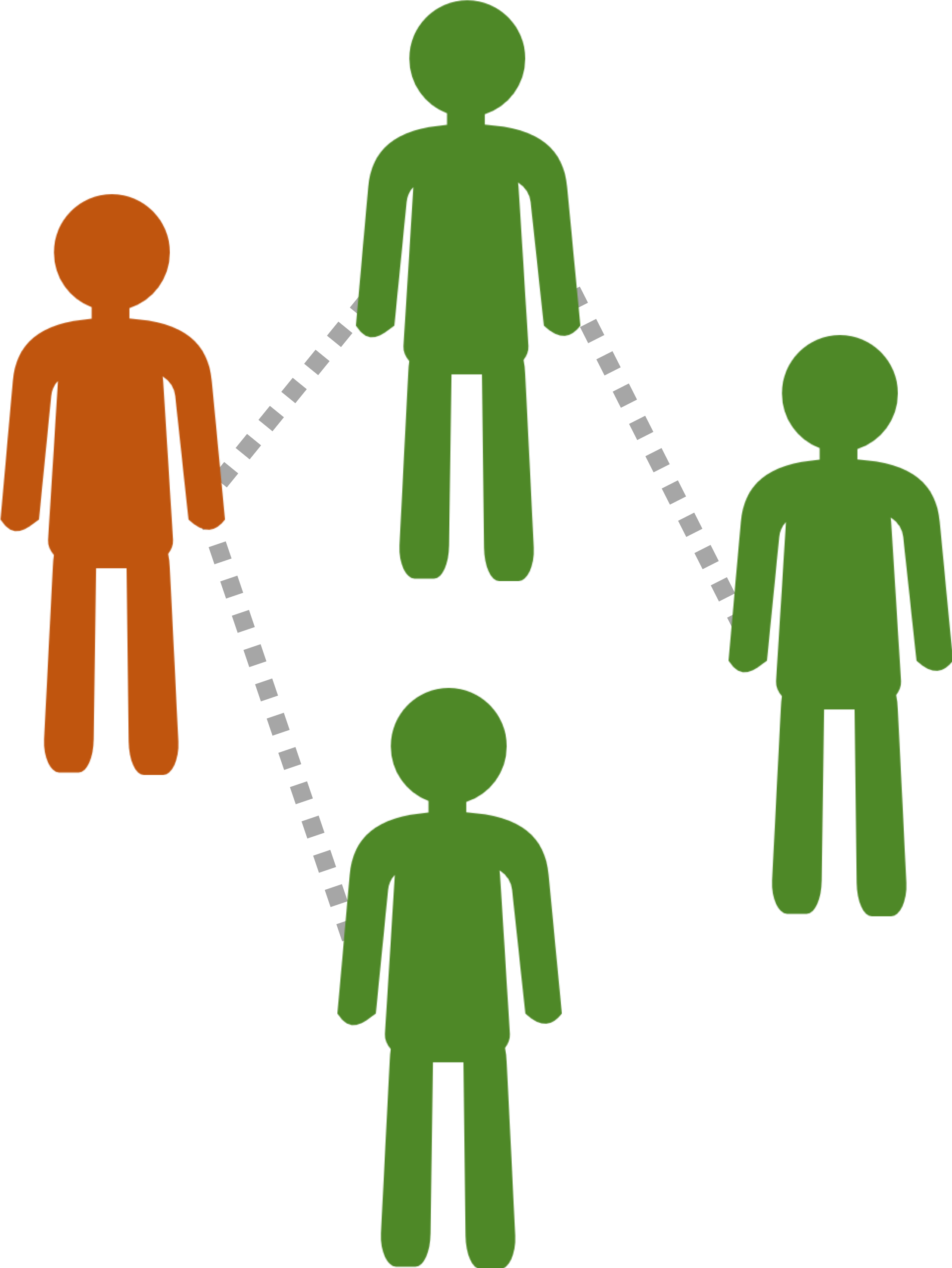}}
    \subfigure[]{\hspace{0.4cm} 
    \includegraphics[scale=0.07] {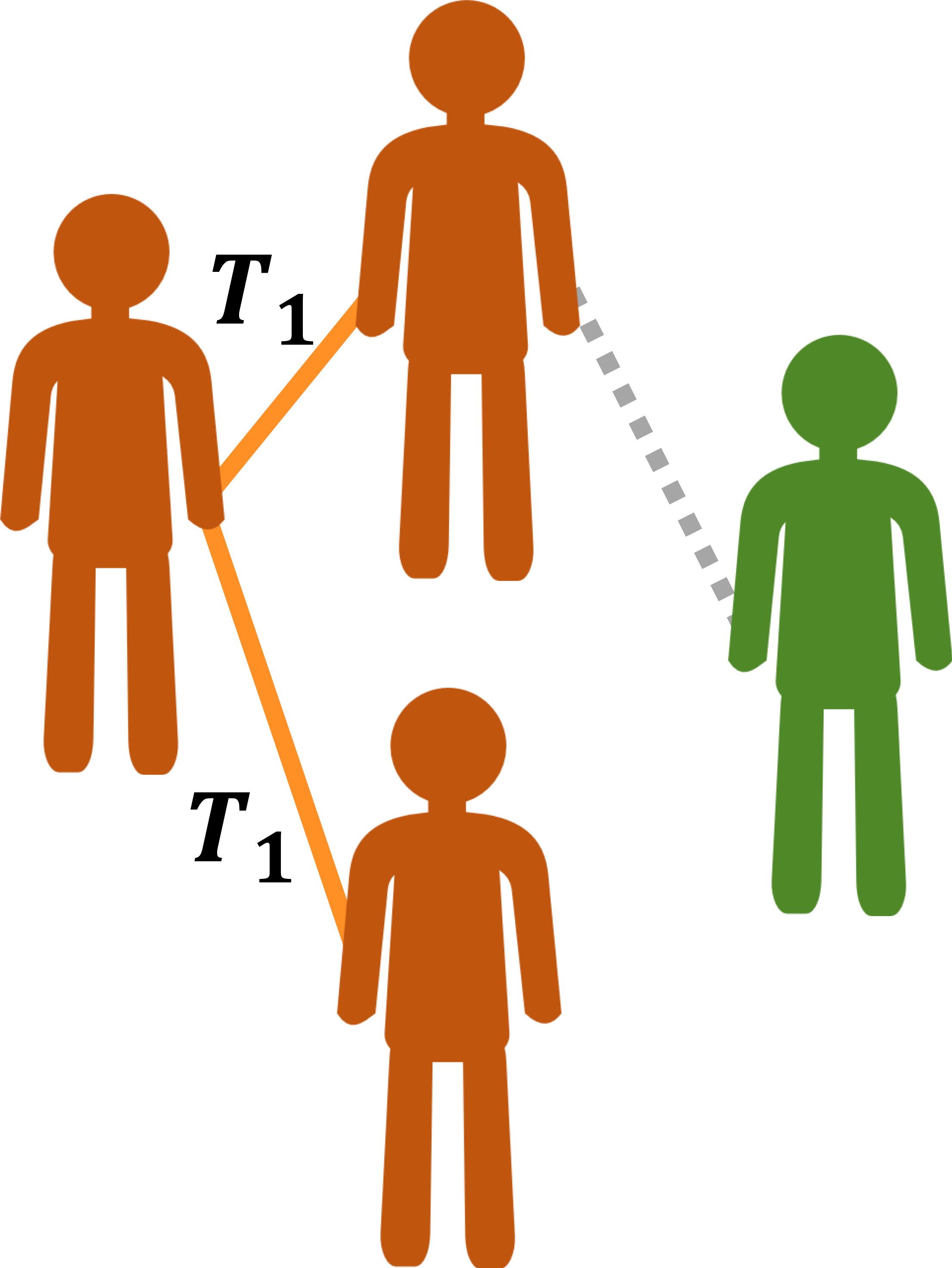}}
    \subfigure[]{\hspace{0.4cm} 
    \includegraphics[scale=0.07] {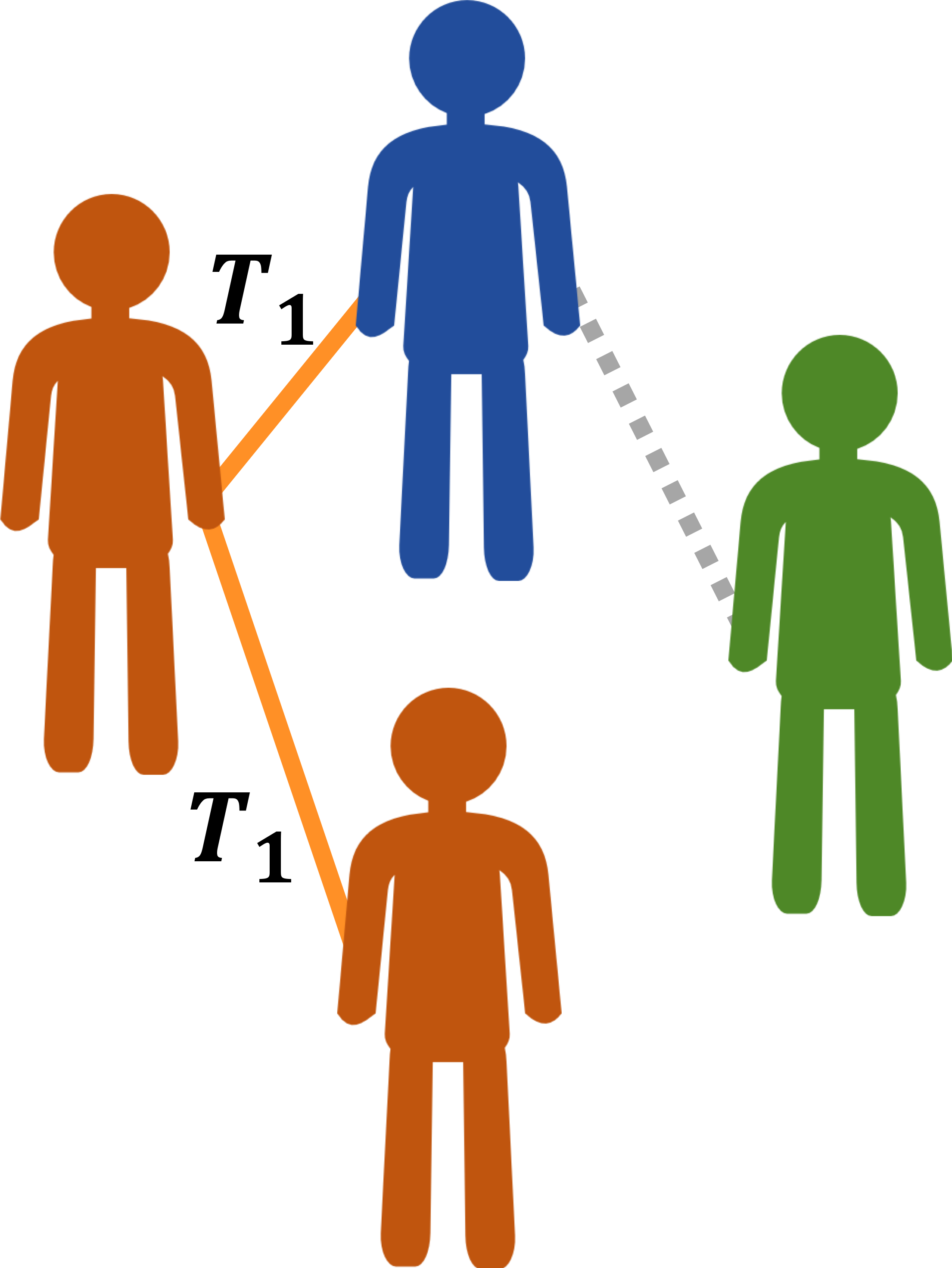}}
    \subfigure[]{\hspace{0.4cm}
    \includegraphics[scale=0.07] {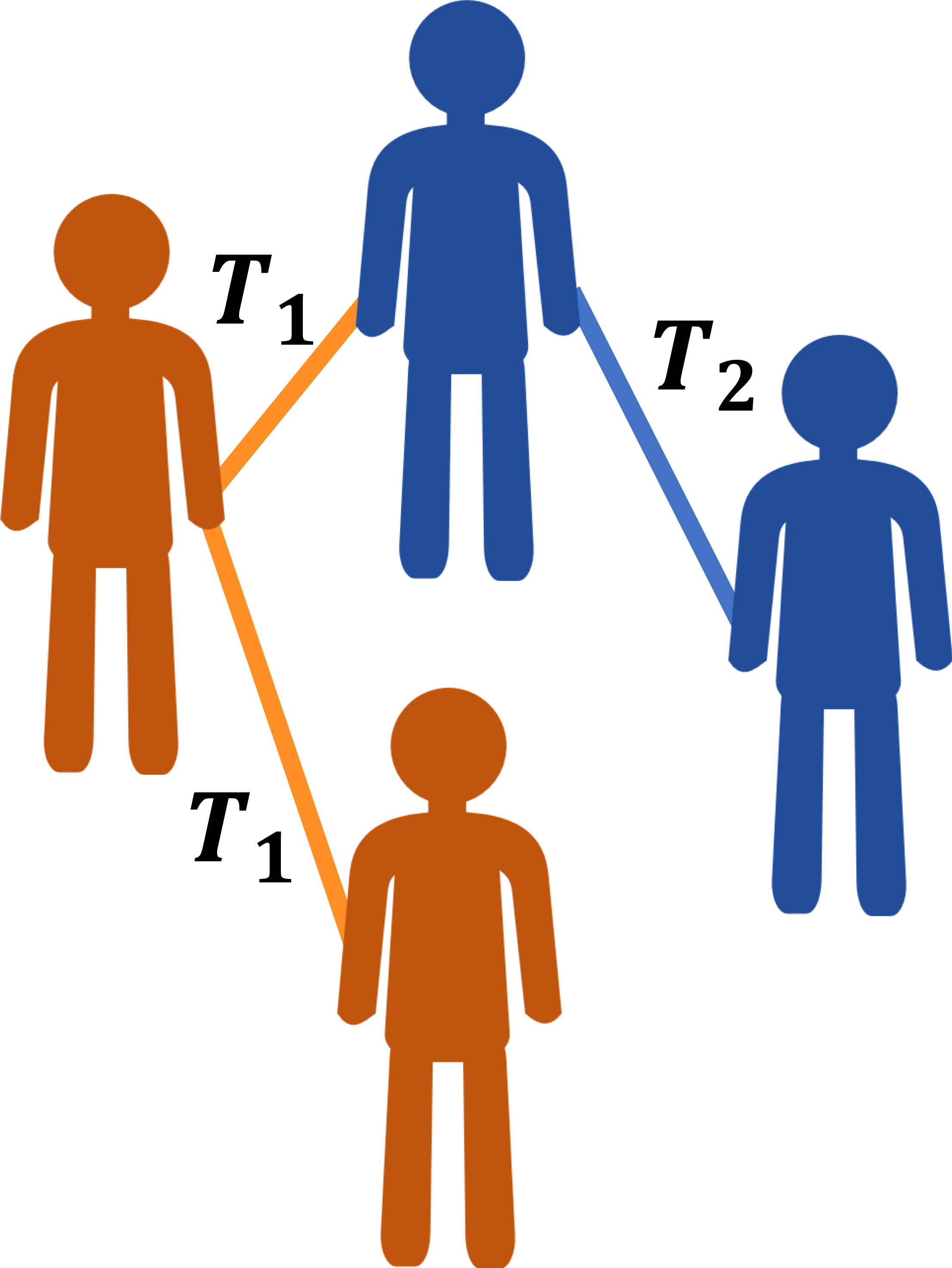}}
    \subfigure[]{\hspace{0.4cm}
    \includegraphics[scale=0.07] {demon4.png}}
    \vspace{4mm}
    \caption{\sl {\bf The multiple-strain model for evolution}.
   (a) The process starts with a single individual, i.e., the seed, receiving infection with strain-$1$ (highlighted in orange) from an external reservoir.
   (b) The seed infects each of her susceptible neighbors (highlighted in green) independently with probability $T_1$.
   (c) The pathogen mutates independently within hosts. The pathogen remains as strain-$1$ with probability $\mu_{11}$ or mutates to strain-$2$ (highlighted in blue) with probability $\mu_{12}$. 
   (d) Individuals whose pathogen has mutated to strain-$i$ infect their neighbors independently with probability $T_i$.
   (e) The pathogen mutates independently within hosts. The pathogen remains as strain-$2$ with probability $\mu_{22}$ or mutates to strain-$1$ with probability $\mu_{21}$. 
    }
\label{fig:demonstration}
\end{figure*}

The multiple-strain model proposed by Alexander and Day \cite{alexander2010risk} works as follows. Consider a spreading process that starts with an individual, i.e., the seed, receiving infection with strain-$1$ from an external reservoir. Since strain-$1$ has transmissibility $T_1$, the seed infects each of her contacts independently with probability $T_1$. Once a susceptible individual receives the infection from the seed, the pathogen may evolve within that new host prior to any subsequent infections. In particular, the pathogen may remain as strain-$1$ with probability $\mu_{11}$ or mutate to strain-$i$ (that has transmissibility $T_i$) with probability $\mu_{1i}$ for $i=2,\ldots,m$. If the pathogen remains as strain-$1$ (respectively, mutates to strain-$i$), then the host infects each of her susceptible neighbors in the subsequent stages independently with probability $T_1$ (respectively, $T_i$). Observe that as the process continues to grow, multiple strains may coexist in the population as governed by the transmissibility matrix $\pmb{T}$ and the mutation matrix $\pmb{\mu}$. At an intermediate stage, if any susceptible individual receives strain-$j$, the pathogen may remain as strain-$j$ with probability $\mu_{jj}$ or mutate to strain-$\ell$ with probability $\mu_{j \ell}$ for $\ell \in \{1,2,\ldots, m\} \setminus \{j\}$ prior to subsequent infections. The process terminates when no additional infections are possible. A graphical illustration for the case when $m=2$ is given in Figure~\ref{fig:demonstration}. In this paper, we focus on the case where $m=2$, however, it is straightforward to extend our theory to handle the general case with $m$ strains. More details are given in Section~\ref{sec:analysis}.

\subsection{Network Model: Random graphs with arbitrary degree distribution}
Let $\mathbb{G}$ denote the underlying contact network, defined on the node set $\mathcal{N}=\{1,\ldots,n\}$. We define the structure of $\mathbb{G}$ through its degree distribution $\{p_k\}$. In particular, $\{p_k, k=0,1,\ldots\}$ gives the probability that an arbitrary node in $\mathbb{G}$ has degree $k$. We generate the network $\mathbb{G}$ according to the {\em configuration model} \cite{molloy1995critical, Bollobas}, i.e., the degrees of nodes in $\mathbb{G}$ are all drawn independently from the distribution $\{p_k, k=0,1,\ldots\}$. Furthermore, we assume that the degree distribution is well-behaved in the sense that all moments of arbitrary order are finite. Of particular importance in the context of the configuration model is the degree distribution of a randomly chosen neighbor of a randomly chosen vertex, denoted by $\{\hat{p}_k, k=1,2,\ldots\}$, and given by
\begin{equation}\nonumber
\hat{p}_k = \frac{k p_k}{ \langle k \rangle}, \quad k=1,2,\ldots
\end{equation}
where $ \langle k \rangle$ denotes the {\em mean degree}, i.e., $ \langle k \rangle = \sum_k k p_k$.

\section{Analysis}
\label{sec:analysis}

\subsection{The Probability of Emergence}
The analysis of the probability of emergence was established by Alexander and Day in \cite{alexander2010risk}. Below, we give a brief summary of their results for completeness. Their approach is based on a multi-type branching process \cite{mode1971multitype, haccou2005branching} that starts with an initial infective of a particular type, e.g., type-$1$, and then proceeds by infecting each of her neighbors independently with some probability that is characterized by the infecting strain. Each of the infected neighbors mutate independently with a probability that is also characterized by the infecting strain. The process proceeds similarly for subsequent stages. Clearly, the process differs from the standard Single-Type Branching Process in that individuals of different types may coexist in any generation (other than generation $0$), with different offspring distribution per each type, hence the notion Multi-Type \cite{mode1971multitype, haccou2005branching}.

Next, we summarize the results given by Alexander and Day in \cite{alexander2010risk}. Let $\gamma_i \left(s_1,s_2,\ldots,s_m \right)$ be the probability generating function (PGF) for the number of infections of each type transmitted by an {\em initial} infective of type-$i$. It holds that 
\begin{align}
\gamma_i \left(s_1,s_2,\ldots,s_m \right) = g \left(1-T_i+T_i\sum_{j=1}^m \mu_{ij} s_j \right),  \nonumber
\end{align} 
for $i=1,\ldots,m$ and with $g \left(s \right)$ denoting the PGF of the degree distribution; i.e.,  $g \left(s \right) = \sum_{k=0}^\infty p_k s^k$. Moreover, with $\Gamma_i \left(s_1,s_2,\ldots,s_m \right)$ denoting the PGF for the number of infections of each type transmitted by a {\em later-generation} infective of type-$i$ (i.e., a typical intermediate host in the process); it holds that
\begin{equation}\nonumber
\Gamma_i \left(s_1,s_2,\ldots,s_m \right) = G \left(1-T_i+T_i\sum_{j=1}^m \mu_{ij} s_j \right),
\end{equation} 
for $i=1,\ldots,m$ and with $G \left(s \right)$ denoting the PGF of the {\em excess degree} distribution; i.e.,  
\begin{equation}\nonumber
G \left(s \right) = \sum_{k=1}^\infty \frac{kp_k}{ \langle k \rangle} s^{k-1}.
\end{equation} 
We remind that $kp_k / \langle k \rangle$ gives the probability that a randomly chosen neighbor of a randomly chosen vertex has degree $k$, and note that the excess degree is $k-1$ since one edge is already traversed to reach the node.

The probability of extinction starting from one later-generation infective of type-$i$, denoted $q_i$, is the smallest non-negative root of the equation $q_i = \Gamma_i \left(q_1,\ldots,q_m \right)$ solved simultaneously for all $i=1,\ldots,m$. Finally, the overall extinction probability is given by $g \left(1-T_i+T_i\sum_{j=1}^m \mu_{ij} q_j \right)$ if the whole process starts with an initial infective of type-$i$. It was shown in \cite{alexander2010risk} that the above process resembles a multi-type branching process with mean matrix \footnote{ The mean matrix $\pmb{M}$ of a multi-type branching process is defined as $\pmb{M}=\left[m_{ij} \right]$, where $m_{ij}$ is the mean number of type-$j$ offspring generated by a type-$i$ parent. Note that $m_{ij}= \frac{\partial \Gamma_i(\pmb{s})}{ \partial s_j} \Big |_{\pmb{s}=\pmb{1}}$ for the multiple-strain model proposed in \cite{alexander2010risk}.} given by
\begin{equation}
\pmb{M} =  \left( \dfrac{ \langle k^2  \rangle -  \langle k  \rangle}{ \langle k  \rangle} \right) \pmb{T} \pmb{\mu}  \label{eq:ProbMatrix}
\end{equation}

The theory of multi-type branching processes states that if the dominant eigenvalue of $\pmb{M}$ is less than or equal to one, then the process goes extinct with probability $1$. Otherwise, there is a positive probability of non-extinction. Hence, the phase transition occurs when
\begin{equation}
\rho \left( \pmb{M} \right) = 1,
\label{eq:condition}
\end{equation}
where $\rho \left( \pmb{M} \right)$ denotes the {\em spectral radius}, i.e., the largest eigenvalue (in absolute value) of $\pmb{M}$.

\subsection{Expected Epidemic Size}
Our objective is to derive the expected epidemic size $S$ and the expected fraction of individuals infected by each strain, i.e., $S_1, S_2,\ldots,S_m$ for $m$ possible strains. Note that $S=\sum_{i=1}^m S_i$. Below, we provide analysis for the case of two strains, but we later show how to extend our analysis to the general case with $m$ strains, for some finite integer $m \geq 2$. We apply a {\em tree-based} approach that is based on the work by Gleeson  \cite{Gleeson2007seed, Gleeson2008cascades}. Their approach draws on the tools developed for analyzing the zero-temperature random-field Ising model on Bethe lattices \cite{Sethna1993}. Note that as we build our network using the configuration model, the network structure is locally tree-like with the fraction of cycles approaching zero in the limit of large network size \cite{molloy1995critical, Bollobas, newman2001random}.

Since $\mathbb{G}$ is locally tree-like, we can replace it by a tree and arrange the vertices in a hierarchical structure, such that at the top level, there is a single node (the {\em root}) that has degree $k$ with probability $p_k$. Note that $\left\{p_k\right\}$ is a proper degree distribution with $\sum_k p_k = 1$. Each of the $k$ neighbors of the root has degree $k'$ with probability $k' p_{k'} / \langle k \rangle$, where $\langle k \rangle$ denotes the mean degree of the network. 
%Observe that this is precisely the {\em excess degree distribution} \cite{XX}, i.e., the degree distribution of a randomly chosen neighbor of a randomly chosen vertex. 
Furthermore, we label the levels of the tree from level $\ell=0$ at the bottom to level $\ell = \infty$ at the top, i.e., the root. 

We assume that nodes update their status starting from the bottom of the tree and proceeding towards the top. This gives rise to a delicate case, where a node at some level $\ell$ may be exposed to {\em simultaneous} infections by both strain-$1$ and strain-$2$ from her neighbors at level $\ell-1$. In the remainder of this section, we assume that {\em co-infection} is not possible, hence a node that receives $x$ infections of strain-$1$ and $y$ infections of strain-$2$ becomes infected by strain-$1$ (respectively, strain-$2$) with probability $x/(x+y)$ (respectively, $y/(x+y)$). In Section~\ref{sec:coinfection}, we {\em empirically} consider the case where co-infection is possible, i.e., a node that receives simultaneous infections by both strains becomes co-infected and starts to spread the {\em co-infection} in the subsequent rounds. In this case, co-infection may be modeled as an additional strain that has transmissibility $T_{co}$ and never mutates back to strain-$1$ or strain-$2$.

Throughout, we say that a node is either {\em inactive} if it has not received any infection (i.e., still susceptible) or {\em active and type-$i$} if it has been infected and then {\em mutated} to strain-$i$, for $i =1,2$. With a slight abuse of notations, let $q_{\ell+1, i}$ be the probability that a node at level $\ell+1$, say node $v$, is active {\em and} type-$i$. Furthermore, let $q_{\ell+1} = q_{\ell+1, 1}+q_{\ell+1, 2}$, i.e., $q_{\ell+1}$ is the total probability that a node at level $\ell+1$ is active. We start by an arbitrary initial distribution for $\{q_{0,1},q_{0,2}\}$ satisfying $q_{0,1}>0, q_{0,2}>0$. Then, we update the distribution properly until we reach the root. Note that if the degree of node $v$ is $k$, then node $v$ is using one edge to connect to her parent at level $\ell+2$, and $k-1$ edges to connect to her neighbors at level $\ell$. We can condition on the {\em excess degree} ($\Tilde{d}$) of node $v$ to get
\begin{align} 
&q_{\ell+1, i} = \sum_{k=1}^{\infty} \frac{k p_k}{\langle k \rangle} \mathbb{P}\bigg[\text{node } v \text{ becomes active and type-i}  \given[\bigg] \Tilde{d}=k-1 \bigg] \nonumber
\end{align}

Next, we further condition on the number of {\em active} neighbors of type-$1$ and type-$2$. Note that we have a Multinomial distribution for the number of active neighbors of both types. In particular, a neighbor at level $\ell$ may be active and type-$1$ with probability $q_{\ell,1}$, active and type-$2$ with probability $q_{\ell,2}$, or inactive with probability $1-q_\ell=1-q_{\ell,1}-q_{\ell,2}$. Let $I_i$ denote the number of active neighbors of type-$i$. Thus,

\begin{align}
    q_{\ell+1, i} &= \sum_{k=1}^{\infty} \frac{k p_k}{\langle k \rangle} \sum_{k_1=0}^{k-1} \sum_{k_2=0}^{k-1-k_1} \binom{k-1}{k_1} \binom{k-1-k_1}{k_2} \left( q_{\ell,1} \right)^{k_1}  \left( q_{\ell,2} \right)^{k_2} \left( 1-q_{\ell,1}-q_{\ell,2} \right)^{k-1-k_1-k_2} \nonumber \\
    & \quad \cdot \mathbb{P}\bigg[\text{node } v \text{ becomes active and type-i} \given[\big] I_1 = k_1, I_2 = k_2 \bigg] \nonumber
\end{align}

Let $X$ and $Y$ denote the number of infections received from type-$1$ and type-$2$ neighbors, respectively. Note that conditioned on having $k_1$ and $k_2$ active neighbors of type-$1$ and type-$2$, respectively, we have
\begin{align}
    X &\sim \text{Binomial}(k_1, T_1) \nonumber \\
    Y &\sim \text{Binomial}(k_2, T_2) \nonumber
\end{align}
where $T_i$ denotes the transmissibility of strain-$i$. Let
\begin{equation} \nonumber
    A := \mathbb{P}\left[\text{node } v \text{ becomes active and type-i} \given[\big] I_1 = k_1, I_2 = k_2 \right]
\end{equation}

Consider a particular realization $(x,y)$ of the random variables $(X,Y)$. Observe that if $x>0, y=0$, then node $v$ becomes infected by strain-$1$ and eventually mutates to type-$i$ with probability $\mu_{1i}$. Similarly, if $x=0, y>0$, then node $v$ becomes infected by strain-$2$ and eventually mutates to type-$i$ with probability $\mu_{2i}$. Finally, if $x>0, y>0$, then node $v$ becomes infected by strain-$1$ (respectively, strain-$2$) with probability $x/(x+y)$ (respectively, $y/(x+y)$) and eventually mutates to type-$i$ with probability $\mu_{1i}$ (respectively, $\mu_{2i}$). Hence, by conditioning on $X$ and $Y$, we have
\begin{align}
    A &= \sum_{x=0}^{k_1} \sum_{y=0}^{k_2} \binom{k_1}{x} \binom{k_2}{y} T_1^x T_2^y (1-T_1)^{k_1-x} (1-T_2)^{k_2-y}  \mathbb{P}\left[A \given[\big] X=x, Y=y \right] \nonumber \\
    & = \sum_{x=0}^{k_1} \sum_{y=0}^{k_2} \binom{k_1}{x} \binom{k_2}{y} T_1^x T_2^y (1-T_1)^{k_1-x} (1-T_2)^{k_2-y}  \nonumber \\
    &\quad \cdot \Bigg( \mu_{1i} \pmb{1}[x>0, y=0] + \mu_{2i} \pmb{1}[x=0, y>0] + \left( \frac{x \mu_{1i}}{x+y} + \frac{y \mu_{2i}}{x+y} \right) \pmb{1}[x>0, y>0] \Bigg) \nonumber
\end{align}

Note that
\begin{align}
    \sum_{x=0}^{k_1} \sum_{y=0}^{k_2} \binom{k_1}{x} \binom{k_2}{y} T_1^x T_2^y (1-T_1)^{k_1-x} (1-T_2)^{k_2-y} \mu_{1i} \pmb{1}[x>0,y=0] &= \mu_{1i} (1-T_2)^{k_2} \left(1-\mathbb{P}(X=0) \right)\nonumber \\
    &= \mu_{1i} a_2 b_1 \nonumber
\end{align}
where $a_i = \left(1-T_i\right)^{k_i}$ and $b_i = 1-a_i$. Similarly, 
\begin{align}
    &\sum_{x=0}^{k_1} \sum_{y=0}^{k_2} \binom{k_1}{x} \binom{k_2}{y} T_1^x T_2^y (1-T_1)^{k_1-x} (1-T_2)^{k_2-y}  \mu_{2i} \pmb{1}[x=0,y>0] =\mu_{2i} a_1 b_2 \nonumber
\end{align}

Thus, we have
\begin{align}
    q_{\ell+1, i} &= \sum_{k=1}^{\infty} \frac{k p_k}{\langle k \rangle} \sum_{k_1=0}^{k-1} \sum_{k_2=0}^{k-1-k_1} \binom{k-1}{k_1} \binom{k-1-k_1}{k_2}  \left( q_{\ell,1} \right)^{k_1} \left( q_{\ell,2} \right)^{k_2} \left( 1-q_{\ell,1}-q_{\ell,2} \right)^{k-1-k_1-k_2} \cdot \nonumber \\
    & \quad \cdot \Bigg( b_1 a_2 \mu_{1i} + a_1 b_2 \mu_{2i} + \nonumber \\
    &\quad  \sum_{x=0}^{k_1} \sum_{y=0}^{k_2} \binom{k_1}{x} \binom{k_2}{y} T_1^x T_2^y (1-T_1)^{k_1-x} (1-T_2)^{k_2-y} \left( \frac{x \mu_{1i}}{x+y} + \frac{y \mu_{2i}}{x+y} \right) \pmb{1}[x>0, y>0] \Bigg),
    \label{eq:recursion}
\end{align}
for $\ell=0,1,\ldots$ and $i =1,2$.

Observe that under the assumption that nodes do not become inactive once they turn active, the quantities $q_{\ell,i}$ appearing in (\ref{eq:recursion}) are non-decreasing in $\ell$, and thus they converge to a limit $q_{\infty,i}$ for $i =1,2$. Finally, the final fraction of nodes that are active and type-$i$ is equal (in expected value) to the probability that the root of the tree (at level $\ell \to \infty$) is active and type-$i$. Note that if the tree root has degree $k$, then all of these $k$ edges will be utilized to connect with her neighbors at the lower level. Hence,  
\begin{align}
    Q_i &= \sum_{k=0}^{\infty} p_k \sum_{k_1=0}^{k} \sum_{k_2=0}^{k-k_1} \binom{k}{k_1} \binom{k-k_1}{k_2} \left( q_{\infty,1} \right)^{k_1} \left( q_{\infty,2} \right)^{k_2} \left( 1-q_{\infty,1}-q_{\infty,2} \right)^{k-k_1-k_2} \cdot \nonumber \\
    & \quad \cdot \Bigg( b_1 a_2 \mu_{1i} + a_1 b_2 \mu_{2i} + \nonumber \\ 
    & \quad \sum_{x=0}^{k_1} \sum_{y=0}^{k_2} \binom{k_1}{x} \binom{k_2}{y} T_1^x T_2^y (1-T_1)^{k_1-x} (1-T_2)^{k_2-y} \left( \frac{x \mu_{1i}}{x+y} + \frac{y \mu_{2i}}{x+y} \right) \pmb{1}[x>0, y>0] \Bigg) 
    \label{eq:finalSize}  
\end{align}
where $Q_i$ for $i=1,2$ denotes the probability that the tree root is active and type-$i$ and $q_{\infty,i}$ for $i=1,2$ is the steady-state solution of the recursive equations (\ref{eq:recursion}). Note that $Q=Q_1+Q_2$ is the total probability that the tree root is active.

Observe that $q_{\infty,1}=q_{\infty,2}=0$ gives a trivial fixed-point of the recursive equations (\ref{eq:recursion}). Indeed, this trivial solution leads to $Q=0$ by virtue of (\ref{eq:finalSize}). Although the trivial fixed point is a valid numerical solution for the recursive equations (\ref{eq:recursion}), we can show that this trivial solution is {\em unstable}. Hence, another solution with $q_{\infty,1}>0$ and $q_{\infty,2}>0$ may exist. To test whether or not the trivial fixed point is stable, we check the spectral radius of the Jacobian matrix $\pmb{J}(q_{\ell,1},q_{\ell,2})$ corresponding to the {\em linearization} of (\ref{eq:recursion}) at $q_{\ell,1}=q_{\ell,2}=0$. If the spectral radius of the $\pmb{J}(q_{\ell,1},q_{\ell,2})$ at $q_{\ell,1}=q_{\ell,2}=0$ is larger than one, then the trivial fixed-point is unstable, indicating that there exists another solution with $q_{\infty,1}>0$ and $q_{\infty,2}>0$ implying the existence of a giant component. The Jacobian matrix is given by
\begin{align} \nonumber
\pmb{J}(q_{\ell,1},q_{\ell,2}) |_{q_{\ell,1}=q_{\ell,2}=0} &= \left[
\begin{matrix}
\frac{\partial q_{\ell+1,1}}{\partial q_{\ell,1}} & \frac{\partial q_{\ell+1,1}}{\partial q_{\ell,2}} \\ 
\frac{\partial q_{\ell+1,2}}{\partial q_{\ell,1}} & \frac{\partial q_{\ell+1,2}}{\partial q_{\ell,2}}
\end{matrix} \right]_{q_{\ell,1}=q_{\ell,2}=0}  \nonumber \\
&= \left( \dfrac{ \langle k^2  \rangle -  \langle k  \rangle}{ \langle k  \rangle} \right)
\left[
\begin{matrix}
T_1 \mu_{11} & T_2 \mu_{21}\\ 
T_1 \mu_{12} & T_2 \mu_{22}
\end{matrix} \right]  \nonumber \\
&= \left( \dfrac{ \langle k^2  \rangle -  \langle k  \rangle}{ \langle k  \rangle} \right)
\left( \pmb{T} \pmb{\mu} \right)^T \nonumber
\end{align}
Note that a square matrix and its transpose have the same set of eigenvalues. Hence, the Jacobian matrix admits the same spectral radius of (\ref{eq:ProbMatrix}) as would be expected, implying the same condition (\ref{eq:condition}) for phase transition. Nevertheless, the generating functions approach used by Alexander and Day \cite{alexander2010risk} is useful in its own right as it enables quantifying the probability of emergence.

We remark that it is straightforward to extend our analysis to the general case with $m$ strains, for some finite integer $m \geq 2$ as long as the underlying process is {\em indecomposable} \cite{mode1971multitype, haccou2005branching, alexander2010risk}. At a high level, indecomposable processes are those for which each pathogen strain $i$ eventually gives rise to strain-$j$ at some generation $n_{ij} \geq 1$ for $i,j = 1,2,\ldots,m$. In other words, if an indecomposable process starts with an infection with strain-$i$, then as the process continues to grow, all other strains will eventually emerge. Such a property is established if, for every pair of strains $(i,j)$, there exists a positive integer $n_{ij}$ such that $\pmb{M}^{n_{ij}}(i,j)>0$  \cite{alexander2010risk}. If the underlying process is {\em decomposable}, then there exist classes of strain types such that strain types belonging to the same class can eventually give rise to one another, but not to other strain types. Indeed, the existence of multiple classes leads to multiple solutions for the set of equations (\ref{eq:finalSize}) depending on the initial distribution of $\{q_{0,1},q_{0,2}, \ldots, q_{0,m}\}$. Hence, to guarantee the uniqueness of the solution of (\ref{eq:finalSize}) and for mathematical tractability, we limit our formalism to the case when the underlying process is indecomposable. 

%A process is said to be indecomposable if, for every pair of strains $(i,j)$, there exists a positive integer $n$ such that $\pmb{M}^n(i,j)>0$  \cite{alexander2010risk}.

\section{Numerical results}
\label{sec:numerical}

\subsection{The Structure of the Contact Network}
In this section, we consider synthetic contact networks generated randomly by the configuration model, while real-world networks are considered in Section~\ref{sec:realworld}. In particular, we consider contact networks with {\em Poisson} degree distribution as well as {\em Power-law} degree distribution.

\subsubsection{Poisson degree distribution} 
We start by considering contact networks with Poisson degree distribution. Namely, with $\lambda$ denoting the mean degree, i.e., $\lambda = \langle k  \rangle$, we have
\begin{equation}\nonumber
p_k = e^{- \lambda} \frac{\lambda ^k}{k!}, \qquad k=0,1,\ldots
\end{equation}

In this case, condition (\ref{eq:condition}) implies that phase transition occurs when
\begin{equation}
\lambda \times \rho \left( \pmb{T \mu} \right) = 1 
\label{eq:PoissonPhase}
\end{equation}
where $\rho \left( \pmb{T \mu} \right)$ denotes the spectral radius of the matrix multiplication $\pmb{T \mu} $. Observe that condition (\ref{eq:PoissonPhase}) embodies the structure of the contact network (represented by $\lambda$ for a contact network with Poisson degree distribution), the characteristics of propagation (represented by the matrix $\pmb{T}$) and the process of evolution (represented by $\pmb{\mu}$), hence it unravels how these properties interact together to yield an epidemic.

\subsubsection{Power-law degree distribution}
Poisson degree distribution provides a formalism for {\em homogeneous} networks, where the degree sequence of the graph is highly concentrated around the mean degree. However, degree sequences in real-world networks were observed to be heavily skewed to the right \cite{barabasi2016network, moreno2002epidemic, newman2002spread}, meaning that the distribution is {\em heterogeneous}, or heavy-tailed. We consider Power-law degree distribution with exponential cutoff since they are relevant to a variety of real-world networks \cite{newman2002spread,leicht2009percolation}. In particular, we set
\begin{equation}\nonumber
p_k = 
\begin{cases}
0 & \text{if }k=0\\
\left( \mathrm{Li}_\gamma \left( e^{-1/\Gamma} \right) \right)^{-1} k^{-\gamma} e^{-k / \Gamma} & \text{if }k=1,2,\ldots.
\end{cases}
\end{equation}
where $\gamma$ and $\Gamma$ are positive constants and $\mathrm{Li}_m (z)$ is the $m$th polylogarithm of $z$, i.e., $\mathrm{Li}_m (z) = \sum_{k=1}^\infty \frac{z^k}{k^m}$. Observe that condition~(\ref{eq:condition}) now translates to 
\begin{equation} 
\left( \frac{\mathrm{Li}_{\gamma-2} \left( e^{-1/\Gamma} \right) - \mathrm{Li}_{\gamma-1} \left( e^{-1/\Gamma} \right)}{\mathrm{Li}_{\gamma-1} \left( e^{-1/\Gamma} \right)} \right)  \times \rho \left( \pmb{T \mu} \right) = 1 
\label{eq:PlawPhase}
\end{equation}

Similar to (\ref{eq:PoissonPhase}), condition~(\ref{eq:PlawPhase}) indicates how the structure of the underlying network, the characteristics of propagation, and the process of evolution are intertwined together, and under what conditions their relationship would induce an epidemic.

\begin{figure}[t!]
    \centering
    \subfigure[]{\hspace{-0.4cm} 
    \includegraphics[scale=0.39] {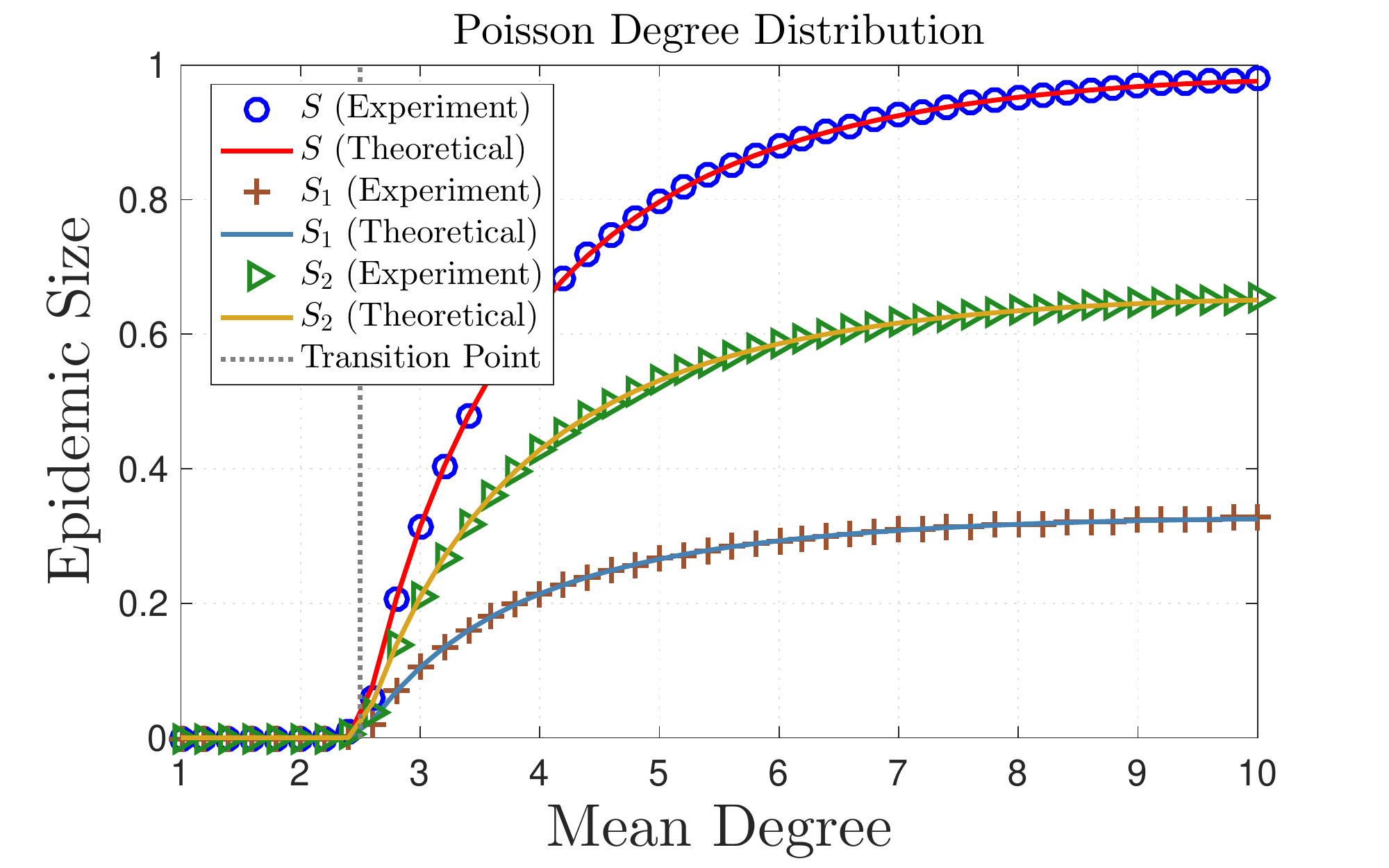}}
    \subfigure[]{\hspace{0.4cm} 
    \includegraphics[scale=0.39] {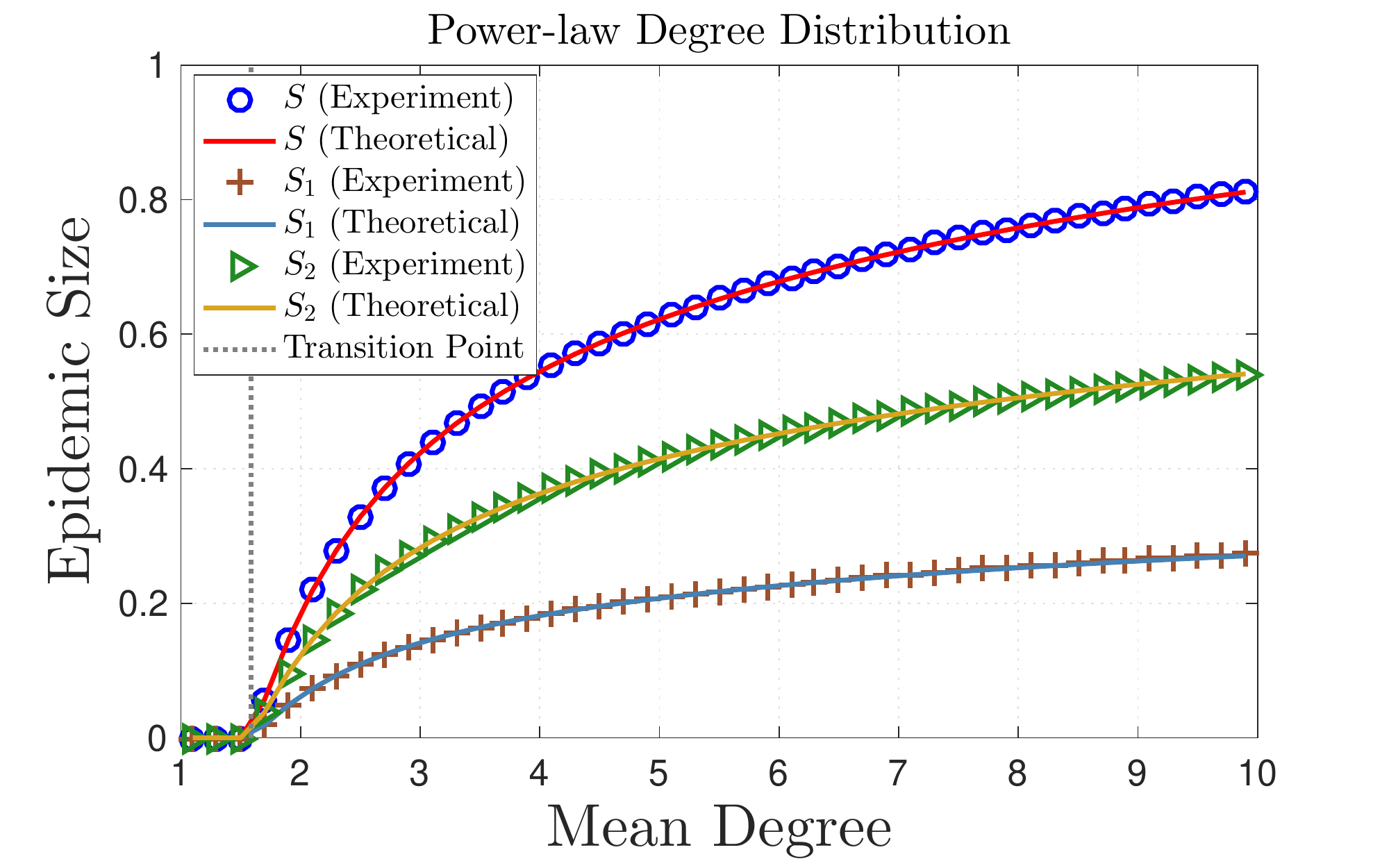}}
    \subfigure[]{\hspace{-0.4cm} 
    \includegraphics[scale=0.39] {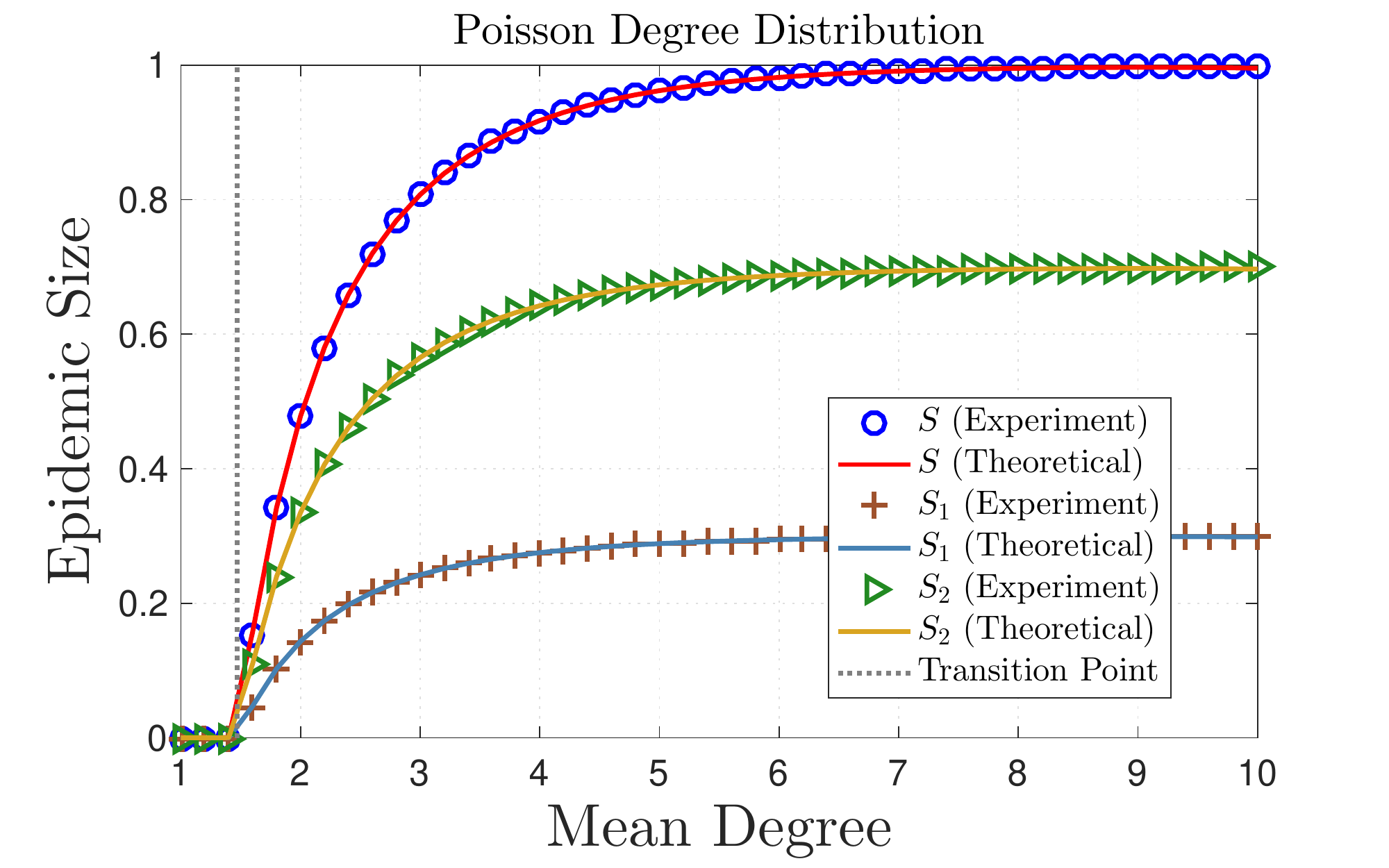}}
    \subfigure[]{\hspace{0.4cm} 
    \includegraphics[scale=0.39] {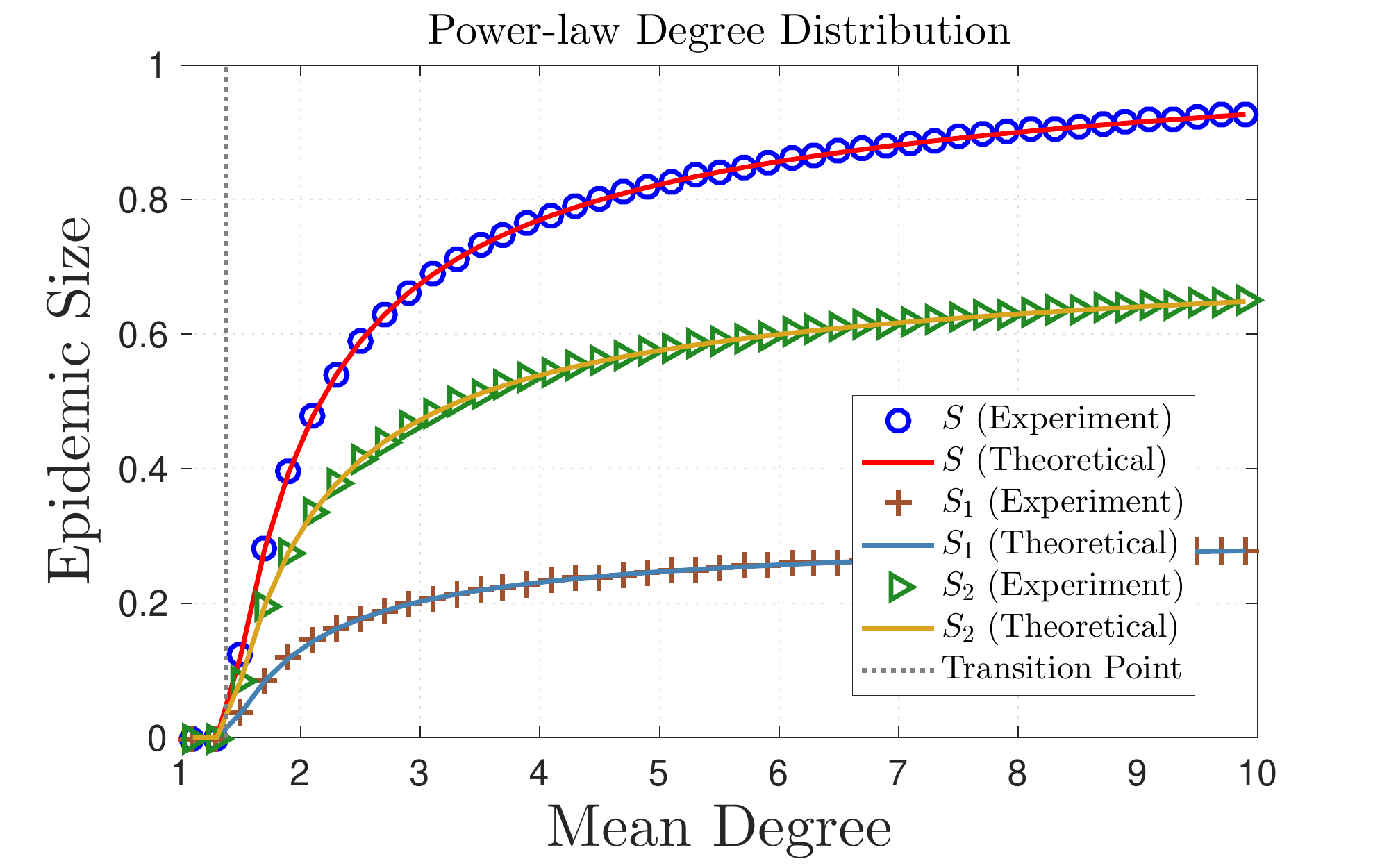}}
    \vspace{4mm}
    \caption{\sl 
   {\bf Evolution on Poisson and Power-law contact networks}. The network size $n$ is $2 \times 10^5$ and the number of independent experiments for each data point is $500$. Blue circles, brown plus signs, and green triangles denote the empirical average epidemic size, average fraction of nodes infected with strain-$1$, and average fraction of nodes infected with strain-$2$, respectively. The red, blue, and yellow lines denote the theoretical average total epidemic size, average fraction of nodes infected with strain-$1$, and average fraction of nodes infected with strain-$2$, respectively. Theoretical results are obtained by solving the system of equations (\ref{eq:finalSize}) with the corresponding parameter set. (a)-(b) We set $T_1 = 0.2$, $T_2 = 0.5$, $\mu_{11}=\mu_{22}=0.75$. (c)-(d) We set $T_1 = 0.4$, $T_2 = 0.8$, and $\mu_{11}=0.3$, and $\mu_{22}=0.7$ implying that an infected node, regardless of what type of infection it has, mutates to strain-$1$ (respectively, strain-$2$) with probability $0.3$ (respectively, $0.7$), independently. In all cases, we observe good agreement with our theoretical results.}.
\label{fig:EvoSize}
\end{figure}

\subsection{Notations and Methods}
{\em Notations:} In what follows, we use $S$, $S_1$ and $S_2$ to denote the {\em total} expected epidemic size, the expected fraction of nodes infected with strain-$1$, and the expected fraction of nodes infected with strain-$2$, respectively and all at the steady state, i.e., when the process terminates. We use $P_1^{\mathrm{BP}}$ and $P_2^{\mathrm{BP}}$ to denote the probability of emergence on a single-strain bond-percolated network with $T_1$ and  the probability of emergence on a single-strain bond-percolated network with $T_2$, respectively. 

{\em Methods:} We use the configuration model to create random random graphs with particular degree distributions. In particular, we sample a degree sequence from the corresponding distribution, then we use the configuration model to construct a random graph with that degree sequence. We use igraph \cite{igrpah} on both C++ and Python for simulations. Our simulation codes are available online \footnote{https://github.com/reletreby/evolution.git}. Unless otherwise stated, we start the process by selecting a node uniformly at random and infecting it with strain-$1$. The node infects each neighbor independently with probability $T_1$. Each of the infected neighbors mutate independently to strain-$1$ with probability $\mu_{11}$, or to strain-$2$ with probability $\mu_{12}$. As the process continues to grow, both strains might exist in the population. An intermediate node that becomes infected with strain-$i$ would mutate to strain-$1$ with probability $\mu_{i1}$, or strain-$2$ with probability $\mu_{i2}$, for $i=1,2$. When cycles start to appear, a susceptible node could be exposed to multiple infections at once. If a node is exposed to $x$ infections of strain-$1$ and $y$ infections of strain-$2$ simultaneously, the node becomes infected with strain-$1$ (respectively, strain-$2$) with probability $x/(x+y)$ (respectively, $y/(x+y)$) for any non-negative constants $x$ and $y$. A node that receives infection at round $i$ mutate first (by the end of round $i$) before it attempts to infect her neighbors at round $i+1$. The node is considered {\em recovered} at round $i+2$, i.e., a node is infective for only one round. 

%To distinguish between epidemics and self-limited outbreaks, we set a threshold of $0.005n$ for a network of $n$ nodes. If the number of recovered nodes at steady state is larger than $0.005n$, we classify the outbreak as an epidemic, otherwise as a self-limited outbreak \footnote{{lor{\mycolorRevv} The fraction $0.005$ was determined empirically by {\em averaging} the fractional size of the giant cluster over $500$ independent Erd\H{o}s-R\'enyi graphs, each with $5,000,000$ nodes and mean degree $1$}.}.

\subsection{Epidemic Size}
We start by focusing on the total epidemic size and the expected fraction of nodes that were infected with strain-$1$ and strain-$2$. The network size $n$ is set to $2 \times 10^5$. We consider two parameter sets that emphasize the correlations between a node's eventual type (after mutation) and the type of infection it has originally received. In particular, we have
\begin{itemize}
\item[-] {\bf Parameter set 1:} $T_1 = 0.2$, $T_2=0.5$, $\mu_{11} = 0.75$, and $\mu_{22}=0.75$.
\item[-] {\bf Parameter set 2:} $T_1 = 0.4$, $T_2=0.8$, $\mu_{11} = 0.3$, and $\mu_{22}=0.7$.
\end{itemize}
Observe that we have $ \mu_{11}=\mu_{21}$ and $\mu_{22} = \mu_{12}$ for the second parameter set. Hence, an infected node, regardless of what type of infection it has, mutates to strain-$1$ (respectively, strain-$2$) with probability $0.3$ (respectively, $0.7$), independently. This is a special case that can easily be treated by our formalism given in Section~\ref{sec:analysis}.

In Figure~\ref{fig:EvoSize}a and Figure~\ref{fig:EvoSize}b, we use the first parameter set and run $500$ independent experiments for each data point. We demonstrate our results on contact networks with Poisson degree distribution (Figure~\ref{fig:EvoSize}a) and Power-law degree distribution with exponential cutoff (Figure~\ref{fig:EvoSize}b). For Figure~\ref{fig:EvoSize}b, we set $\Gamma = 15$, and vary $\gamma$ with the mean degree. In particular, the mean degree $\lambda$ is given by
\begin{equation}
\lambda = \frac{\mathrm{Li}_{\gamma-1} \left( e^{-1/\Gamma} \right)}{\mathrm{Li}_{\gamma} \left( e^{-1/\Gamma} \right)}.
\label{eq:PlawMeanDeg}
\end{equation}
Hence, we can numerically solve (\ref{eq:PlawMeanDeg}) to obtain the particular value of $\gamma$ corresponding to a given value of $\lambda$.

In order to establish the validity of our analytic results given in Section~\ref{sec:analysis}, we plot the theoretical values of $S$, $S_1$, and $S_2$ obtained by solving the system of equations (\ref{eq:finalSize}) with the corresponding parameter set. We also plot a vertical line at the critical mean degree that corresponds to a phase transition (see (\ref{eq:PoissonPhase}) and (\ref{eq:PlawPhase})). Clearly, our experimental results are in perfect agreement with our theoretical results on both contact networks. In Figure~\ref{fig:EvoSize}c and Figure~\ref{fig:EvoSize}d, we repeat the same procedure, but with the second parameter set. Similarly, we observe perfect agreement with our theoretical results on both contact networks.

%In Figure~\ref{fig:EvoSize}c and Figure~\ref{fig:EvoSize}d, we repeat the same procedure, but with the second parameter set. The agreement between the experimental results and the theoretical results indicate the validity of our general results in the special case where the matrix $\pmb{\mu}$ is not full rank, hence no correlations exists between the node's eventual type after mutation and the type of infection it has originally received.

\subsection{Probability of Emergence}\
In \cite{alexander2010risk}, Alexander and Day investigated the probability of emergence for the multiple strain model presented in Section~\ref{sec:model}. However, authors did not provide a comprehensive simulation study to validate their formalism on random or real-world networks. Instead, in \cite[Section~3]{alexander2010risk}, authors only evaluated their equations {\em numerically}. In this subsection, we aim to establish the validity of the results presented in  \cite{alexander2010risk} on random networks generated by the configuration model. For brevity, we limit our scope to contact networks with Poisson degree distribution. However, similar patterns are observed for contact networks with Power-law degree distribution.  

\begin{figure}[t!]
    \centering
    \subfigure[]{\hspace{-0.4cm} 
    \includegraphics[scale=0.39] {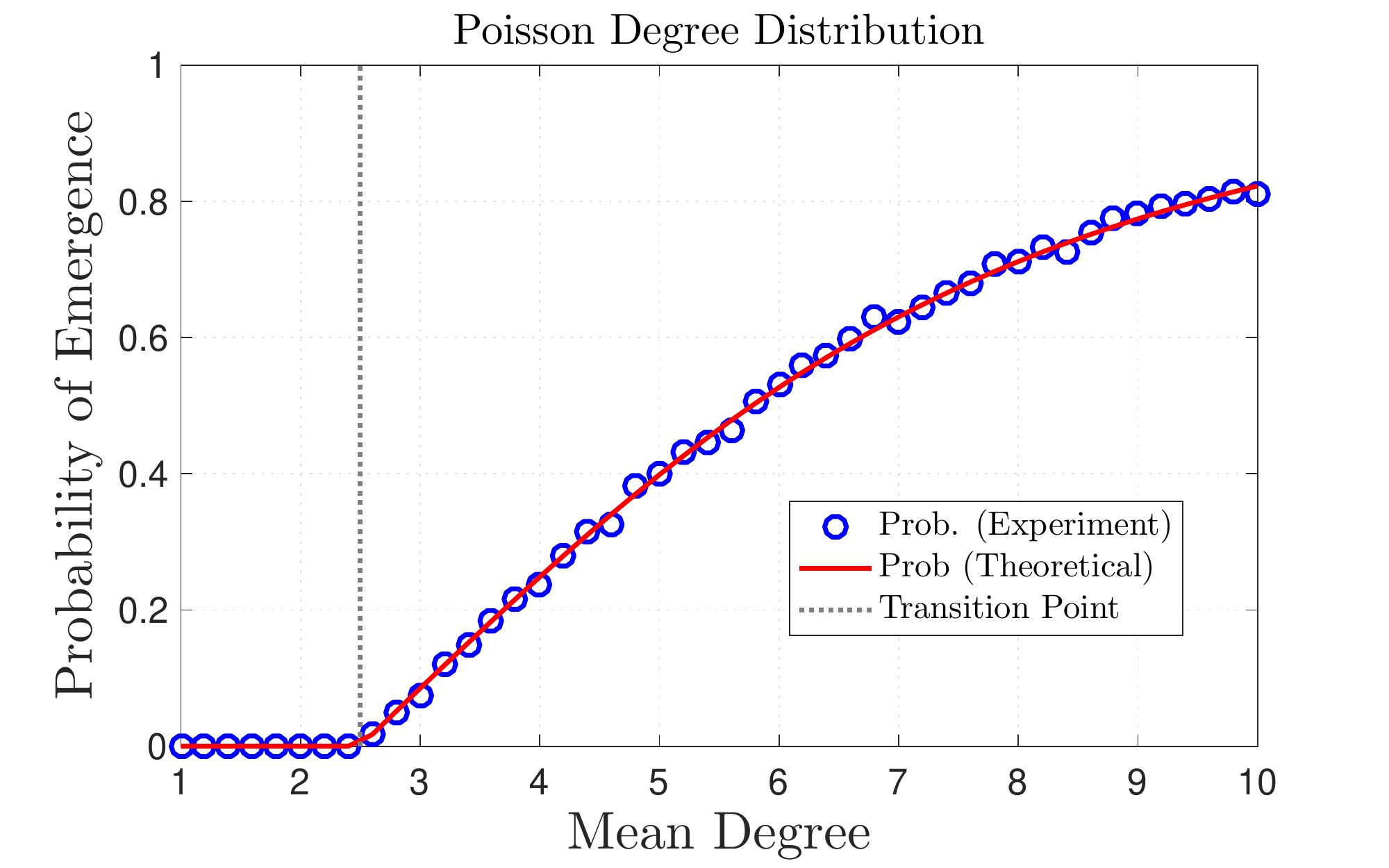}}
    \subfigure[]{\hspace{0.4cm} 
    \includegraphics[scale=0.39] {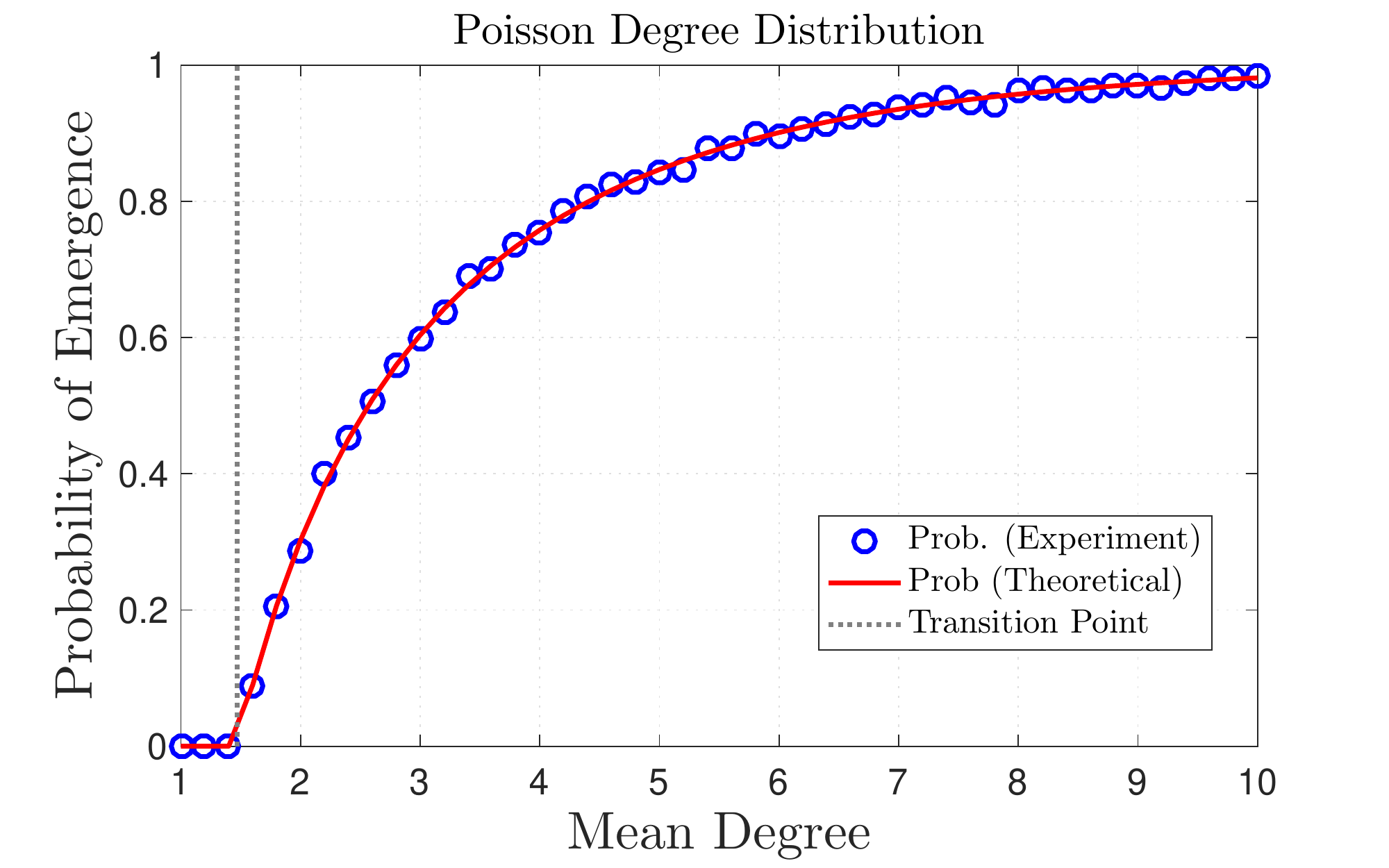}}
    \vspace{4mm}
    \caption{\sl 
    		{\bf The probability of emergence on contact networks with Poisson degree distribution}. The network size $n$ is $5 \times 10^5$ and the number of independent experiments for data point is $10^4$. Blue circles denote the empirical probability of emergence while the red line denotes the theoretical probability of emergence according to \cite{alexander2010risk}. (a) We set $T_1 = 0.2$, $T_2 = 0.5$, $\mu_{11}=\mu_{22}=0.75$. (b) We set $T_1 = 0.4$, $T_2 = 0.8$, and $\mu_{11}=0.3$, and $\mu_{22}=0.7$. Our experimental results prove the validity of the formalism presented by Alexander and Day in \cite{alexander2010risk}}.
\label{fig:ProbSim}
\end{figure}

In Figure~\ref{fig:ProbSim}, we set the network size $n = 5 \times 10^5$ and run a computer simulation with $10^4$ independent experiment for each data point. We use the two parameter sets given in Section~\ref{sec:analysis}.C. Namely, we set
\begin{itemize}
\item[-] $T_1 = 0.2$, $T_2 = 0.5$, and $\mu_{11}=\mu_{22}=0.75$ for Figure~\ref{fig:ProbSim}.a, and
\item[-] $T_1 = 0.4$, $T_2 = 0.8$, $\mu_{11}=0.3$ and $\mu_{22}=0.7$ for Figure~\ref{fig:ProbSim}.b.
\end{itemize} 

Note that in Figure~\ref{fig:ProbSim}, we plot the probability of emergence conditioned on the initial node receiving infection with strain-$1$ \footnote{We remark that the formalism provided by Alexander and Day allows for computing the probability of emergence given any arbitrary initial type.}. We observe an agreement between our experimental results and the theoretical results given in \cite{alexander2010risk}. The reasoning behind this is intuitive; the multi-type branching framework assumes that the underlying graph is tree-like, an assumption that works best for networks with vanishingly small clustering coefficient, e.g., networks which are generated by the configuration model.

\subsection{Reduction to Single-Type Bond-Percolation}
An important question to ask is whether the classical single-type bond percolation models could predict the threshold, probability, and final size of epidemics that entail {\em evolution}, i.e., information or diseases that propagate according to the multiple-strain model given in Section~\ref{sec:model}. In pursing an answer to this question, we start by establishing a {\em matching condition} between single-strain models and multiple-strain models for epidemics.

In \cite{newman2002spread}, Newman proposed a stochastic SIR model for the propagation of a single-strain pathogen on a contact network. Newman showed that, under some conditions, the SIR model is isomorphic to a bond-percolation model on the underlying contact network. Specifically, with the {\em average transmissibility} of the pathogen (denoted $T_{\mathrm{BP}}$) as the bond-percolation parameter, if we are to occupy each edge of the network with probability $T_{\mathrm{BP}}$, then the probability of emergence as well as the final size of the epidemic are precisely given by the fraction of nodes in the giant component of the {\em percolated} graph. Finally, it was shown that a phase transition occurs when 
\begin{equation}
\left( \dfrac{ \langle k^2  \rangle -  \langle k  \rangle}{ \langle k  \rangle} \right) T_{\mathrm{BP}} = 1
\label{eq:BPphase}
\end{equation}
In other words, if the left hand side of (\ref{eq:BPphase}) is strictly larger than $1$, a giant component emerges indicating an epidemic. Otherwise, we have self-limited outbreaks.

Comparing (\ref{eq:condition}) to (\ref{eq:BPphase}) suggests the proposal of a matching that results in the same condition for phase transition. More precisely, if we are to set
\begin{equation}
T_{\mathrm{BP}} = \rho \left(\pmb{T} \pmb{\mu} \right)
\label{eq:matchingCondition}
\end{equation}
then, both (\ref{eq:condition}) and (\ref{eq:BPphase}) collapse to the same condition for a given contact network. In what follows, we explore the extent to which classical, single-type bond-percolation models (under the matching condition (\ref{eq:matchingCondition})) may predict the threshold, probability, and final size of epidemics that entail evolution, i.e., information or diseases that propagate according to the multiple-strain model given in Section~\ref{sec:model}. We focus on contact networks with Poisson degree distribution, generated by the configuration model, while we devote Section~\ref{sec:realworld} for real-world networks. 

\begin{figure}[t!]
    \centering
    \subfigure[]{\hspace{-0.4cm} 
    \includegraphics[scale=0.39] {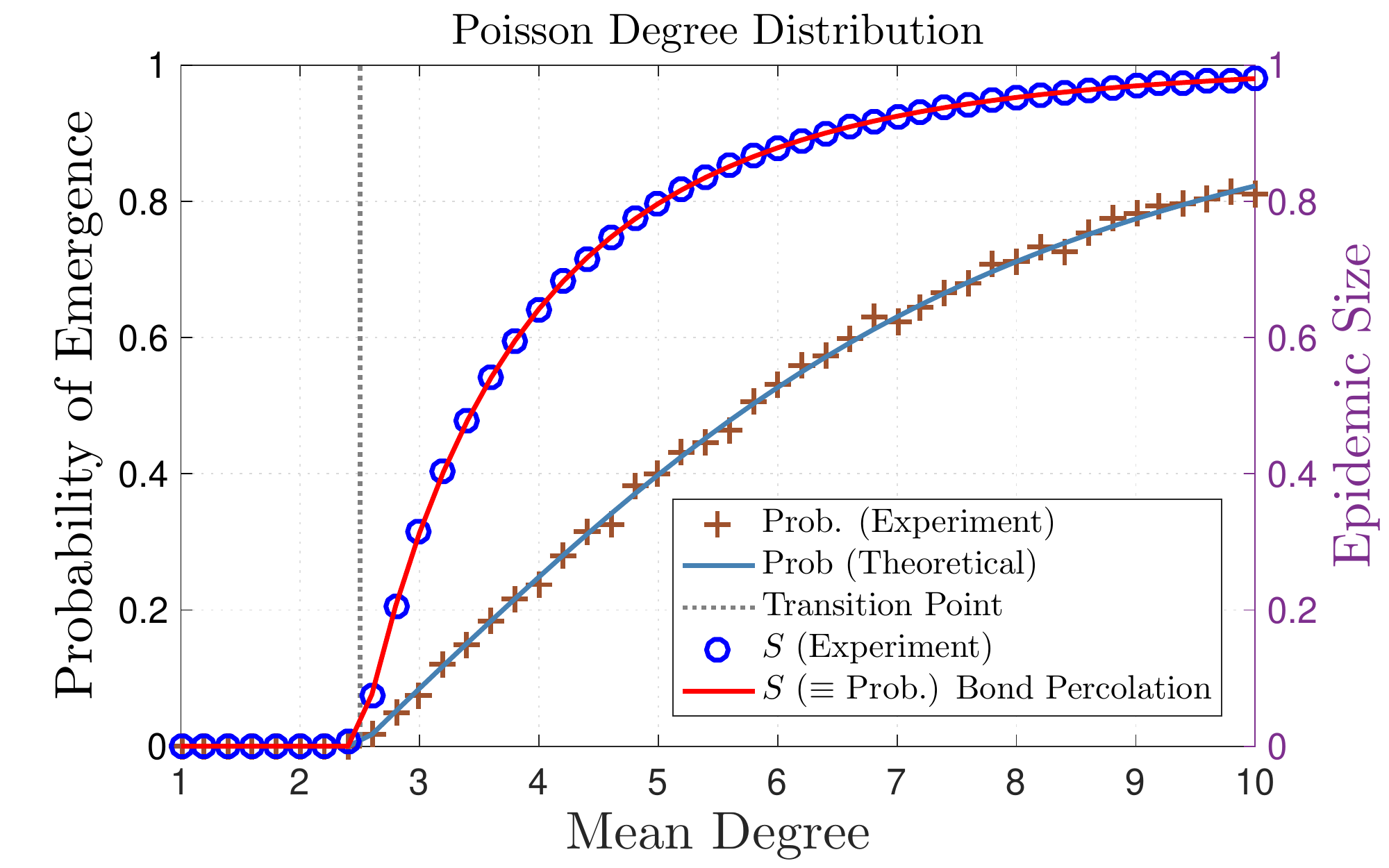}}
    \subfigure[]{\hspace{0.4cm} 
    \includegraphics[scale=0.39] {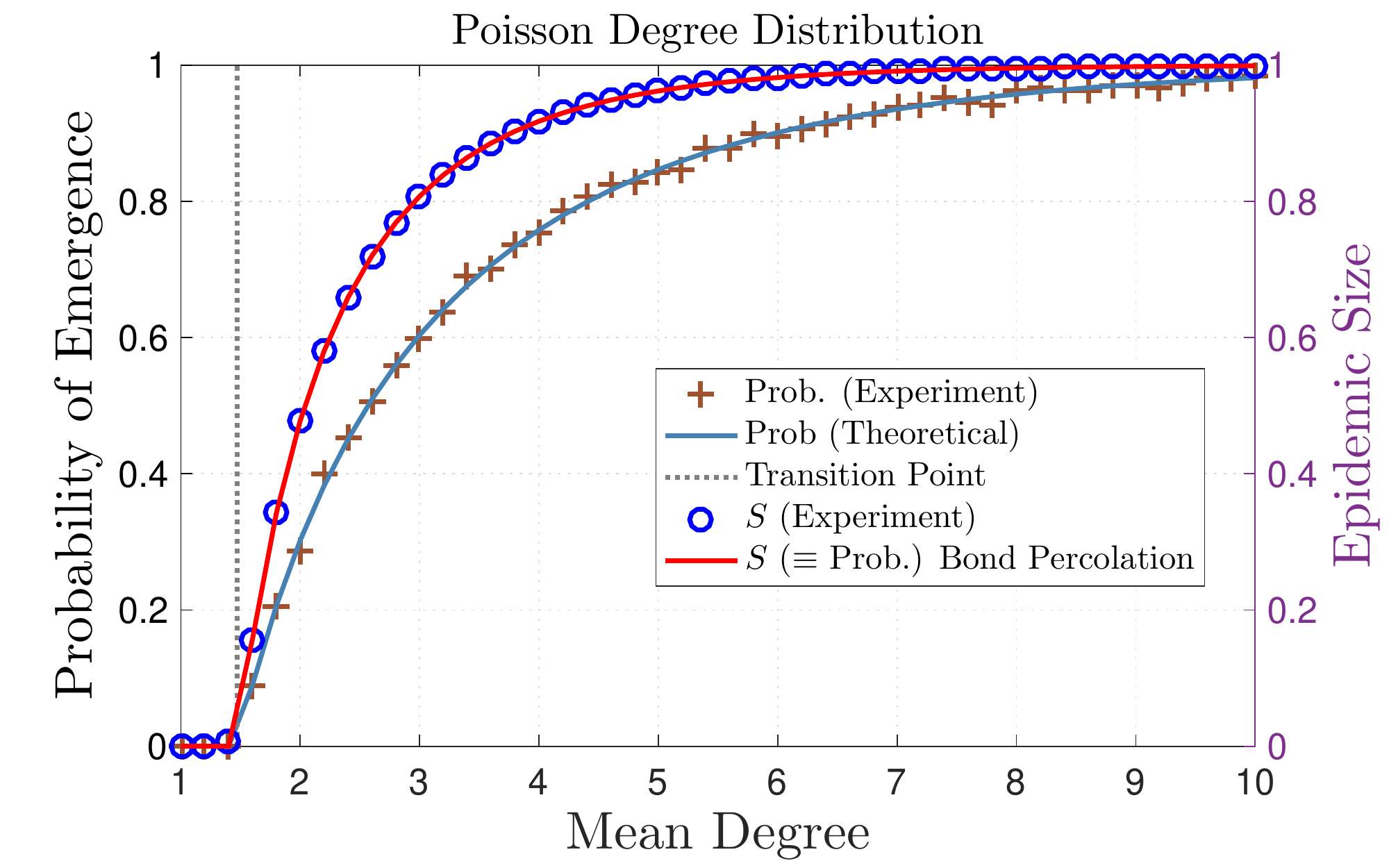}}
    \vspace{4mm}
    \caption{\sl  
    		{\bf Reduction to single-type bond-percolation}. The network size $n$ is $5 \times 10^5$ and the number of independent experiments for each data point is $10^4$. Blue circles and brown plus signs denote the empirical average epidemic size and the probability of emergence, respectively. The navy blue line denotes the theoretical probability of emergence according to \cite{alexander2010risk} while the red line denotes the theoretical average epidemic size (as well as the probability of emergence) predicted by the single-type bond-percolation framework under the matching condition (\ref{eq:matchingCondition}). (a) We set $T_1 = 0.2$, $T_2 = 0.5$, $\mu_{11}=\mu_{22}=0.75$. (b) We set $T_1 = 0.4$, $T_2 = 0.8$, and $\mu_{11}=0.3$, and $\mu_{22}=0.7$. The classical, single-type bond percolation models may accurately predict the threshold and final size of epidemics, but their predictions on the probability of emergence are clearly inaccurate.
    		}
\label{fig:ProbSimAll}
\end{figure}

In Figure~\ref{fig:ProbSimAll}, we extend Figure~\ref{fig:ProbSim} by further adding the experimental results for the final epidemic size as well as the corresponding theoretical values for the probability of emergence on a bond-percolated network under the matching condition (\ref{eq:matchingCondition}). Note that the probability of emergence is equivalent to the final epidemic size for single-type, bond-percolated networks \cite{newman2002spread}. Observe that the classical single-type bond-percolation model accurately captures the threshold and final size of epidemic but provides significantly inaccurate predictions when it comes to the probability of emergence. Similar pattern will be observed in Section~\ref{sec:realworld} for real-world networks. This inaccuracy sheds the light on a fundamental disconnect between the classical, single-type bond-percolation models and real-life spreading processes that entail evolution. We explain the intuition behind our findings in Appendix~\ref{app:explanation}.

\subsection{Effect of Mutation}
When only a single evolutionary pathway is available, mutations have to occur in a particular order \cite{gokhale2009pace}. In \cite{antia2003role}, Antia et al. considered the case where the fitness landscape consists of $m$ strains such that $R_{0,i}<1$ for $i=1,\ldots,m-1$, while $R_{0,m}>1$. Hence, an introduced pathogen (with $R_{0,1}<1$) must acquire $m-1$ successive mutations in order for the disease to emerge. Antia et al. derived a set of recursive equations whose solution characterizes the probability of emergence under some conditions; see \cite{antia2003role} for more details. To gain further insights on the effect of mutation, Antia et al. proposed a theoretical approximation of the probability of emergence as a product of the probability of mutation, i.e., the probability that the introduced pathogen would eventually mutate to strain-$m$, and the probability of emergence of strain-$m$. Indeed, the probability of mutation plays a key role in the overall extinction probability. After all, if the introduced pathogen does not gain $m-1$ successive mutations, the disease would eventually die out. 

Recall that the mathematical theory developed by Alexander and Day \cite{alexander2010risk} defines the probability of emergence as a function of the evolutionary dynamics of the pathogen (i.e., the mutation matrix $\pmb{\mu}$), the characteristics of the spreading process (i.e., the transmissibility matrix $\pmb{T}$), and the structure of the underlying contact network (i.e., the degree distribution $\{p_k, \quad k=0,1,\ldots\}$). All of these factors are intertwined together in a way that makes it difficult to predict how the probability of mutation influences the probability of emergence. In what follows, we provide a theoretical approximation to the probability of emergence in a way that clearly distinguishes the role of mutation and shows how it strongly influences the probability of emergence. 

Consider the case when the fitness landscape consists of two strains with transmissibility matrix $\pmb{T}$ and mutation matrix $\pmb{\mu}$ given by
\begin{equation}\nonumber
\pmb{T} = \left[
\begin{matrix}
T_1 & 0  \\
 0 & T_2 \\ 
\end{matrix} 
\right]
\quad
\mathrm{and}
\quad
\pmb{\mu}=
\left[
\begin{matrix}
1-\mu & \mu  \\
 0 & 1 \\ 
\end{matrix} 
\right].
\end{equation}
Assume also that $T_1<T_2$. Note that the process starts by picking a random individual uniformly at random and infecting her with strain-$1$. Fix the mean degree of the underlying network to $\lambda$. Let $\lambda_1$ and $\lambda_2$ denote the phase transition points (i.e., critical mean degrees) for a single-strain, bond-percolated network with $T_1$ and $T_2$, respectively. Observe that $\rho\left( \pmb{T \mu} \right) = T_2$, hence, in view of (\ref{eq:condition}), the phase transition is entirely controlled by the parameters of strain-$2$, i.e., the phase transition occurs at $\lambda_2$. Indeed, we can conclude from (\ref{eq:condition}) that for $\lambda<\lambda_2$, the probability of emergence is zero (in the limit of large network size). We can write
\begin{align}
&\mathbb{P}\left[\mathrm{emergence} \right]  = \mathbb{P}\left[\mathrm{emergence} \given[\big] \text{at least one mutation} \right] \times P_\mu \nonumber \\
& \quad + \mathbb{P}\left[\mathrm{emergence} \given[\big] \text{no mutation} \right] \times \left( 1- P_\mu \right)  \label{eq:approxFull}
\end{align}
where $P_\mu$ denotes the probability that at some point along the chain of infections (starting from the type-$1$ seed), a node would be infected by strain-$1$, but then mutate to strain-$2$. In other words, $P_\mu$ captures the probability that at some point during the propagation, a type-$2$ node would emerge.

Observe that for $\lambda < \lambda_1$, we have $ \mathbb{P}\left[\mathrm{emergence} \given[\big] \text{no mutation} \right] = 0$ in the limit of large network size (since $P_1^{\mathrm{BP}}=0$ on this interval), while for $\lambda \geq \lambda_1$, we have $P_\mu = 1$ in the limit of large network size \footnote{When $\lambda \geq \lambda_1$, a giant component of type-1 nodes emerges. Now, since $\mu>0$, and the number of nodes in the giant component tends to infinity in the limit of large network size, the probability that none of the nodes mutate to strain-$2$ is zero.}. Hence, the second term in (\ref{eq:approxFull}) is always zero in the limit of large network size, leading to
\begin{align}
&\mathbb{P}\left[\mathrm{emergence} \right] = \mathbb{P}\left[\mathrm{emergence} \given[\big] \text{at least one mutation} \right] \times P_\mu \nonumber 
\end{align}

Note that on the range $\lambda_2 \leq \lambda < \lambda_1$, we have $ \mathbb{P}\left[\mathrm{emergence} \given[\big] \text{at least one mutation} \right] = P_2^{\mathrm{BP}}$. However, on the range $\lambda \geq \lambda_1$, strain-$1$ nodes are able to form a giant component on their own. Hence, in the cases where a strain-$2$ node emerges at some point, but fails to infect any of her neighbors, strain-$1$ nodes could still trigger the emergence of the disease. It follows that $ \mathbb{P}\left[\mathrm{emergence} \given[\big] \text{at least one mutation} \right] \geq P_2^{\mathrm{BP}}$ on the range $\lambda \geq \lambda_2$. Note that the bound is tight whenever $T_2$ is significantly larger than $T_1$. The reasoning behind this can be explained as follows. Whenever $T_2$ is significantly larger than $T_1$, the average number of secondary infections of strain-$2$ would be much larger than that of strain-$1$. Hence, infections with strain-$2$ would propagate much faster and block potential pathways for strain-$1$ to propagate. In this case, the overall probability of emergence becomes tightly controlled by $P_2^{\mathrm{BP}}$. Next, we turn our attention to deriving $P_\mu$.

Consider a tree of infections that starts with a single node infected with strain-$1$. Let $H$ be the probability that strain-$2$ never appears throughout the tree, i.e., $H$ is the probability that the tree of infections starting from the seed does not give rise to strain-$2$ at any intermediate point. Similarly, let $h$ be the probability that a {\em subtree} of infections starting from a type-$1$ host does not give rise to strain-$2$ at any intermediate point. Recall that $G(.)$ gives the PGF of the excess degree distribution while $g(.)$ gives the PGF of the degree distribution. By conditioning on the {\em excess} degree as well the number of secondary infections, we get
\begin{align}
h &= \sum_{k=1}^\infty \frac{k p_k}{\langle k \rangle} \sum_{x=0}^{k-1} \binom{k-1}{x} \left( T_1  \left(1-\mu \right) \right)^x  \left( 1-T_1 \right)^{k-1-x}  h^x \nonumber \\
& =  \sum_{k=1}^\infty \frac{k p_k}{\langle k \rangle} \left(1-T_1 + T_1 \left(1-\mu\right) h \right)^{k-1} \nonumber \\
& = G \left( 1-T_1 + T_1 \left(1-\mu\right) h  \right) 
\label{eq:probApprox_h}
\end{align}

The validity of (\ref{eq:probApprox_h}) can be explained as follows. Note that the root of any subtree, say node $v$, has already used an edge to receive an infection with strain-$1$ from her parent. Hence, if the degree of node $v$ is $k$, then node $v$ is only using $k-1$ edges to infect her offspring,  leading us to use the excess degree distribution. Furthermore, conditioned on the excess degree being $k-1$, the number of secondary infections of each type generated by node $v$ is given by a multinomial distribution characterized by $(k-1, T_1 (1-\mu), T_1 \mu, 1-T_1)$. In particular, conditioned on node $v$ being type-$1$ and having an excess degree of $k-1$, the probability of generating $x$ infections of type-$1$ and $y$ infections of type-$2$ is given by 
\begin{equation}\nonumber
\binom{k-1}{x} \binom{k-1-x}{y} \left(T_1 \left(1-\mu \right) \right)^x \left(T_1 \mu \right)^y \left( 1- T_1 \right)^{k-1-x-y}
\end{equation}
However, the only relevant term for the computation of $h$ is the one with $y=0$, as all other terms with $y>0$ are contributing with a zero probability to $h$ by definition. Finally, $h^x$ denotes the probability that the subtrees emanating from the current $x$ offspring are themselves free of any strain-$2$ node. 

Recall that $H$ denotes the probability that strain-$2$ never appears throughout the tree (starting from the root) and note that if the tree root has degree $k$, then all of these $k$ edges will be utilized to connect with her neighbors at the lower level. Hence, in view of (\ref{eq:probApprox_h}), we can write
\begin{equation}\nonumber
H = g \left( 1-T_1 + T_1 \left(1-\mu\right) h_\infty  \right) 
\end{equation}
where $h_\infty$ denotes the steady-state solution of (\ref{eq:probApprox_h}). It is now immediate that $P_\mu = 1-H$, leading to
\begin{equation}
\mathbb{P}\left[\mathrm{emergence} \right] \geq \left(1-H\right) P_2^{\mathrm{BP}}
\label{eq:probApprox_final}
\end{equation}

\begin{figure}[t!]
    \centering
    \subfigure[]{\hspace{-0.4cm} 
    \includegraphics[scale=0.39] {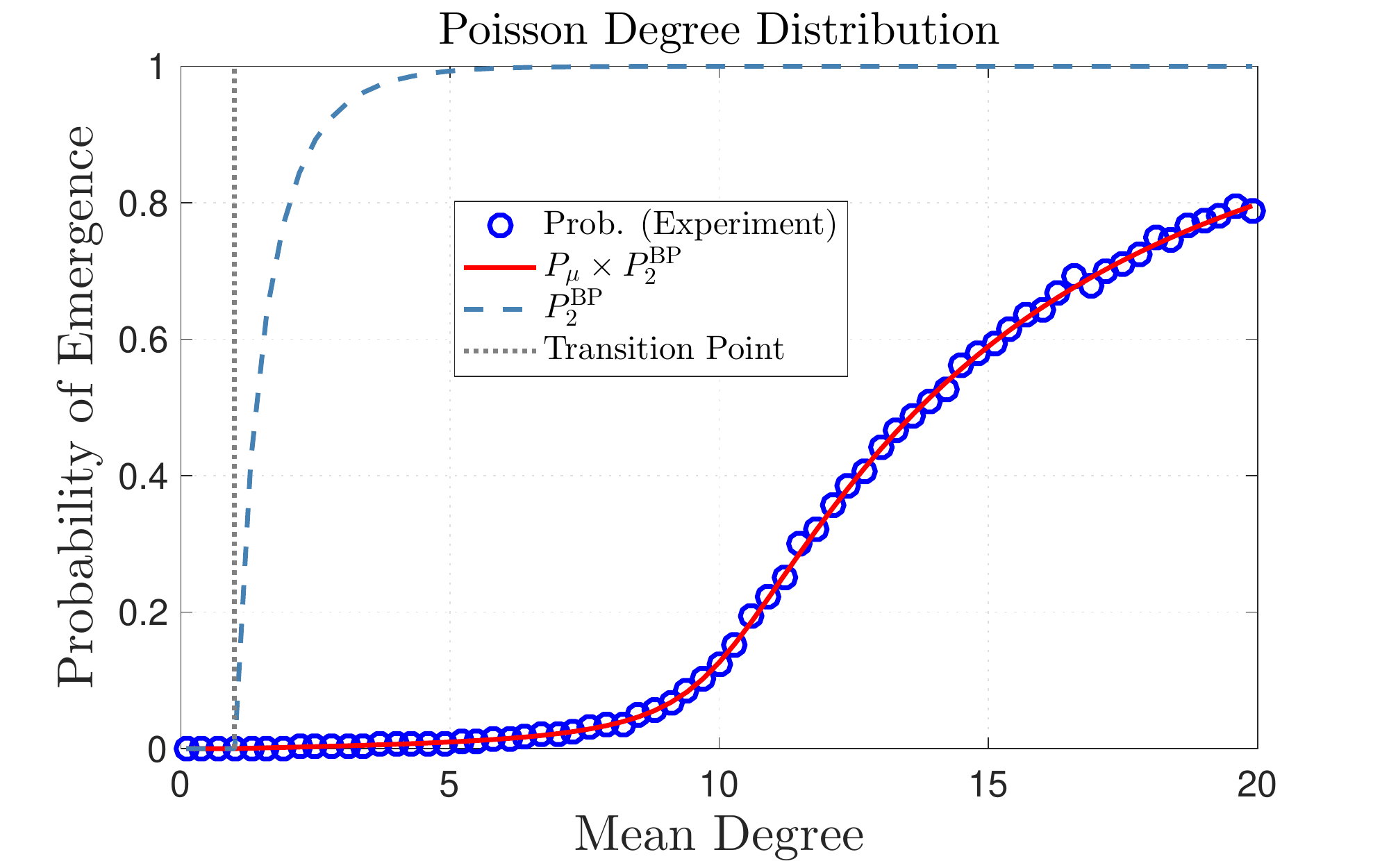}}
    \subfigure[]{\hspace{0.4cm} 
    \includegraphics[scale=0.39] {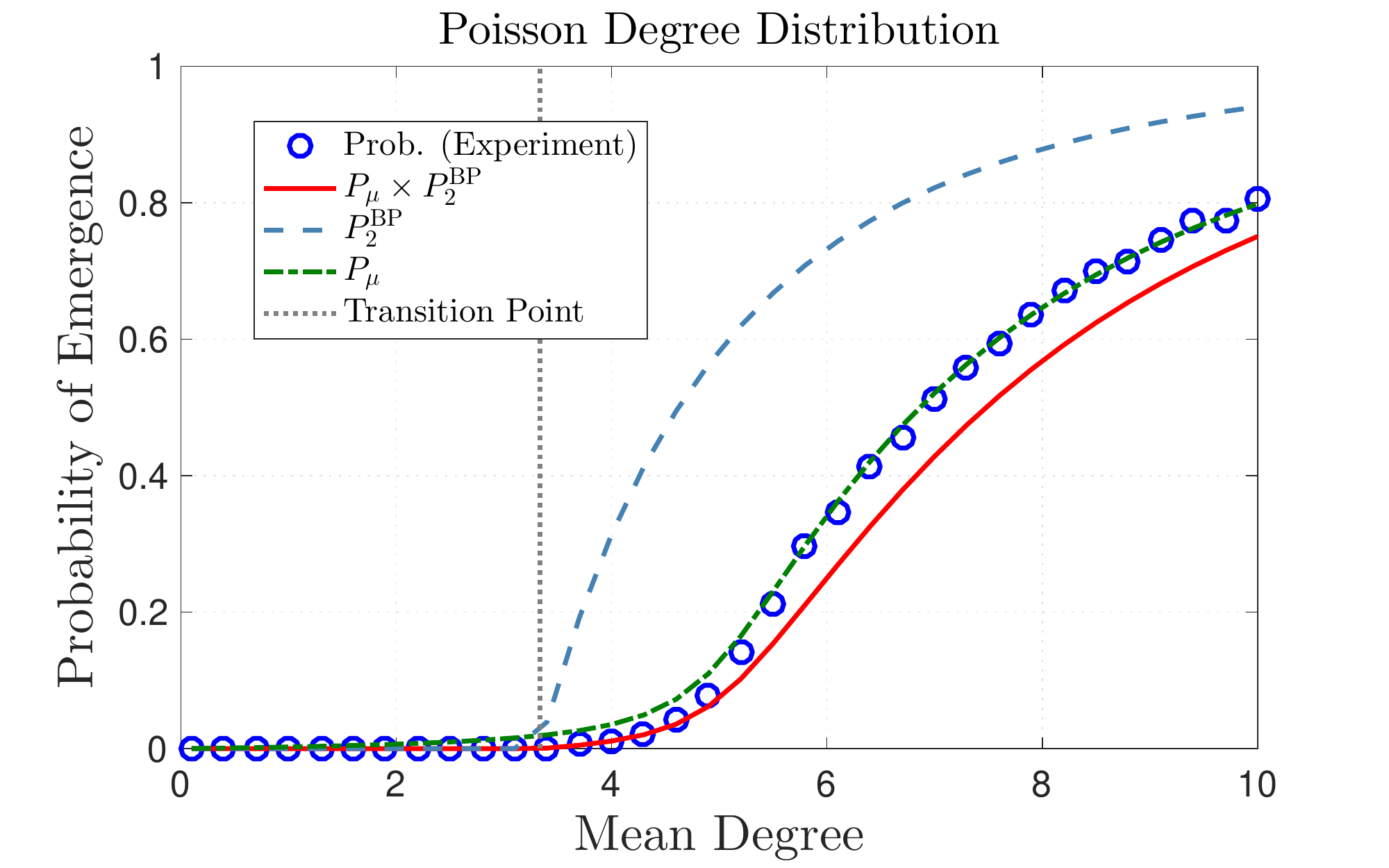}}
    \vspace{4mm}
    \caption{\sl {\bf Approximating the probability of emergence:} The network size $n$ is $2 \times 10^5$ and the number of independent experiments for each data point is $10^4$. Blue circles denote the empirical probability of emergence while the red line denotes the theoretical approximation of the probability of emergence according to (\ref{eq:probApprox_final}). The light blue dashed line denotes the probability of emergence for a single-strain, bond-percolated network with $T_2$. (a) We set $T_1=0.1$, $T_2 = 1$, and $\mu=0.01$. (b) We set $T_1=0.2$, $T_2 = 0.3$, and $\mu=0.01$. We observe good agreement between the experimental results and the theoretical approximation given by (\ref{eq:probApprox_final}) whenever $\lambda_2 \leq \lambda < \lambda_1$ or whenever $T_2$ is significantly larger than $T_1$.
    }
\label{fig:ProbApprox1}
\end{figure}

To confirm the validity of (\ref{eq:probApprox_final}), we run a computer simulation on random networks generated by the configuration model with Poisson degree distribution. In Figure~\ref{fig:ProbApprox1}, we set the network size $n =2 \times 10^5$ and perform $10^4$ independent experiments for each data point. 
In Figure~\ref{fig:ProbApprox1}a, we set $T_1 = 0.1$, $T_2 = 1$, and $\mu=0.01$. Observe that the bound given by (\ref{eq:probApprox_final}) is {\em tight}, as $T_2$ is significantly larger than $T_1$. In general, we would expect a tight bound whenever $\lambda_2 \leq \lambda < \lambda_1$ (i.e., $1 \leq \lambda < 10$ for the given parameter set). As $\lambda$ increases beyond $\lambda_1$, the tightness of the bound depends on the ratio between $T_2$ to $T_1$. This is illustrated in Figure~\ref{fig:ProbApprox1}b for the case when $T_1=0.2$ and $T_2=0.3$.

\begin{figure}[t!]
    \centering
    \hspace*{-10mm}
    \includegraphics[scale=0.5] {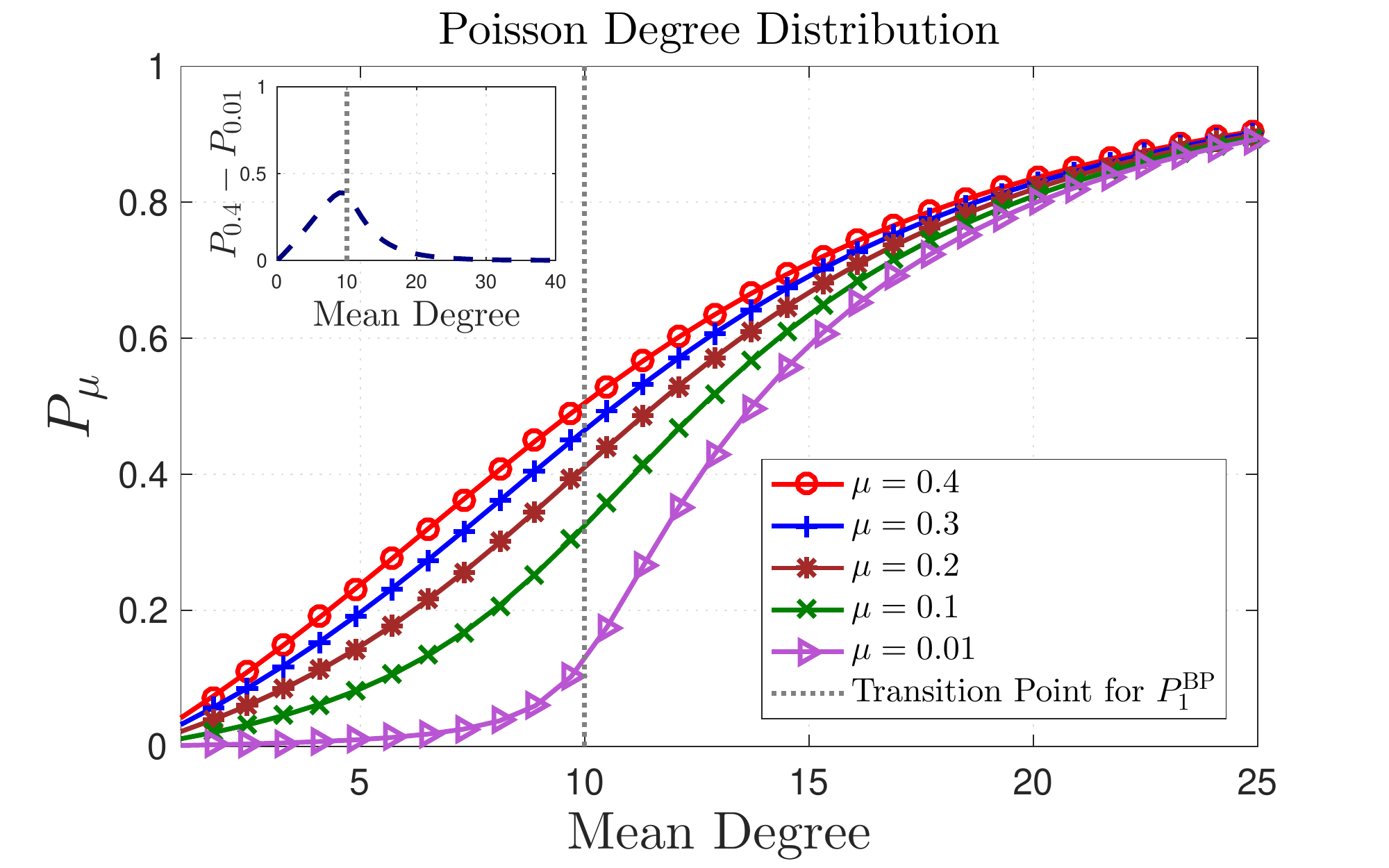}
    \caption{\sl {\bf Effect of Mutation:} We set $T_1=0.1$ and plot the behavior of $P_\mu$ against the mean degree for a network with Poisson degree distribution. Intuitively, different values of $\mu$ have different impact on $P_\mu$. The impact is pronounced before the critical mean degree corresponding to a single-strain, bond-percolated network with $T_1$. Inset: The difference between the value of $P_\mu$ when $\mu=0.4$ and the value of $P_\mu$ when $\mu=0.01$ as a function of the mean degree of the underlying contact network.
    }
\label{fig:ProbApprox2}
\end{figure}

The availability of an explicit expression for the probability of mutation allows for exploring the effects of mutation on the overall probability of emergence. Indeed, the way the probability of emergence behaves with respect to changes in the mean degree resembles, to a great extent, the way $P_\mu$ behaves, as illustrated in Figure~\ref{fig:ProbApprox1}. Hence, in what follows, we focus on the behavior of $P_\mu$ with respect to changes in the mean degree.
In Figure~\ref{fig:ProbApprox2}, we set $T_1=0.1$ and plot $P_\mu$ against the mean degree for a network with Poisson degree distribution. We observe that different values for $\mu$ impacts the shape of $P_\mu$ (hence, the probability of emergence) in a remarkable way. Firstly, for all values of $\mu \in (0,1)$, the behavior of $P_\mu$ appears to be strikingly different than the universality class of percolation models, e.g., see the shape of the probability of emergence (respectively, $P_2^{\mathrm{B}}$) in Figure~\ref{fig:ProbSim} (respectively, Figure~\ref{fig:ProbApprox1}). Secondly, the effect of mutation probabilities on $P_\mu$ appears to be significant as the mean degree increases from small values, reaches its peak right before the critical mean degree corresponding to $P_1^{\mathrm{BP}}$, then decays as the mean degree increases further.

The reasoning behind the aforementioned observation is intuitive. Recall that the process starts with a single infection with strain-$1$ and note that $P_\mu$ is influenced by the structure of the underlying contact network, the transmissibility of strain-$1$, and the particular value of $\mu$. As the mean degree $\lambda$ increases towards $\lambda_1$, the length of the tree of infections starting from the seed \footnote{The length of the tree of infections can be interpreted as the size of the component (of a bond percolated network with $T_1$) that contains the seed.} also increases, however, no cycles appear and the epidemic propagates on a finite, tree-like percolated network (since $\lambda<\lambda_1$). Increasing the length of the tree increases the probability that at least one intermediate node would mutate to strain-$2$, but the fact that the tree is finite makes the particular value of $\mu$ very crucial to $P_\mu$. Namely, a small value of $\mu$ makes it less likely that a mutant emerges before the chain of infections is terminated, while a relatively larger value could drive the emergence of strain-$2$ and lead the epidemic to escape extinction. Put differently, the finiteness of the chain of infections when $\lambda < \lambda_1$ creates a limited number of opportunities for mutation, causing the particular value of $\mu$ to {\em bear the burden} of generating a mutant and driving the whole process to emergence. However, as $\lambda$ increases beyond $\lambda_1$, cycles start to appear and a giant component of nodes infected with strain-$1$ emerges. In this case, the chain of infections is no longer finite, and any positive value of $\mu$ results in a mutation almost surely in the limit of large network size. Put differently, when $\lambda \geq \lambda_1$, the structure of the underlying network starts to facilitate the emergence of strain-$2$, hence reducing the dependence on $\mu$.

\section{Evolution in real-world networks}
\label{sec:realworld}
In Section~\ref{sec:numerical}.F, we explored the validity of analyzing the multiple-strain model for evolution with the available tools from the classical, single-type bond-percolation framework. We focused on {\em random} networks generated by the configuration model and demonstrated that a reduction to the classical, single-type bond percolation framework leads to accurate results with respect to the threshold and final size of epidemics, but significantly inaccurate results with respect to the probability of emergence. In this section, we aim to examine the universality of our findings by analyzing the probability of emergence on {\em real-world} contact networks obtained from SNAP data sets \cite{snapnets}. Our objective is twofold. Firstly, we would like to validate the multi-type branching formalism of Alexander and Day (see Section~\ref{sec:analysis}.A) on real-world networks. Secondly, we seek to highlight and confirm the limitations of the single-type bond-percolation framework in predicting the probability of emergence on real-world networks.

\begin{figure}[t!]
\centering
 \includegraphics[scale=0.9] {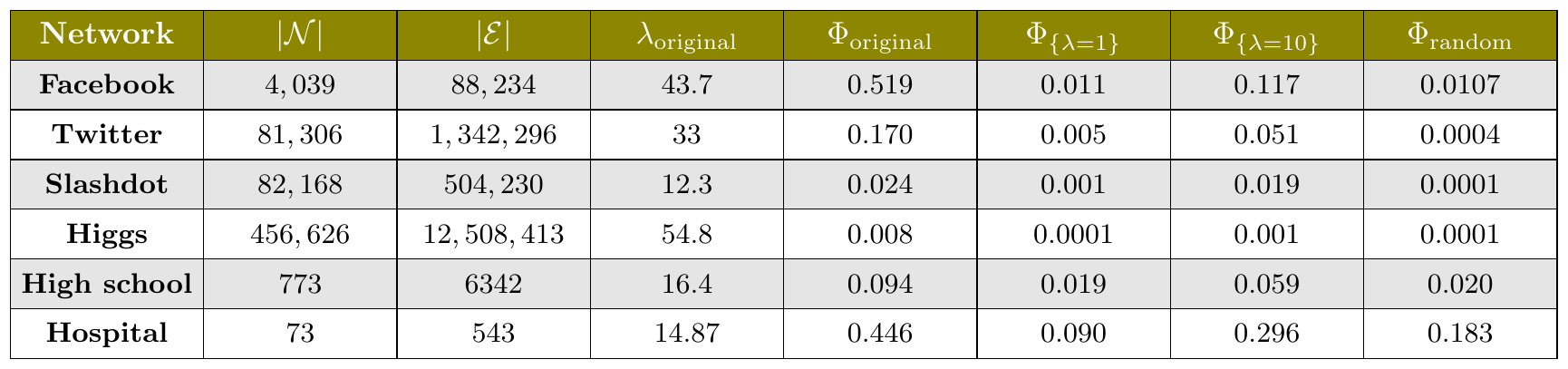}
\caption{\sl {\bf Real-world contact networks}. We consider four real-world contact networks in the context of information propagation, namely, Facebook, Twitter, Slashdot, and Higgs networks from SNAP \cite{snapnets} dataset. We also consider two real-world contact networks in the context of infectious disease propagation, namely, a contact network among students, teachers, and staff at a US high school \cite{Salath22020} and a contact network among professional staff and patients in a hospital in Lyon, France \cite{Vanhems2013}. For each network, we indicate the number of nodes $|\mathcal{N}|$, the number of edges $|\mathcal{E}|$, the mean degree of the original network $\lambda_{\mathrm{original}}$, and the clustering coefficient of the original network $\Phi_{\mathrm{original}}$. $\Phi_{ \{\lambda = 1 \}}$ (respectively, $\Phi_{ \{\lambda = 10 \}}$) denotes the clustering coefficient of the original network after removing a random subset of edges such that the resulting mean degree is $1$ (respectively, $10$). $\Phi_{\mathrm{random}}$ denotes the average clustering coefficient (over $200$ independent realizations) of a random network generated by the configuration model with Poisson degree distribution. The random network has the same number of nodes and the same (original) mean degree of the corresponding real-world network.}
 \label{fig:table}
\end{figure}

\begin{figure}[t!]
    \centering
    \subfigure[]{\hspace{-0.4cm} 
    \includegraphics[scale=0.38] {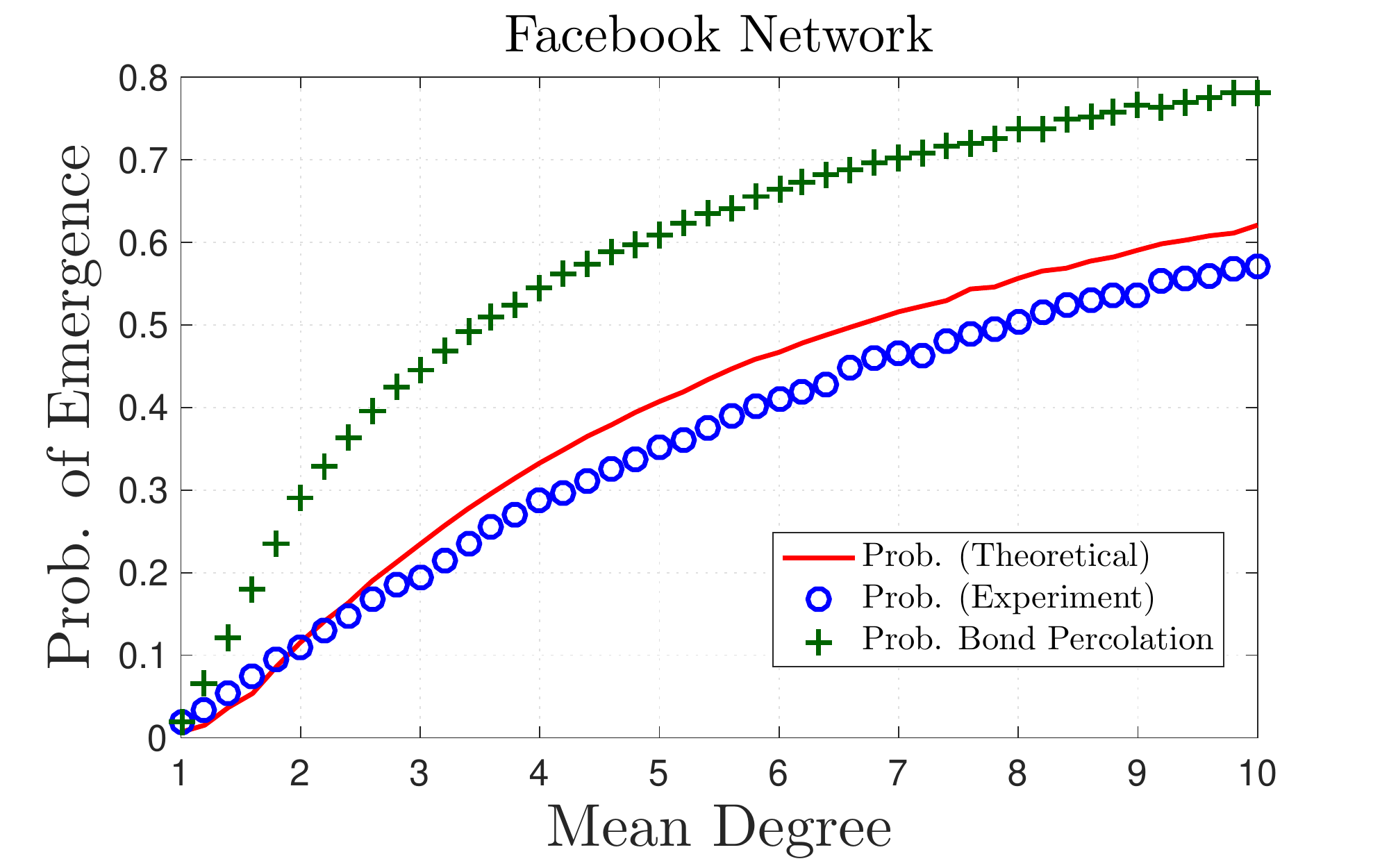}}
    \subfigure[]{\hspace{-0.4cm} 
    \includegraphics[scale=0.38] {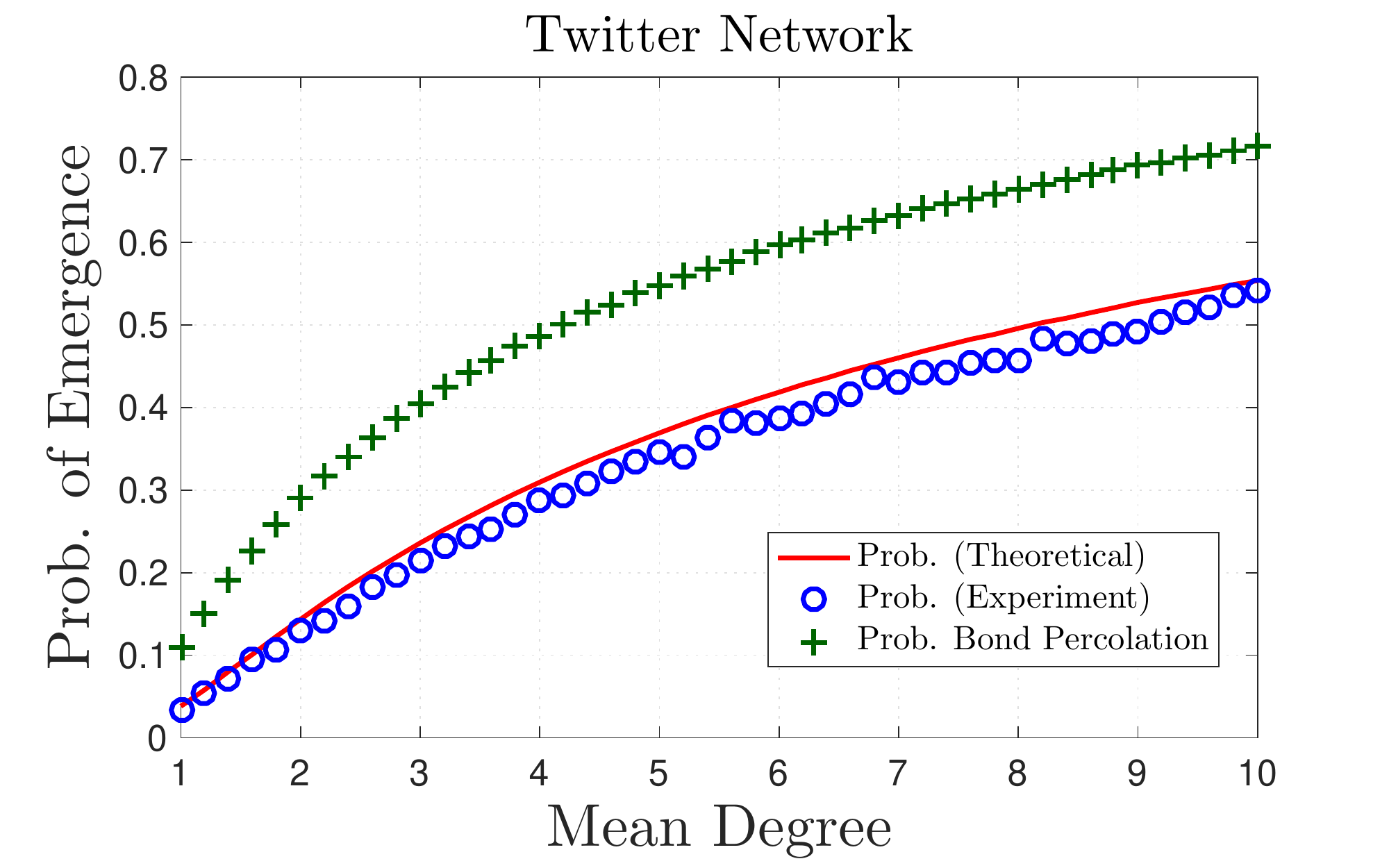}}
     \subfigure[]{\hspace{-0.4cm} 
    \includegraphics[scale=0.38] {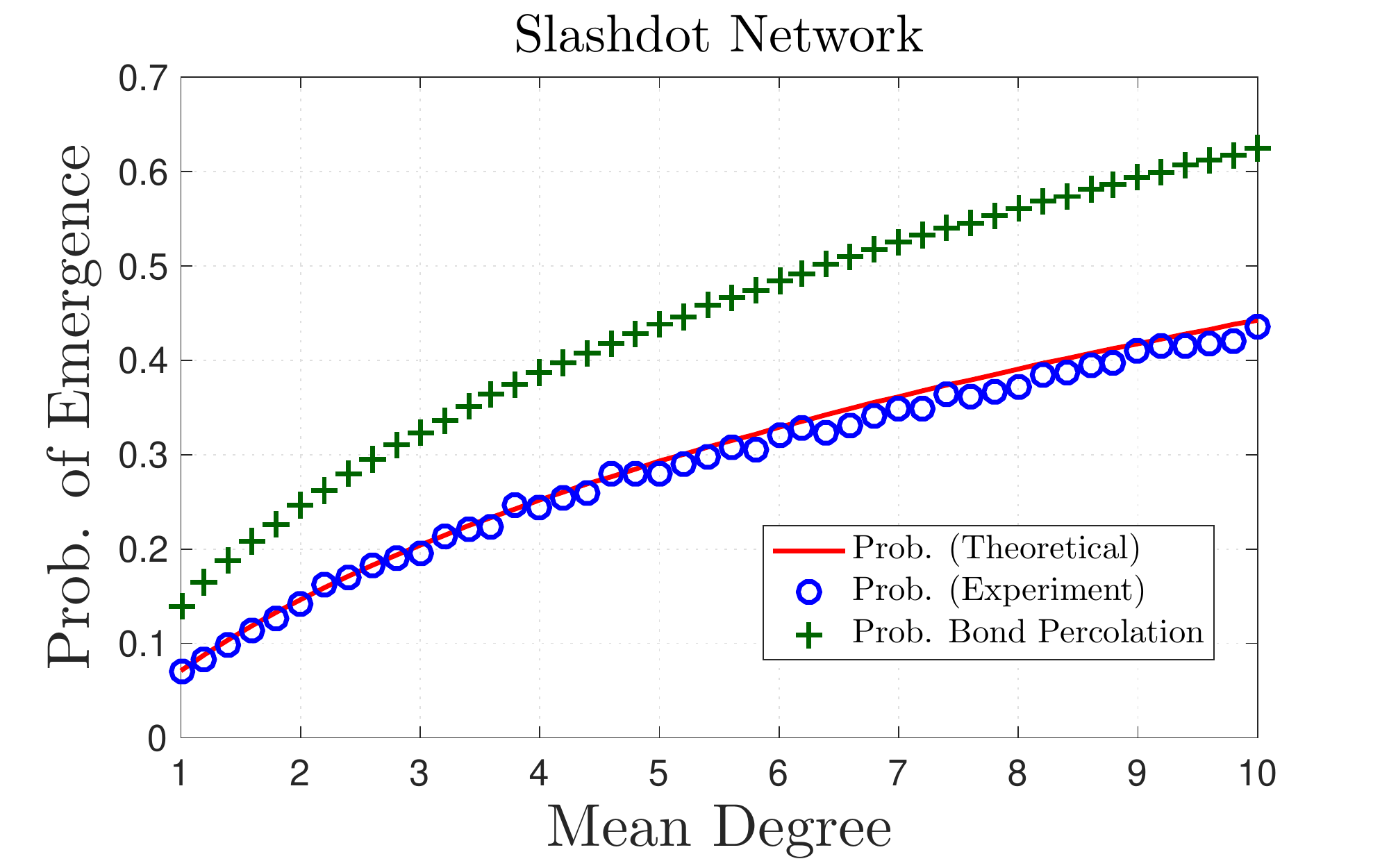}}
     \subfigure[]{\hspace{-0.4cm} 
    \includegraphics[scale=0.38] {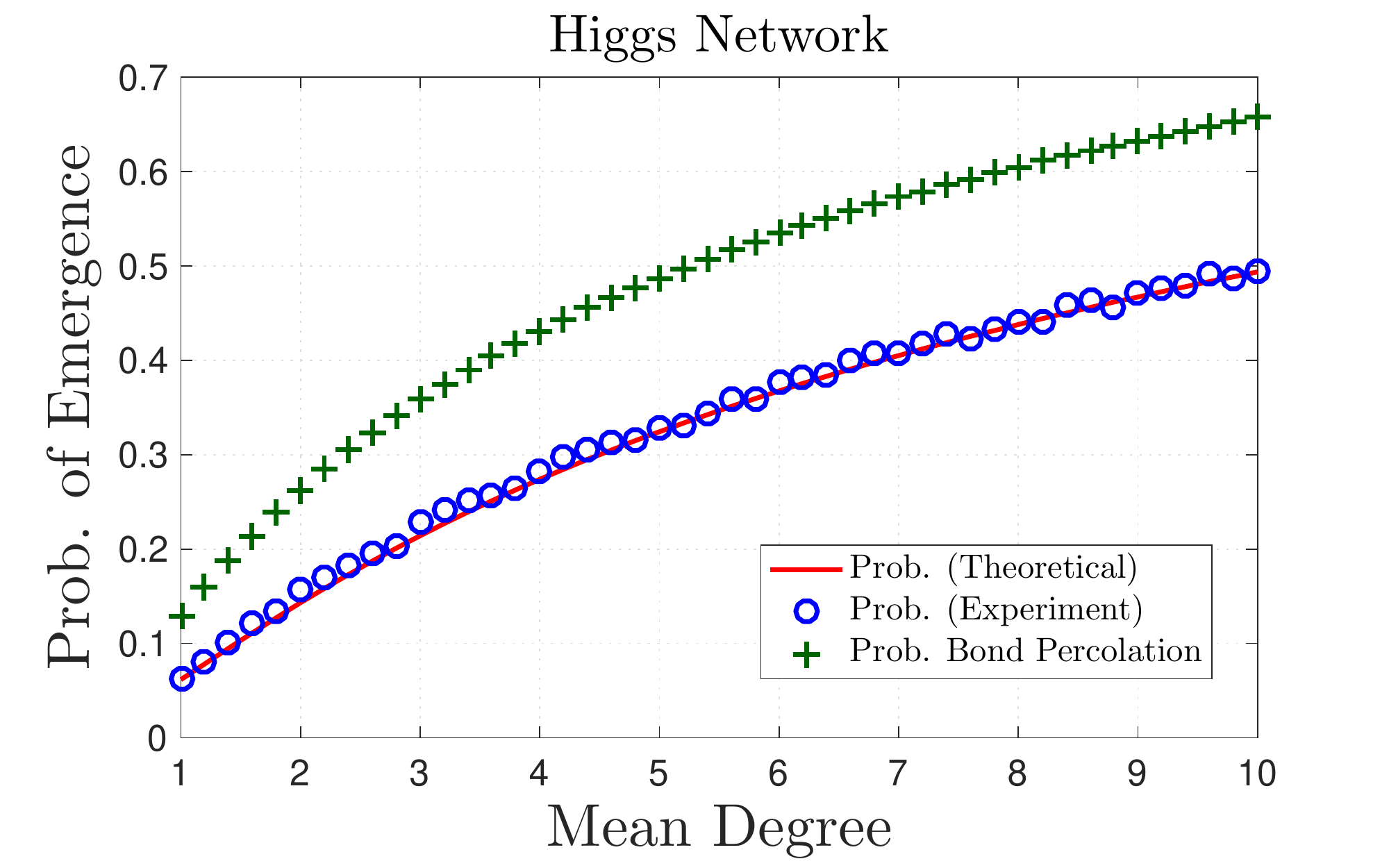}}
    \subfigure[]{\hspace{-0.4cm} 
    \includegraphics[scale=0.38] {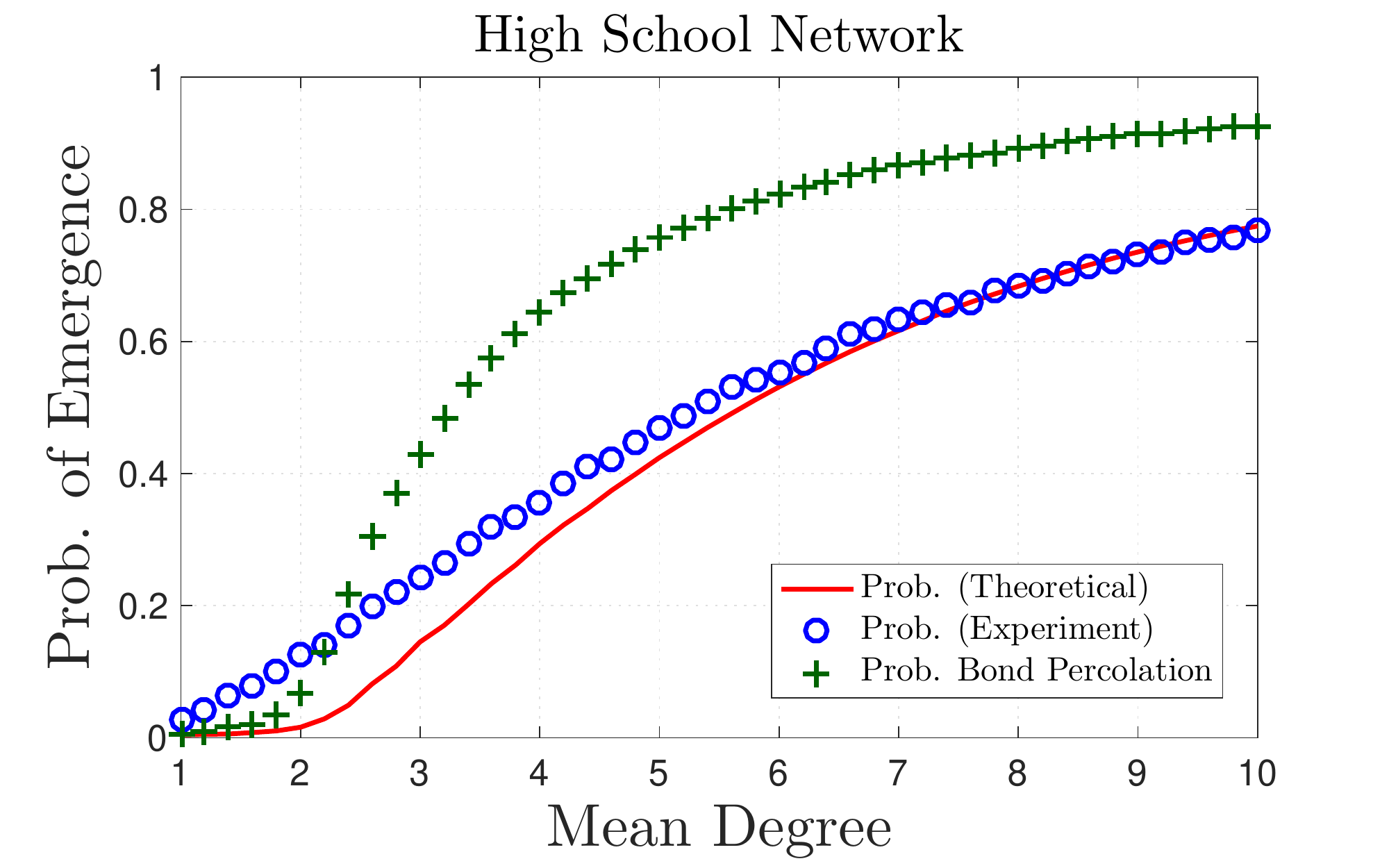}}
    \subfigure[]{\hspace{-0.4cm} 
    \includegraphics[scale=0.38] {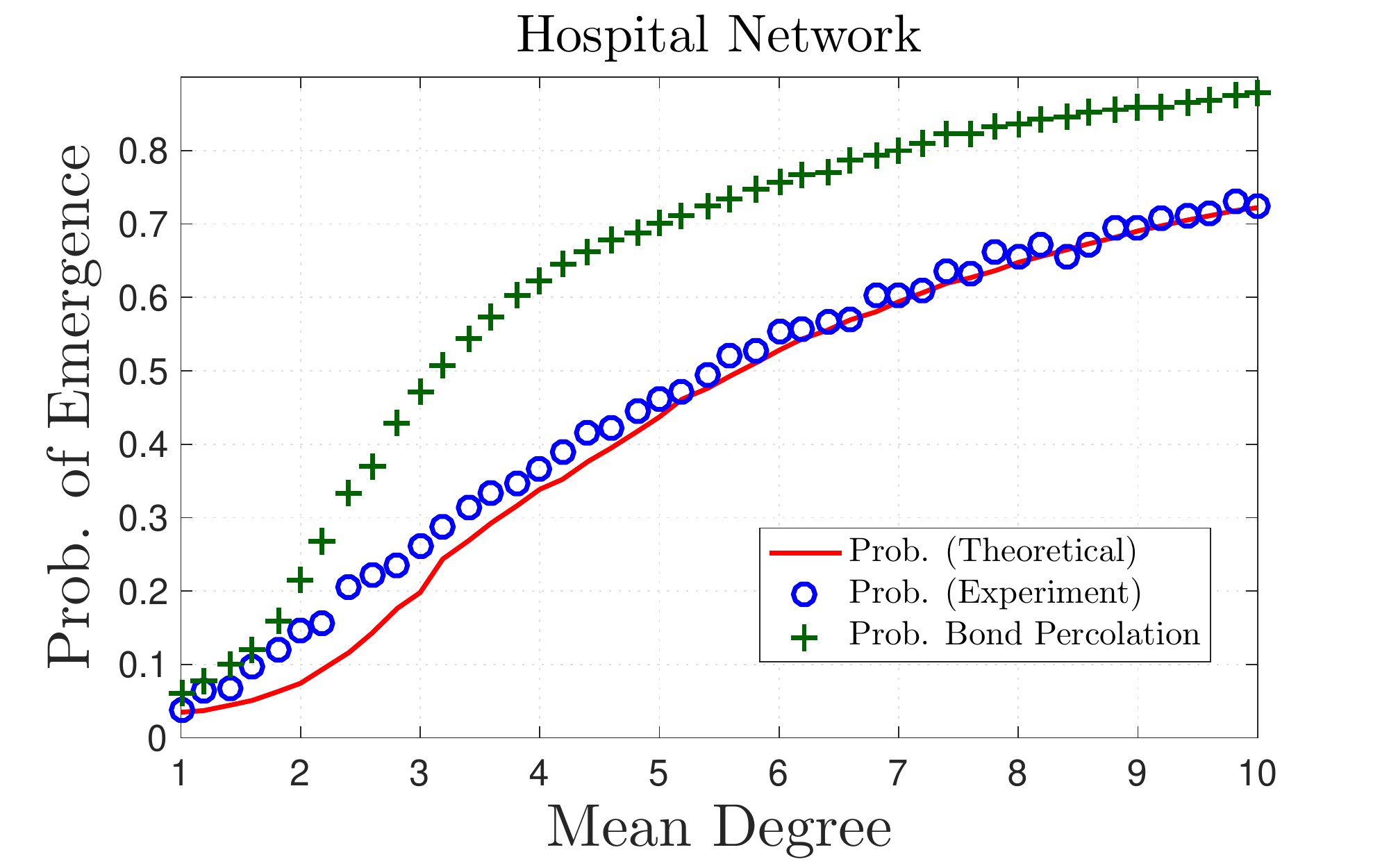}}
    \vspace{4mm}
    \caption{\sl 
             {\bf The probability of emergence on real-world contact networks.} We set $T_1=0.2$, $T_2=0.5$, $\mu_{11}=\mu_{22}=0.75$ (hence $T_{\mathrm{BP}}=0.4$) and vary the mean degree, denoted $\lambda$, from $1$ to $10$. For each value of $\lambda$, we remove a random subset of edges such that the resulting graph is of mean degree $\lambda$ (approximately). The sampled networks still exhibit higher clustering coefficient as compared to random networks with the same mean degree. The single-type bond-percolation framework provides inaccurate predictions on the probability of emergence, in contrast to the multiple-strain formalism given by Alexander and Day \cite{alexander2010risk}. The multiple-strain formalism offers remarkably accurate predictions on a class of real-world networks with low clustering coefficient.
        }
\label{fig:RealWorld}
\end{figure}

{\bf Dataset:}
In the context of information propagation, we consider four different contact networks obtained from SNAP \cite{snapnets}. In particular, we consider the following contact networks:
\begin{itemize}
\item[-] \textsc{Facebook} \cite{snapnets, snapFB}: The contact network among the friends of $10$ users (including those $10$ users). 
\item[-] \textsc{Twitter} \cite{snapnets, snapFB}: The contact network among the friends of $1000$ users (including those $1000$ users). 
\item[-] \textsc{Slashdot} \cite{snapnets, snapSlashdot}:  The network contains friend/foe links between the users of Slashdot.
\item[-] \textsc{Higgs} \cite{snapnets, snapHiggs}: The Higgs data set has been collected upon monitoring the spreading processes on Twitter before, during and after the announcement of the discovery of a new particle with the features of the elusive Higgs boson on July 4, 2012. Nodes correspond to the authors of the collected tweets and edges represent the followee/follower relationships between them.
\end{itemize}

In the context of infectious disease propagation, we consider the following two contact networks:
\begin{itemize}
\item[-] High school network \cite{Salath22020}: The contact network observed at a US high school during a typical school day. The dataset covers $762,868$ interactions between students, teachers, and staff. Each interaction between two individuals is characterized by their identification numbers as well as the duration of the interaction. Two individuals could have multiple interactions throughout the day, and we sum the durations of these interactions to calculate the total contact time between these two individuals over the whole day. We proceed by sampling a {\em static} graph out of this dataset, by assigning an edge between nodes $u$ and $v$ with probability $t_{uv}/t_{\mathrm{max}}$ where $t_{uv}$ denotes the total contact time between nodes $u$ and $v$ throughout the day and $t_{\mathrm{max}}$ denotes the maximum total contact time observed in the dataset.

\item[-] Hospital network \cite{Vanhems2013}: The contact network observed in a short stay geriatric unit of a university hospital in Lyon, France. The dataset covers five days of interactions between professional staff members and patients. Similar to the high school network, we compute the total contact time between two individuals (over the span of five days), then we sample a static graph out of the dataset, by assigning an edge between nodes $u$ and $v$ with probability $t_{uv}/t_{\mathrm{max}}$.
\end{itemize}

More details on the networks, including their clustering coefficients are given in Figure~\ref{fig:table}. We assume that all edges are unidirectional.

\subsection{Methods}
To conduct a fair comparison between the formalism given in Section~\ref{sec:analysis}.A and the single-type bond percolation framework, we fix the parameters of the transmissibility matrix $\pmb{T}$ and the mutation matrix $\pmb{\mu}$, hence fixing $\rho \left( \pmb{T \mu} \right) $ and $T_{\mathrm{BP}}$ (according to (\ref{eq:matchingCondition})). We vary the mean degree, denoted $\lambda$, for each of the contact networks between $1$ and $10$. For each value of $\lambda$, we remove a random subset of edges such that the resulting network is of mean degree $\lambda$ (approximately). Note that the random removal of edges would indeed lower the clustering coefficient of the network, however, the resulting subgraph would remain highly clustered compared to random networks with the same mean degree (see Figure~\ref{fig:table}). In other words, the sampled networks still exhibit specific structural properties that distinguish them from synthetic contact networks generated randomly by the configuration model (with Poisson degree distribution of the same mean degree). After the mean degree is adjusted, the process proceeds similar to Section~\ref{sec:numerical}.B.

\subsection{Results}
In Figure~\ref{fig:RealWorld}, we plot the probability of emergence for the four contact networks shown in Figure~\ref{fig:table}. We compare the results obtained by computer simulations with those obtained by the multiple-strain formalism (Section~\ref{sec:analysis}.A) and the single-type bond-percolation framework. We set $T_1 = 0.2$, $T_2 = 0.5$, and $\mu_{11}=\mu_{22}=0.75$. It follows that $T_{\mathrm{BP}} = 0.4$ according to (\ref{eq:matchingCondition}). 

Similar to our observations on random networks (Section~\ref{sec:numerical}.E), the single-type, bond-percolation framework provides significantly inaccurate predictions on the probability of emergence, should the underlying process entail evolution. The limitation is universal as it applies to both random and real-world networks. Appendix~\ref{app:explanation} explains the intuition behind our observations. In contrast, the multiple-strain formalism provides remarkably accurate predictions, especially on contact networks with low clustering coefficient. Note that the multi-type branching framework assumes that the underlying graph is tree-like; an assumption that holds for networks with small clustering coefficient. Hence, one could reasonably argue that the multiple-strain formalism would provide high prediction accuracy on such networks.

%\begin{figure}[t]
%	\centering
%    \includegraphics[width=0.45\textwidth]{facebookVisual}
%    \caption{\sl
%    A
%        }
%	\label{fig:facebookVisual}
%\end{figure}

\section{Co-infection controls the order of phase transition}
\label{sec:coinfection}

The preceding discussion considers the case when {\em co-infection} is not possible, hence each infected host either carries strain-$1$ or strain-$2$, but not both. However, humans, animals, plants, and other organisms may become {\em co-infected} with multiple pathogen strains, causing major consequences for both within- and between-host disease dynamics \cite{susi2015co, balmer2011prevalence, read2001ecology, cohen2012mixed, alizon2013multiple, de2005dynamics}. For instance, in the case of human malaria, the majority of infected adults are {\em simultaneously} infected by more than five strains of {\em Plasmodium falciparum} \cite{alizon2013multiple, lord1999aggregation}. The competition and interaction patterns between the resident strains trigger significant ramifications of the disease dynamics. Also, the aggregate virulence experienced by the co-infected host could be higher than the most virulent strain, or lower than the least virulent strain, or anywhere in between \cite{alizon2013multiple, cezilly2014cooperation, tollenaere2016evolutionary, lass2013generating}. {
Co-infection also applies in the context of information propagation. Observe that with the growing number of news outlets, we may come across various variants of information on social media platforms. Similar to the case of infectious diseases, these variants may reinforce or weaken each other based on whether they share the same bias or not.}

In this section, we seek to shed the light on the effects of co-infection on { information/disease} propagation. In particular, we investigate the extent to which co-infection dynamics could enhance or suppress the scale of epidemics. Of particular interest is whether co-infection could change the order of phase transition from second-order (as it is the case with most epidemic models) to first-order, leading to a phenomenon that is commonly described as {\em avalanche outbreaks} \cite{cai2015avalanche}. To that end, we extend the multiple-strain model given in Section~\ref{sec:model} to account for co-infection. In particular, a susceptible individual who comes into infectious contacts with type-$1$ and type-$2$ hosts {\em simultaneously} becomes {\em co-infected} and starts to spread the {\em co-infection}. Henceforth, we consider the case when the co-infection has its own transmissibility $T_{co}$ and does not mutate back to either strain-$1$ or strain-$2$. In other words, a co-infected host infects each of her neighbors independently with probability $T_{co}$, and infected neighbors are deemed {\em co-infected} with probability $1$.

\begin{figure}[t!]
    \centering
    \subfigure[]{\hspace{-0.4cm} 
    \includegraphics[scale=0.38] {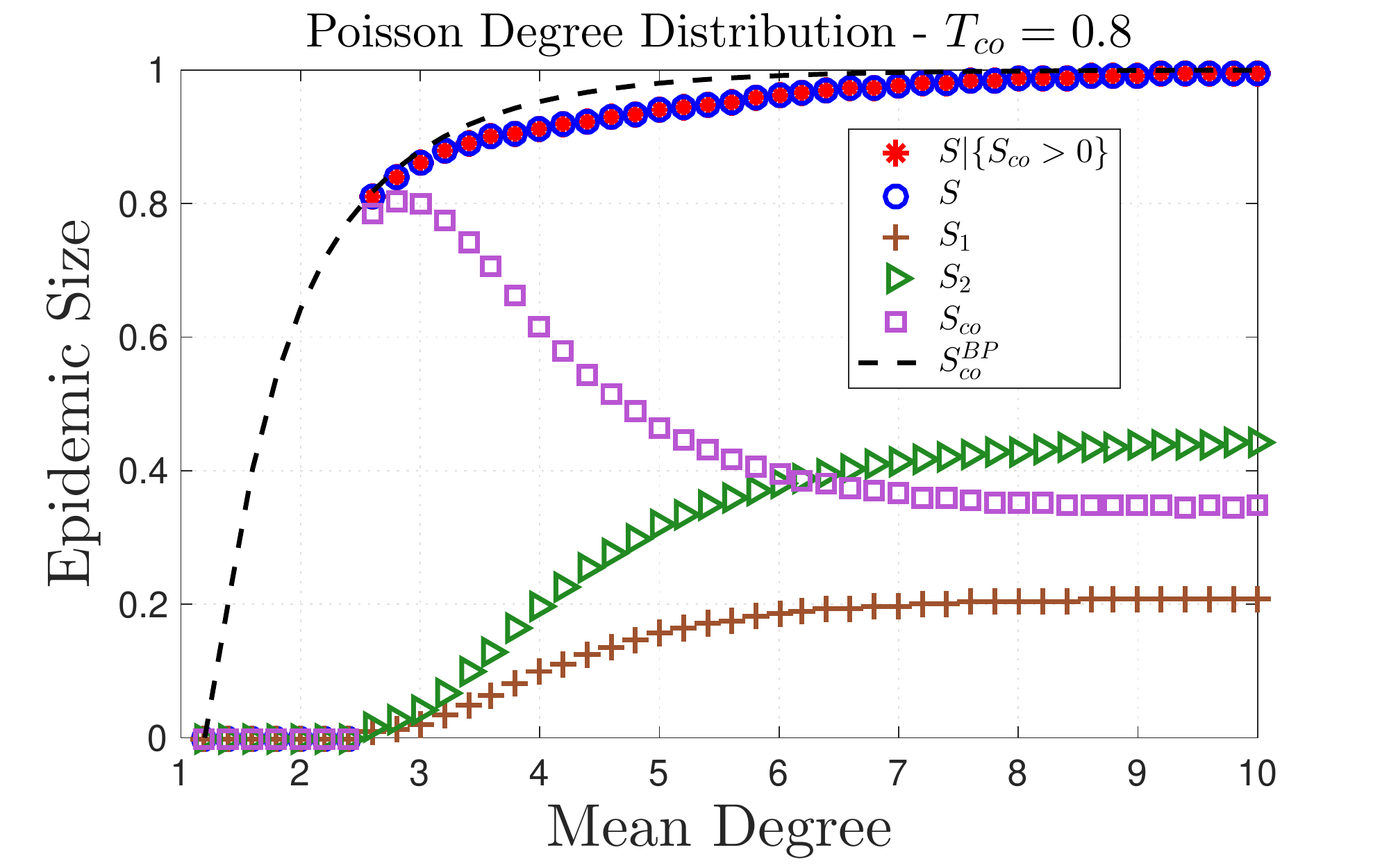}}
    \subfigure[]{\hspace{0.4cm} 
    \includegraphics[scale=0.38] {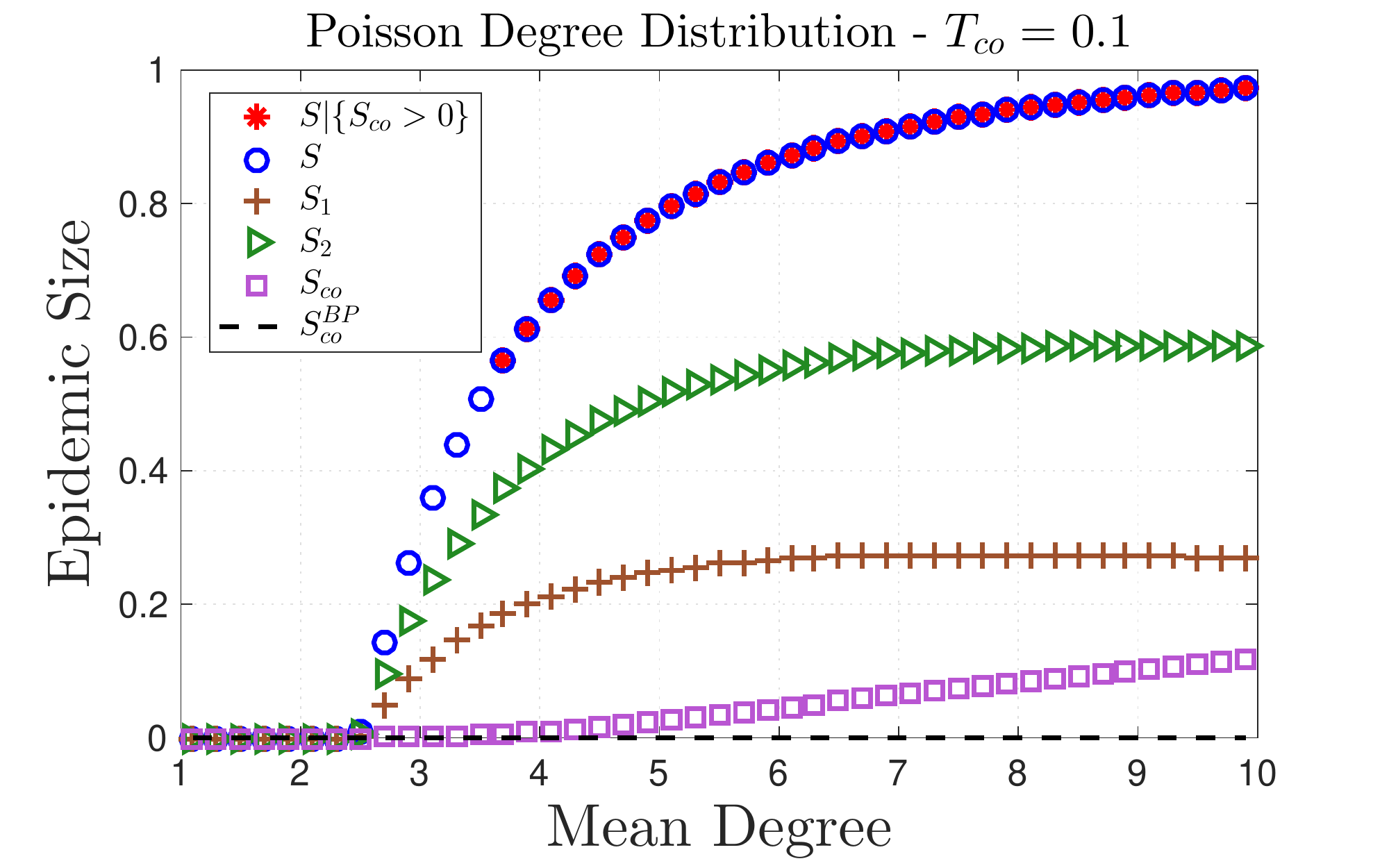}}
     \subfigure[]{\hspace{-0.4cm} 
    \includegraphics[scale=0.38] {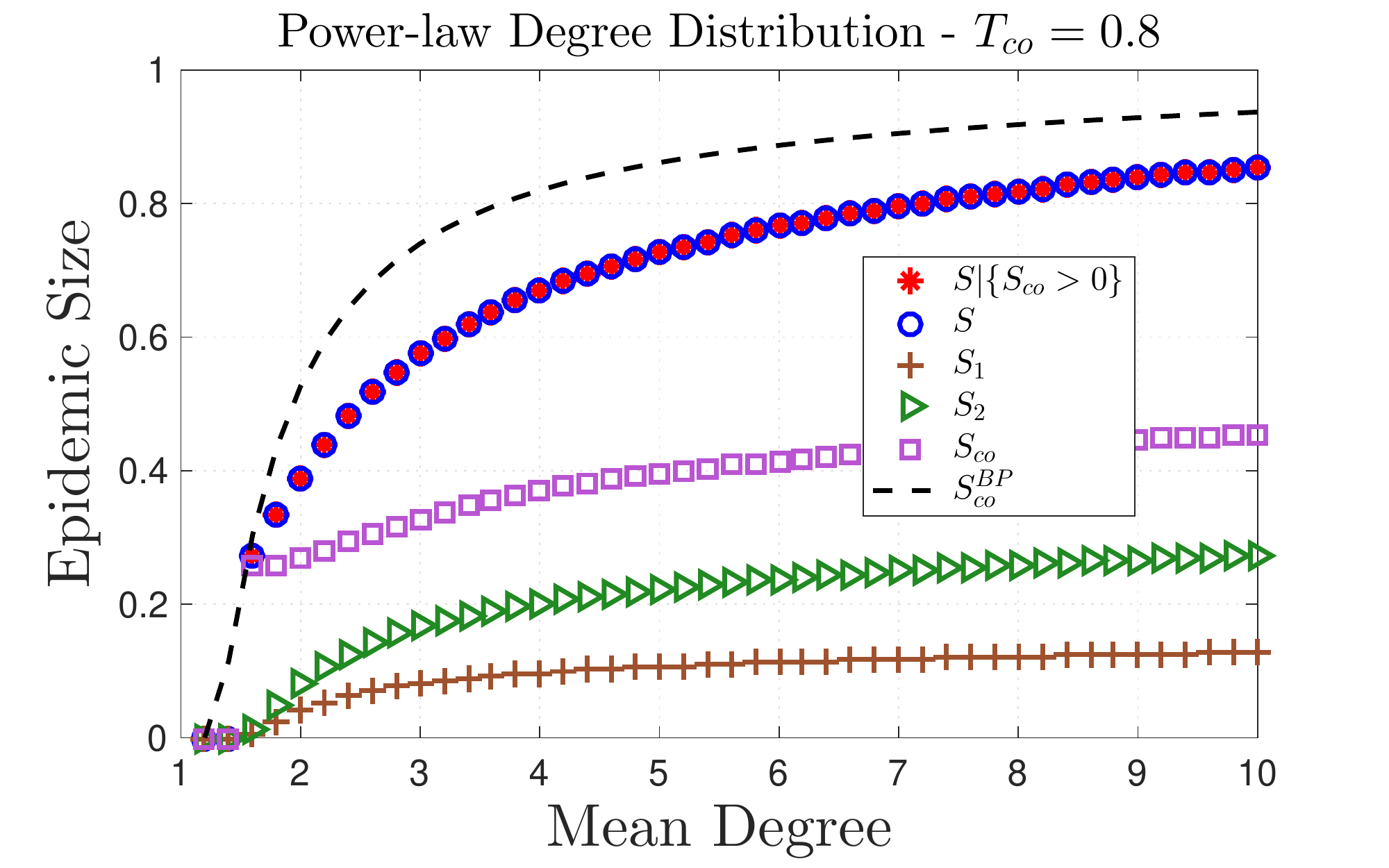}}
     \subfigure[]{\hspace{0.4cm} 
    \includegraphics[scale=0.38] {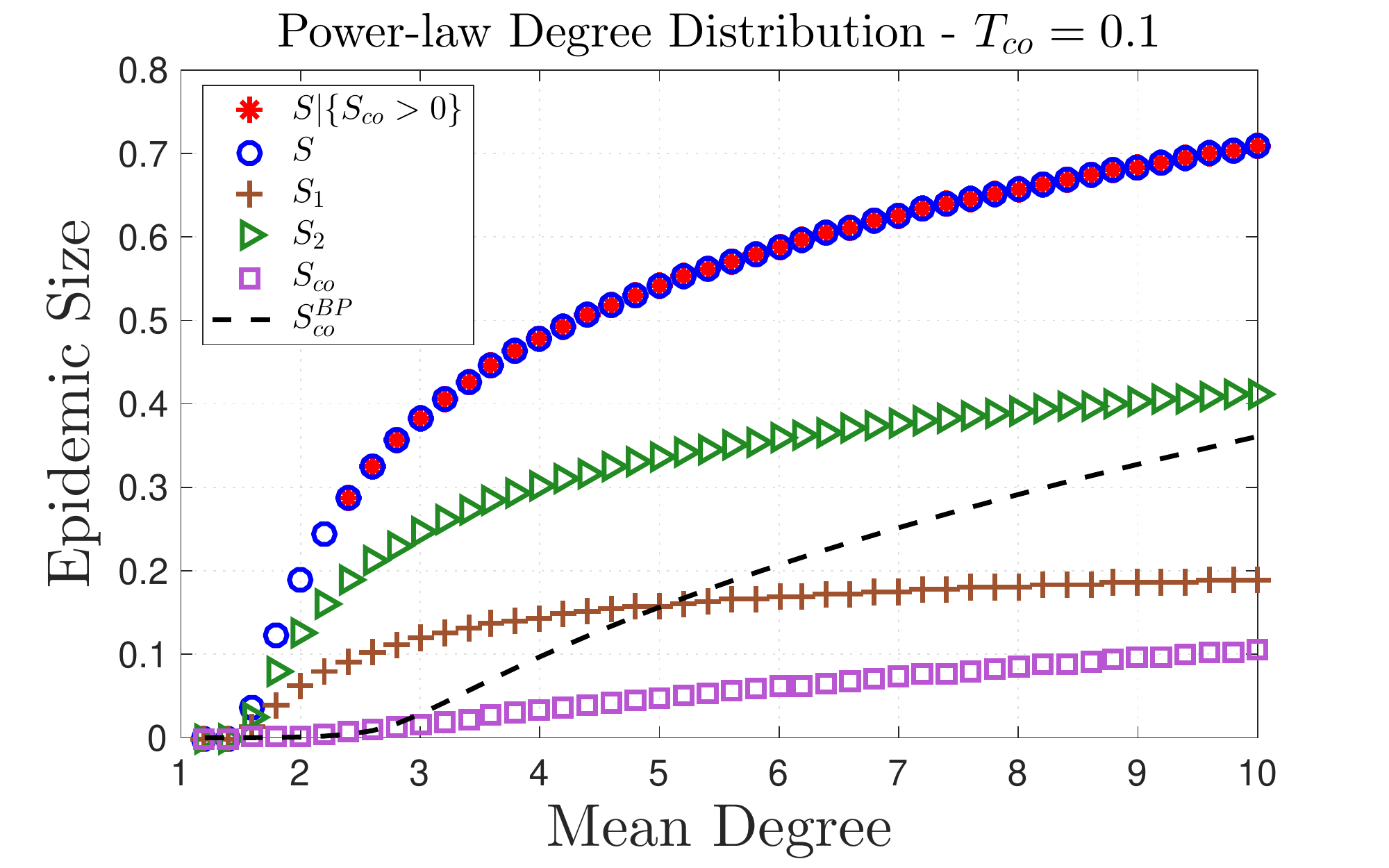}}
    \vspace{4mm}
    \caption{\sl 
    {\bf Co-infection dynamics determine the order of phase transition}. We set $T_1=0.2$, $T_2 = 0.5$, and $\mu_{11}=\mu_{22}=0.75$ for all subfigures.  The network size $n$ is $2\times10^6$ and the number of independent experiments for each data point is $5 \times 10^3$. Blue circles denote the average total epidemic size $S$ and red stars denote the average total epidemic size $S$ conditioned on $S_{co}$ being greater than zero, i.e., conditioned on the existence of a positive fraction of co-infected nodes. Blue plus signs, orange triangles, and yellow squares denote the fraction of nodes infected with strain-$1$, strain-$2$, and co-infection, respectively. The black dashed-line denotes the epidemic size for a single-strain, bond-percolated network with $T_{co}$, i.e., $S^{BP}_{co}$. (a) and (c): A first order phase transition is observed when $T_{co}=0.8$ owing to the corresponding first order transition of $S_{co}$. Co-infection emerges at the phase transition point that characterizes an epidemic of strain-$1$ and strain-$2$. At this point, the value of $S_{co}$ jumps discontinuously to (approximately) the corresponding value of $S^{BP}_{co}$ with $T_{co}=0.8$. Observe that $S^{BP}_{co}>0$ at the transition point, hence, a first-order phase transition is observed. (b) and (d): Co-infection still emerges right at the phase transition point. However, since $T_{co}$ is small, $S^{BP}_{co}=0$ at the transition point. Hence, a second-order phase transition is observed.
        }
\label{fig:CoFull}
\end{figure}

As with Section~\ref{sec:numerical}, we consider contact networks with Poisson degree distribution and Power-law degree distribution with exponential cutoff, respectively. For both cases, we set $T_1=0.2$, $T_2 = 0.5$, and $\mu_{11}=\mu_{22}=0.75$. Moreover, we set the network size to $2 \times 10^6$ and the number of independent experiments for each data point to $5 \times 10^3$. To illustrate how co-infection dynamics control the order of phase transition, we simulate and compare the process for two values of $T_{co}$, namely $T_{co} = 0.1$ and $T_{co} = 0.8$. Finally, we plot the epidemic size, denoted by $s^{BP}_{co}$, for a single-strain, bond-percolated network \cite{newman2002spread}.

In all cases, co-infection emerges at the phase transition point that characterizes an epidemic of strain-$1$ and strain-$2$, i.e., the mean degree for which $\rho(\pmb{M})=1$, where $\pmb{M}$ is given by
\begin{equation}\nonumber
\pmb{M} = 
 \left( \dfrac{ \langle k^2  \rangle -  \langle k  \rangle}{ \langle k  \rangle} \right)
\left[
\begin{matrix}
T_1 & 0 \\
0 & T_2 \\
\end{matrix}
\right]
\left[
\begin{matrix}
\mu_{11} & \mu_{12} \\
\mu_{21} & \mu_{22} \\
\end{matrix}
\right]
\end{equation}

As seen in Figure~\ref{fig:CoFull}, a {\em first-order} phase transition is observed on both contact networks when $T_{co}=0.8$ due to the corresponding first order transition of $S_{co}$. In particular, the value of $S_{co}$ jumps discontinuously from zero to (approximately) the corresponding value of $S^{BP}_{co}$ for a single-strain, bond-percolated network with $T_{co}=0.8$. Hence, a first-order phase transition is observed. In general, we conjecture that a first-order phase transition emerges whenever $T_{co}$ is large enough such that $S^{BP}_{co}>0$ at the critical point $\rho(\pmb{M})=1$. If, however, $T_{co}$ is small such that $S^{BP}_{co}=0$ when $\rho(\pmb{M})=1$, then a second-order phase transition is observed. This is confirmed by our simulation results for the case when $T_{co}=0.1$.

\begin{figure}[t!]
    \centering
    \subfigure[]{\hspace{-0.4cm} 
    \includegraphics[scale=0.38] {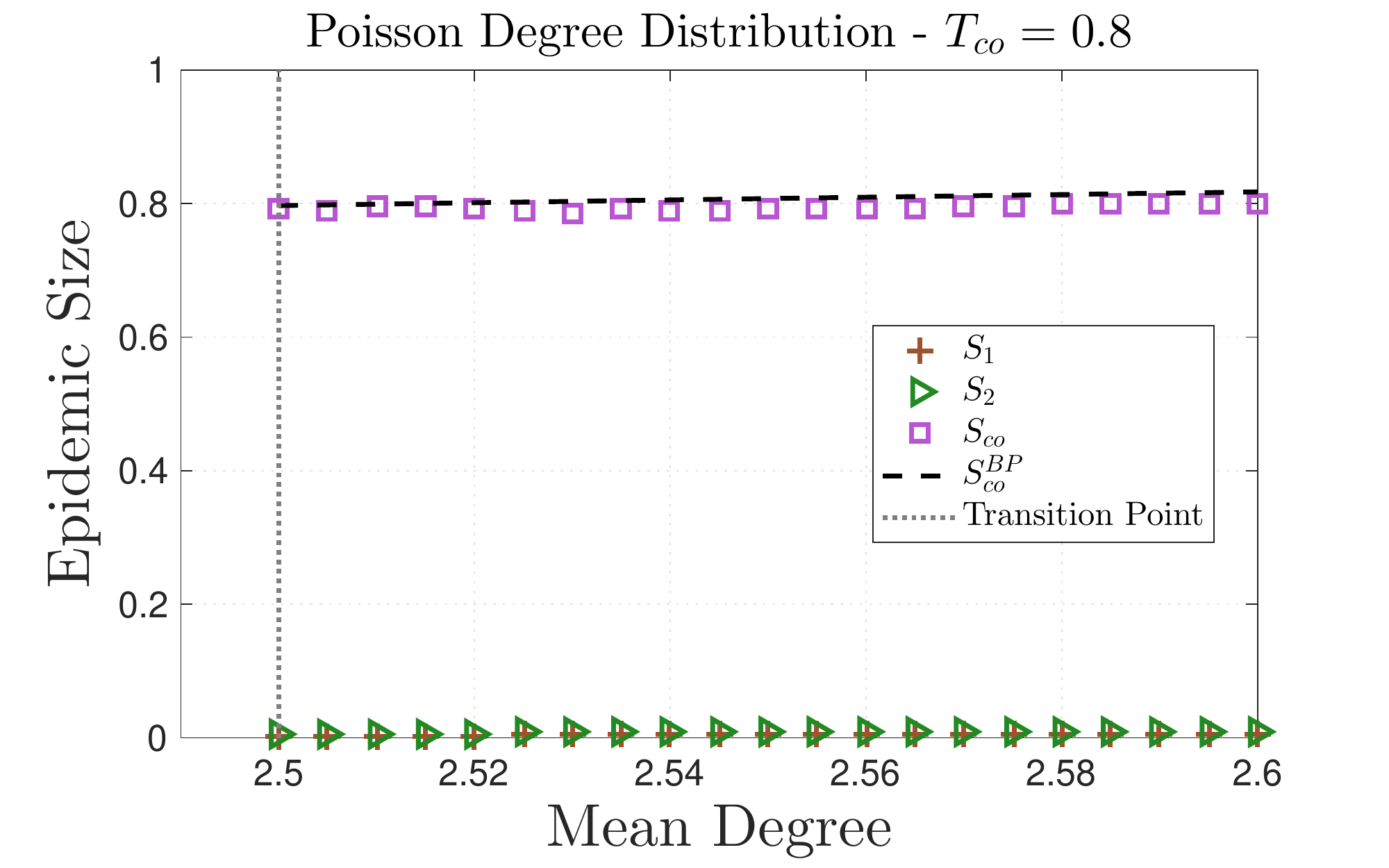}}
    \subfigure[]{\hspace{0.4cm} 
    \includegraphics[scale=0.38] {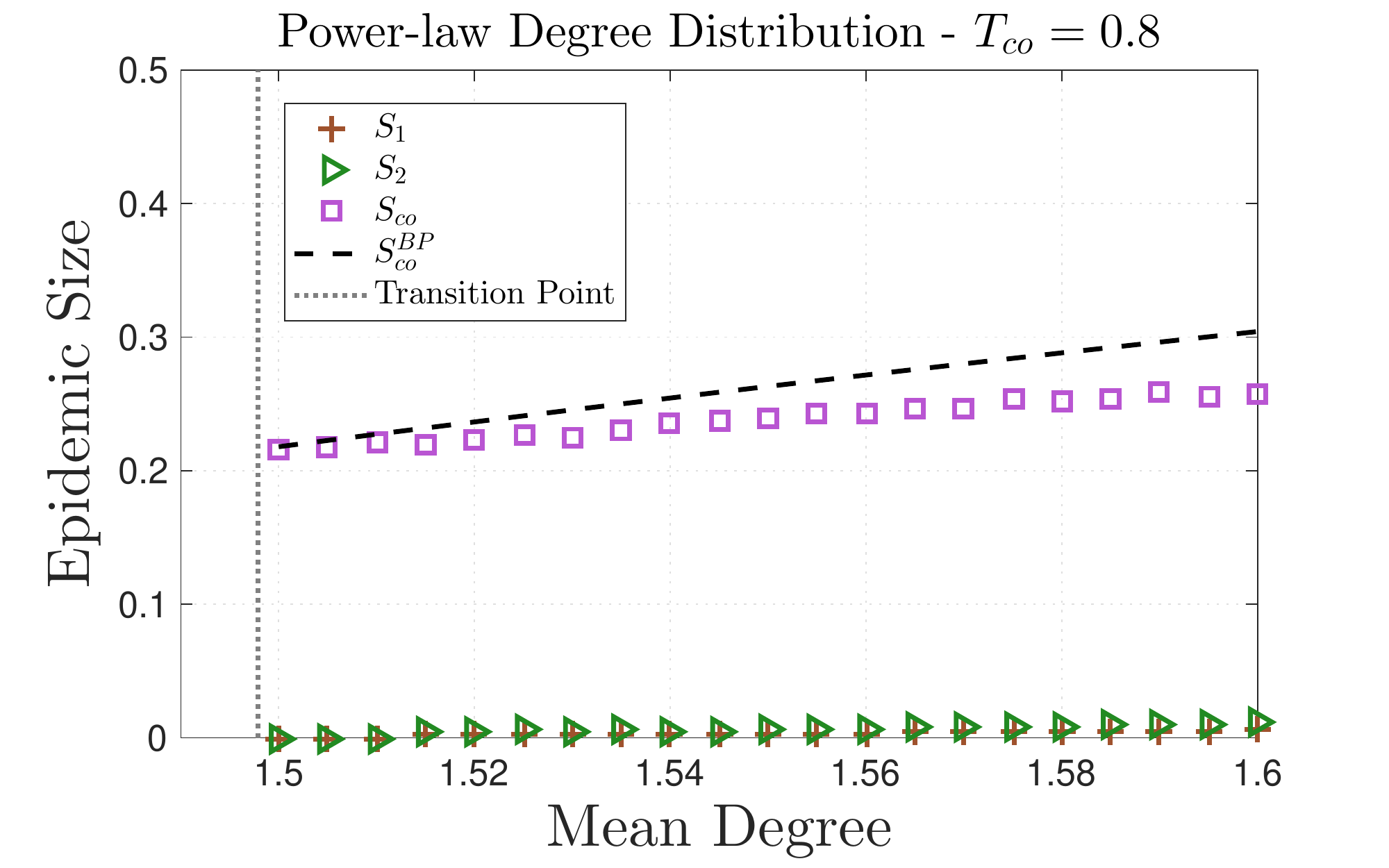}}
    \vspace{4mm}
    \caption{\sl 
    {\bf Validating the order of phase transition}. We set the network size $n$ to $15 \times 10^6$, the number of independent experiments for each data point to $10^4$, $T_1 = 0.2$, $T_2=0.5$, and $\mu_{11}=\mu_{22}=0.75$. Our results confirm that the phase-transition is indeed first order on both contact networks. The value of $S_{co}$ jumps discontinuously to (approximately) the corresponding value of $S^{BP}_{co}$ with $T_{co}=0.8$.
        }
\label{fig:CoZoom}
\end{figure}

In order to validate the order of phase transition when $T_{co}=0.8$, we conduct an extensive simulation study around the phase transition point on both contact networks. In Figure~\ref{fig:CoZoom}, we set the number of nodes $n$ to $15 \times 10^6$ (to alleviate finite size effects) and the number of experiments to $10^4$ for each data point. We use the same parameters that were used to generate Figure~\ref{fig:CoFull}, i.e., $T_1 = 0.2$, $T_2 = 0.5$, and $\mu_{11}=\mu_{22}=0.75$. Our results confirm that the phase-transition is indeed first order on both contact networks. In fact, the value of $S_{co}$ jumps discontinuously to (approximately) the corresponding value of $S^{BP}_{co}$ with $T_{co}=0.8$.

\section{Conclusion}
\label{sec:conclusion}
In this paper, we have investigated the {\em evolution} of spreading processes on complex networks and developed a mathematical theory that unravels the relationship between the characteristics of the spreading process, evolution, and the structure of the contact network on which the process spreads. Our mathematical theory was complemented by an extensive simulation study on both random and real-world contact networks. The simulation results proved the validity of our theory and revealed the significant shortcomings of the classical mathematical models that do not capture evolution. A matching condition between single- and multiple-strain models was proposed and evaluated in the context of probability of emergence, epidemic size, and epidemic threshold. Under the proposed matching condition, our results revealed that the classical bond-percolation models may accurately predict the threshold and final size of epidemics that entail evolution, but their predictions on the probability of emergence are {\em significantly inaccurate} on both random and real-world networks. Hence, our formalism is necessary to bridge the disconnect between how spreading processes propagate and evolve on complex networks, and the current mathematical models that do not capture evolution. 

We proceeded by deriving a lower bound on the probability of emergence to gain further insights on the effects of mutation. The bound was derived for the special case of one-step irreversible mutation. Our results revealed that the probability of mutation plays a key role in determining the shape and behavior of the probability of emergence. Moreover, the way the particular value of $\mu$ influences the probability of mutation varies according to the connectivity of the underlying contact network. Finally, we considered the case when {\em co-infection} is possible and showed that co-infection dynamics control the order of phase transition in an interesting way. In particular, depending on co-infection dynamics, the order of phase transition of the epidemic size could change from second-order to {\em first-order}, in contrast to the universality class of percolation models that are typically second-order.

\vspace{5mm}
\section*{Acknowledgement}
This work has been supported (in part) by the National Science Foundation through grant CCF-1813637, (in part) by the Army Research Office through grant W911NF-17-1-0587, and (in part) by the Office of Naval Research through grants N0001418SB001 and N000141512797. The first author was funded in part by the Dowd Fellowship from the College of Engineering at Carnegie Mellon University. The authors would like to thank Philip and Marsha Dowd for their financial support and encouragement. The first author would like to thank Ms. Mary Turocy from the School of Medicine at University of California San Francisco for her helpful and constructive comments.

\bibliographystyle{IEEEtran}
\bibliography{main_article}

\appendix
\section{Correlations of infection events}
\label{app:explanation}

We have shown that the inability of the single-type bond-percolation framework to predict the probability of emergence is universal; it is observed on both random and real-world contact networks. The universality of the behavior suggests that single-type bond-percolation framework does not properly capture a fundamental property of spreading processes that entail evolution. Below, we argue that this property is stemming from the underlying {\em correlations} between the infection events of the multiple-strain model. For reasons that will become apparent soon, it is useful to draw parallels between the multiple strain model proposed by Alexandar and Day \cite{alexander2010risk} and the single-strain model proposed by Newman in \cite{newman2002spread}.

In \cite{newman2002spread}, Newman proposed a stochastic SIR model where the probability that an infected node $i$ infects a susceptible node $j$ is given by $T_{i j}=1-\mathrm{exp}(- \beta_{i j} \tau_{i})$, where $\beta_{ij}$ denotes the rate of infectious contacts from node $i$ to node $j$ and $\tau_i$ denotes the infectious period of node $i$, i.e., the period of time during which node $i$ remains infective. The infectious period $\tau_i$ is a random variable with a Cumulative Distribution Function (CDF) $F_\tau(u)$, and the infectious contact rate $\beta_{ij}$ is also a random variable with a CDF $F_\beta(v)$. Newman claimed that under the assumptions that i) the infectious contact rates between individuals are independent and identically distributed (i.i.d) and that ii) the infectious periods for all individuals are also i.i.d., the spread of a diseases on a contact network is isomorphic to a bond-percolation model on the contact network with a bond percolation parameter given by
\begin{equation}\nonumber
T = \langle T_{ij} \rangle = 1 - \int_{0}^\infty e^{- \beta \tau} dF_\beta(\beta) dF_\tau(\tau)
\end{equation}
where $T$ was called the {\em transmissibility} of the disease. The isomorphism to a bond-percolation problem allowed for the use of generating functions to derive the threshold, probability, and final size of epidemics on a contact network with arbitrary degree distributions.  

Later on, Kenah and Robins \cite{kenah2007second} proved that this isomorphism to a bond-percolation problem is valid only when the distribution of the infectious periods is {\em degenerate}, i.e., $\tau_i = \tau_0$ for all $i=1,2,\ldots$, where $\tau_0$ is a constant. Kenah and Robins showed that when the distribution of the infectious periods is non-degenerate, there is no bond-percolation probability that will make the bond-percolation model isomorphic to the SIR model. The fundamental reason behind their findings is the fact that the infection events across edges emanating from node $i$ are {\em conditionally} independent given $\tau_i$, but {\em marginally} dependent unless $\tau_i = \tau_0$ with probability one. That said, Kenah and Robins showed that even when the distribution of the infectious periods is non-degenerate, the mapping to a bond-percolation process can still be used to accurately predict the epidemic threshold and epidemic size.

The multiple-strain model presented by Alexander and Day exhibits a similar form of correlations between infection events. In particular, infection events are conditionally independent given the type of the infective node. Namely, conditioned on node $i$ being infected with strain-$\ell$, node $i$ infects each of her neighbors {\em independently} with probability $T_\ell$. However, infection events are marginally dependent, unless $T_i=T_0$ for all $i$ with probability one; a condition that essentially reduces the dynamics to that of single-strain processes without evolution. To give an example, consider a {\em regular} network, where each node has exactly $2$ neighbors. Let $T_1 =1$ and $\mu_{11}=\mu_{21}=\mu$. In this case, we have $T_{\mathrm{BP}}=\mu + T_2 \left(1-\mu \right)$ by virtue of (\ref{eq:matchingCondition}). Now, we can easily compute the probability that an infection of a
randomly selected node results in an outbreak of size {\em one}. Under the bond percolation framework, this is given by $ \left(1- T_{\mathrm{BP}}\right)^2= \left(1-\mu-T_2 \left(1-\mu \right) \right)^2$. However the multiple-strain formalism predicts a zero probability for this event, should the initial node be infected with strain-$1$. Indeed, the probability predicted by the bond percolation framework will match the one predicted by the multiple-strain formalism only if $T_2=1$ or $\mu=1$; a condition that diminishes the role of evolution and reduces the dynamics into that of single-strain processes.

\end{document}